\makeatletter
\newcommand{\dontusepackage}[2][]{%
  \@namedef{ver@#2.sty}{9999/12/31}%
  \@namedef{opt@#2.sty}{#1}}
\makeatother
\dontusepackage{subfigure}

\documentclass[]{article}
\pdfoutput=1
\usepackage{lmodern}
\usepackage{amssymb,amsmath}
\usepackage{ifxetex,ifluatex}
\usepackage[usenames,dvipsnames]{color}
\usepackage{fixltx2e} 
\ifnum 0\ifxetex 1\fi\ifluatex 1\fi=0 
  \usepackage[T1]{fontenc}
  \usepackage[utf8]{inputenc}
\else 
  \ifxetex
    \usepackage{mathspec}
    \usepackage{xltxtra,xunicode}
  \else
    \usepackage{fontspec}
  \fi
  \defaultfontfeatures{Mapping=tex-text,Scale=MatchLowercase}
  
\fi
\IfFileExists{upquote.sty}{\usepackage{upquote}}{}
\IfFileExists{microtype.sty}{%
\usepackage{microtype}
\UseMicrotypeSet[protrusion]{basicmath} 
}{}
\usepackage[margin=1.0in,bottom=1.5in]{geometry}
\usepackage[numbers]{natbib}
\usepackage{listings}
\usepackage{longtable,booktabs}
\usepackage{graphicx}
\makeatletter
\def\maxwidth{\ifdim\Gin@nat@width>\linewidth\linewidth\else\Gin@nat@width\fi}
\def\maxheight{\ifdim\Gin@nat@height>\textheight\textheight\else\Gin@nat@height\fi}
\makeatother
\setkeys{Gin}{width=\maxwidth,height=\maxheight,keepaspectratio}
\usepackage{caption}
\usepackage{float}
\usepackage{subfig}
\usepackage{algorithm} 
\let\scholmdAlgorithm\algorithm
\let\endscholmdAlgorithm\endalgorithm
\let\algorithm\relax \let\endalgorithm\relax

{
 \catcode`\*=11\relax
 \global\let\scholmdAlgorithm*\algorithm*
 \global\let\endscholmdAlgorithm*\endalgorithm*
 \global\let\algorithm*\relax
 \global\let\endalgorithm*\relax
}
\ifxetex
  \usepackage[setpagesize=false, 
              unicode=false, 
              xetex]{hyperref}
\else
  \usepackage[unicode=true]{hyperref}
\fi
\hypersetup{breaklinks=true,
            bookmarks=true,
            pdfauthor={},
            pdftitle={A Unified 2D/3D Large Scale Software Environment for Nonlinear Inverse Problems},
            colorlinks=true,
            citecolor=black,
            urlcolor=blue,
            linkcolor=black,
            pdfborder={0 0 0}}
\DeclareMathOperator*{\argmin}{arg\,min}

\usepackage[framed,autolinebreaks]{mcode}

\title{A Unified 2D/3D Large Scale Software Environment for Nonlinear Inverse
Problems}
\author{Curt Da Silva\textsuperscript{1,2}, Felix
Herrmann\textsuperscript{2}\\\textsuperscript{1}Department of
Mathematics, University of British Columbia\\\textsuperscript{2}Seismic
Laboratory for Imaging and Modeling (SLIM), University of British
Columbia\\}
\date{}

\begin{document}
\maketitle

\section{Abstract}\label{abstract}

Large scale parameter estimation problems are among some of the most
computationally demanding problems in numerical analysis. An academic
researcher's domain-specific knowledge often precludes that of software
design, which results in inversion frameworks that are technically
correct, but not scalable to realistically-sized problems. On the other
hand, the computational demands for realistic problems result in
industrial codebases that are geared solely for high performance, rather
than comprehensibility or flexibility. We propose a new software design
for inverse problems constrained by partial differential equations that
bridges the gap between these two seemingly disparate worlds. A
hierarchical and modular design allows a user to delve into as much
detail as she desires, while exploiting high performance primitives at
the lower levels. Our code has the added benefit of actually reflecting
the underlying mathematics of the problem, which lowers the cognitive
load on user using it and reduces the initial startup period before a
researcher can be fully productive. We also introduce a new
preconditioner for the 3D Helmholtz equation that is suitable for
fault-tolerant distributed systems. Numerical experiments on a variety
of 2D and 3D test problems demonstrate the effectiveness of this
approach on scaling algorithms from small to large scale problems with
minimal code changes.

\section{Introduction}\label{pdeintroduction}

Large scale inverse problems are challenging for a number of reasons,
not least of which is the sheer volume of prerequisite knowledge
required. Developing numerical methods for inverse problems involves the
intersection of a number of fields, in particular numerical linear
algebra, nonlinear non-convex optimization, numerical partial
differential equations, as well as the particular area of physics or
biology the problem is modelled after, among others. As a result, many
software packages aim for a complete general approach, implementing a
large number of these components in various sub-modules and interfaced
in a hierarchical way. There is often a danger with approaches that
increase the cognitive load on the user, forcing them to keep the
conceptual understanding of many components of the software in their
minds at once. This high cognitive load can result in prolonging the
initial setup time of a new researcher, delaying the time that they are
actually productive while they attempt to comprehend how the code
behaves. Moreover, adhering to a software design model that does not
make intuitive sense can disincentivize modifications and improvements
to the codebase. In an ideal world, a researcher with a general
knowledge of the subject area should be able to sit in front of a
well-designed software package and easily associate the underlying
mathematics with the code they are presented. If a researcher is
interested in prototyping high level algorithms, she is not necessarily
interested in having to deal with the minutia of compiling a large
number of software packages, manually managing memory, or writing low
level code in C or Fortran in order to implement, for example, a simple
stochastic optimization algorithm. Researchers are at their best when
actually performing research and software should be designed to
facilitate that process as easily as possible.

Academic software environments for inverse problems are not necessarily
geared towards high-performance, making use of explicit modeling
matrices or direct solvers for 3D problems. Given the enormous
computational demands of solving such problems, industrial codes focus
on the performance-critical aspect of the problem and are often written
in a low-level language without focusing on proper design. These
software engineering decisions results in code that is hard to
understand, maintain, and improve. Fortran veterans who have been
immersed in the same software environment for many years are perfectly
happy to squeeze as much performance out of their code as possible, but
cannot easily integrate higher-level algorithms in to an existing
codebase. As a result of this disparity, the translation of higher-level
academic research ideas to high-performance industrial codes can be
lost, which inhibits the uptake of new academic ideas in industry and
vice-versa.

One of the primary examples in this work is the seismic inverse problem
and variants thereof, which are notable in particular for their large
computational requirements and industrial applications. Seismic inverse
problems aim to reconstruct an image of the subsurface of the earth from
multi-experiment measurements conducted on the surface. An array of
pressure guns inject a pressure differential in to the water layer,
which in turn generates a wave that travels to the ocean floor. These
waves propagate in to the earth itself, reflect off of various
discontinuities, before traveling back to the surface to be measured at
an array of receivers. Our goal in this problem, as well as many other
boundary-value problems, is to reconstruct the coefficients of the model
(i.e., the wave equation in the time domain or the Helmholtz equation in
the frequency domain) that describes this physical system such that the
waves generated by our model agree with those in our measured data.

The difficulty in solving industrial-scale inverse problems arises from
the various constraints imposed by solving a real-world problem.
Acquired data can be noisy, lack full coverage, and, in the seismic
case, can miss low and high frequencies \citep{ten2013broadband} as a
result of equipment and environmental constraints. Particularly in the
seismic case, missing low frequencies results in a highly-oscillatory
objective function with multiple local minima, requiring a practitioner
to estimate an accurate starting model, while missing high frequencies
results in a loss of detail \citep{virieux2009overview}. Realistically
sized problems involve the propagation of hundreds of wavelengths in
geophysical \citep{gray2001seismic} and earthquake settings
\citep{kanamori1978quantification}, where wave phenomena require a
minimum number of points per wavelength to model meaningfully
\citep{holberg1987computational}. These constraints can lead to large
models and the resulting system matrices become too large to store
explicitly, let alone invert with direct methods.

Our goal in this work is to outline a software design approach to
solving partial differential equation (PDE) constrained optimization
problems that allows users to operate with the high-level components of
the problem such as objective function evaluations, gradients, and
Hessians, irrespective of the underlying PDE or dimensionality. With
this approach, a practitioner can design and prototype inversion
algorithms on a complex 2D problem and, with minimal code changes, apply
these same algorithms to a large scale 3D problem. The key approach in
this instance is to structure the code in a hierarchical and modular
fashion, whereby each module is responsible for its own tasks and the
entire system structures the dependencies between modules in a tiered
fashion. In this way, the entire codebase becomes much easier to test,
optimize, and understand. Moreover, a researcher who is primarily
concerned with the high level `building blocks' of an inversion
framework can simply work with these units in a standalone fashion and
rely on default configurations for the lower level components. By using
a proper amount of information hiding through abstraction, users of this
code can delve as deeply in to the code architecture as they are
interested in. We also aim to make our `code look like the math' as much
as possible, which will help our own development as well as that of
future researchers, and reduce the cognitive load required for a
researcher to start performing research.

There are a number of existing software frameworks for solving inverse
problems with varying goals in mind. The work of
\citep{symes2011simulator} provides a C++ framework built upon the
abstract Rice Vector Library \citep{padula2009} for time domain modeling
and inversion, which respects the underlying Hilbert spaces where each
vector lives by automatically keeping track of units and grid spacings,
among other things. The low-level nature of the language it is built in
exposes too many low level constructs at various levels of its
hierarchy, making integrating changes in to the framework cumbersome.
The Seiscope toolbox \citep{metivier2016} implements high level
optimization algorithms in the low-level Fortran language, with the
intent to interface in to existing modeling and derivative codes using
reverse communication. As we highlight below, this is not necessarily a
beneficial strategy and merely obfuscates the codebase, as low level
languages should be the domain of computationally-intensive code rather
than high-level algorithms. The Trilinos project
\citep{heroux2005overview} is a large collection of packages written in
C++ by a number of domain-specific experts, but requires an involved
installation procedure, has no straightforward entrance point for
PDE-constrained optimization, and is not suitable for easy prototyping.
Implementing a modelling framework in PETSc \citep{balay2012petsc}, such
as in \citep{knepley2006developing}, let alone an inversion framework,
exposes too many of the unnecessary details at each level of the
hierarchy given that PETSc is written in C. The Devito framework
\citep{lange2016devito} offers a high-level symbolic Python interface to
generate highly optimized stencil-based C- code for time domain
modelling problems, with extensions to inversion. This is promising work
that effectively delineates the high-level mathematics from the
low-level computations. We follow a similar philosophy in this work.
This work builds upon ideas in \citep{van2012parallel}, which was a
first attempt to implement a high-level inversion framework in Matlab.

Software frameworks in the finite-element regime have been successfully
applied to optimal control and other PDE-constrained optimization
problems. The Dolfin framework \citep{farrell2013automated} employs a
high-level description of the PDE-constrained problem written in the UFL
language for specifying finite elements, which is subsequently compiled
in to lower level finite element codes with the relevant adjoint
equations derived and solved automatically for the objective and
gradient. For the geophysical examples, the finite element method does
not easily lend itself to applying a perfectly-matched layer to the
problem compared to finite differences, although some progress has been
made in this front, i.e., see \citep{cimpeanu2015parameter}. In general,
finite difference methods are significantly easier to implement,
especially in a matrix-free manner, than finite element methods,
although the latter have a much stronger convergence theory. Moreover,
for systems with oscillatory solutions such as the Helmholtz equation,
applying the standard 7 point stencil to the problem is inadvisable due
to the large amount of numerical dispersion introduced, resulting in a
system matrix with a large number of unknowns. This fact, along with the
indefiniteness of the underlying matrix, makes it very challenging to
solve with standard Krylov methods. More involved approaches are needed
to adequately discretize such equations, see, e.g.,
\citep{turkel2013compact, operto2007fd, chen201427}. The SIMPEG package
\citep{cockett2015simpeg} is designed in a similar spirit to the
considerations in this work, but does not fully abstract away
unnecessary components from the user and is not designed with
large-scale computations in mind as it lacks inherent parallelism. The
Jinv package \citep{ruthotto2016jinv} is written in Julia in a similar
spirit to this work with an emphasis on finite element discretizations
using the parallel MUMPS solver \citep{amestoy2000multifrontal} for
computing the fields and is parallelized over the number of source
experiments.

When considering the performance-understandability spectrum for
designing inverse problem software, it is useful to consider Amdahl's
law \citep{patterson2011computer}. Roughly speaking, Amdahl's law states
that in speeding up a region of code, through parallelization or other
optimizations, the speedup of the overall program will always be limited
by the remainder of the program that does not benefit from the speedup.
For instance, speeding up a region where the program spends 50\% of its
time by a factor of 10 will only speed up the overall program by a
maximum factor of 1.8. For any PDE-based inverse problem, the majority
of the computational time is spent solving the PDEs themselves. Given
Amdahl's law and a limited budget of researcher time, this would imply
that there is virtually no performance benefit in writing both the
`heavy lifting' portions of the code as well as the auxiliary operations
in a low level language, which can impair readability and obscure the
role of the individual component operations in the larger framework.
Rather, one should aim to use the strengths of a high level language to
express mathematical ideas cleanly in code and exploit the efficiency of
a low level language, at the proper instance, to speed up primitive
operations such as a multi-threaded matrix-vector product. These
considerations are also necessary to manage the complexity of the code
and ensure that it functions properly. Researcher time, along with
computational time, is valuable and we should aim to preserve
productivity by designing these systems with these goals in mind.

It is for this reason that we choose to use Matlab to implement our
parallel inversion framework as it offers the best balance between
access to performance-critical languages such as C and Fortran, while
allowing for sufficient abstractions to keep our code concise and loyal
to the underlying mathematics. A pure Fortran implementation, for
instance, would be significantly more difficult to develop and
understand from an outsider's perspective and would not offer enough
flexibility for our purposes. Python would also potentially be an option
for implementing this framework. At the time of the inception of this
work, we found that the relatively new scientific computing language
Julia \citep{bezanson2014julia} was in too undeveloped of a state to
facilitate all of the abstractions we needed; this may no longer be the
case as of this writing.

\subsection{Our Contributions}\label{pdecontributions}

Using a hierarchical approach to our software design, we implement a
framework for solving inverse problems that is flexible, comprehensible,
efficient, scalable, and consistent. The flexibility arises from our
design decisions, which allow a researcher to swap components
(parallelization schemes, linear solvers, preconditioners,
discretization schemes, etc.) in and out to suit her needs and the needs
of her local computational environment. Our design balances efficiency
and understandability through the use of object oriented programming,
abstracting away the low-level mechanisms of the computationally
intensive components through the use of the SPOT framework \citep{spot}.
The SPOT methodology allows us to abstract away function calls as
matrix-vector multiplications in Matlab, the so-called matrix-free
approach. By abstracting away the lower level details, our code is clean
and resembles the underlying mathematics. This abstraction also allows
us to swap between using explicit, sparse matrix algebra for 2D problems
and efficient, multi-threaded matrix-vector multiplications for 3D
problems. The overall codebase is then agnostic to the dimensionality of
$m$, which encourages code reuse when applying new algorithms to large
scale problems. Our hierarchical design also decouples parallel data
distribution from computation, allowing us to run the same algorithm as
easily on a small 2D problem using a laptop as on a large 3D problem
using a cluster. We also include unit tests that demonstrate that our
code accurately reflects the underlying mathematics in Section
(\#validation). We call this package WAVEFORM (softWAre enVironmEnt For
nOnlinear inveRse probleMs), which can be obtained at
\url{https://github.com/slimgroup/WAVEFORM}.

In this work, we also propose a new multigrid-based preconditioner for
the 3D Helmholtz equation that only requires matrix-vector products with
the system matrix at various levels of discretization, i.e., is
matrix-free at each multigrid level, and employs standard Krylov solvers
as smoothers. This preconditioner allows us to operate on realistically
sized 3D seismic problems without exorbitant memory costs. Our numerical
experiments demonstrate the ease of which we can apply high level
algorithms to solving the PDE-based parameter estimation problem and its
variants while still having full flexibility to swap modeling code
components in and out as we choose.

\section{Preamble: Theory and
Notation}\label{preamble-theory-and-notation}

To ensure that this work is sufficiently self-contained, we outline the
basic structure and derivations of our inverse problem of interest.
Lowercase, letters such as $x,y,z$ denote vectors and uppercase letters
such as $A,B,C$ denote matrices or linear operators of appropriate size.
To distinguish between continuous objects and their discretized
counterparts, with a slight abuse of notation, we will make the spatial
coordinates explicit for the continuous objects, i.e., sampling
$u(x,y,z)$ on a uniform spatial grid results in the vector $u$. Vectors
$u$ can depend on parameter vectors $m$, which we indicate with $u(m)$.
The adjoint operator of a linear mapping $x \mapsto Ax$ is denoted as
$A^*$ and the conjugate Hermitian transpose of a complex matrix $B$ is
denoted $B^H$. If $B$ is a real-valued matrix, this is the standard
matrix transpose.

Our model inverse problem is the multi-source parameter estimation
problem. Given our data $d_{i,j}$ depending on the $i^{\text{th}}$
source and $j^{\text{th}}$ frequency, our measurement operator $P_r$,
and the linear partial differential equation $H(m) u(m) = q$ depending
on the model parameter $m$, find the model $m$ that minimizes the misfit
between the predicted and observed data, i.e.,
\begin{equation*}
\begin{split}
\min_{m,u_{i,j}} \sum_{i}^{N_s} \sum_{j}^{N_f} \phi(P_r u_{i,j},d_{i,j}) \\
\text{subject to } H_j(m) u_{i,j} = q_{i,j}.
\end{split}
\end{equation*}
 Here $\phi(s,t)$ is a smooth misfit function between the inputs $s$ and
$t$, often the least-squares objective
$\phi(s,t) = \dfrac{1}{2}\|s-t\|_2^2$, although other more robust
penalties are possible, see e.g.,
\citep{aravkin2012robust, aravkin2011robust}. The indices $i$ and $j$
vary over the number of sources $N_s$ and number of frequencies $N_f$,
respectively. For the purposes of notational simplicity, we will drop
this dependence when the context permits. Note that our design is
specific to boundary-value problems rather than time-domain problems,
which have different computational and storage challenges.

A well known instance of this setup is the full waveform inversion
problem in exploration seismology, which involves discretizing the
constant-density Helmholtz equation
\begin{equation}
\begin{split}
    (\nabla^2 + m(x,y,z))u(x,y,z) = S(\omega)\delta(x-x_s)\delta(y-y_s)\delta(z-z_s) \\
  \lim_{r\to\infty} r \left(\dfrac{\partial}{\partial r} - i \sqrt{m} \right)u(x,y,z) = 0
\end{split}
\label{helmholtz}
\end{equation}
 where $\nabla^2 = \partial_x^2 + \partial_y^2 + \partial_z^2$ is the
Laplacian, $m(x,y,z) = \frac{\omega^2}{v^2(x,y,z)}$ is the wavenumber,
$\omega$ is the angular frequency and $v(x,y,z)$ is the gridded
velocity, $S(\omega)$ is the per-frequency source weight,
$(x_{s},y_{s},z_{s})$ are the spatial coordinates of the source, and the
second line denotes the Sommerfeld radiation condition
\citep{sommerfeld1949partial} with $r = \sqrt{x^2 + y^2 + z^2}$. In this
case, $H(m)$ is any finite difference, finite element, or finite volume
discretization of~\eqref{helmholtz}. Other examples of such problems
include electrical impedance tomography using a simplified form of
Maxwell's equations, which can be reduced to Poisson's equation
\citep{cheney1999electrical, borcea2002electrical, adler2011electrical},
X-ray tomography, which can be thought of as the inverse problem
corresponding to a transport equation, and synthetic aperture radar,
using the wave equation \citep{natterer2006imaging}. For realistically
sized industrial problems of this nature, in particular for the seismic
case, the size of the model vector $m$ is often $\mathcal{O}(10^9)$ and
there can be $\mathcal{O}(10^5)$ sources, which prevents the full
storage of the fields. Full-space methods, which use a Newton iteration
on the associated Lagrangian system, such as in
\citep[\citep{prudencio2006parallel}]{biros2005parallel}, are infeasible
as a large number of fields have to be stored and updated in memory. As
a result, these large scale inverse problems are typically solved by
eliminating the constraint and reformulating the problem in an
unconstrained or reduced form as
\begin{equation}
  \min_{m} f(m) := \sum_{i}^{N_s} \sum_{j}^{N_f} \phi(P_r H_j(m)^{-1}q_{i,j}, d_{i,j}).
\label{mainprob}
\end{equation}
 We assume that we our continuous PDEs are posed on a rectangular domain
$\Omega$ with zero Dirichlet boundary conditions. For PDE problems that
require sponge or perfectly matched layers, we extend $\Omega$ to
$\Omega' \supset \Omega$ and vectors defined on $\Omega$ are extended to
$\Omega'$ by extension in the normal direction. In this extended domain
for the acoustic Helmholtz equation, for instance, we solve
\begin{equation*}
\begin{split}
(\partial^2_{x} + \partial^2_y + \partial^2_z + \omega^2 m(x,y,z))u(x,y,z) = \delta(x-s_x)\delta(y-s_x)\delta(z-s_z) \quad (x, y, z) \in \Omega\\
(\tilde{\partial_x}^2 + \tilde{\partial_y}^2 + \tilde{\partial_z}^2 + \omega^2 m(x,y,z)) = 0 \quad (x,y,z) \in \Omega'\setminus \Omega,
\end{split}
\end{equation*}
 where $\tilde{\partial_x} = \dfrac{1}{\eta(x)}\partial_x$, for
appropriately chosen univariate PML damping functions $\eta(x)$, and
similarly for $y,z$. This results in solutions $u$ that decay
exponentially for $x \in \Omega'\setminus \Omega$. We refer the reader
to
\citep{berenger1994perfectly, chew1997complex, hastings1996application}
for more details.

We assume that the source functions $q_{i,j}(x)$ are localized in space
around the points $\{x^s_i\}_{i=1}^{n_s},$ which make up our source
grid. The measurement or sampling operator $P_r$ samples function values
defined on $\Omega'$ at the set of receiver locations
$\{x^r_k\}_{k=1}^{n_r}$. In the most general case, the points $x^r_k$
can vary per-source (i.e., as the location of the measurement device is
dependent on the source device), but we will not consider this case
here. In either case, the source grid can be independent from the
receiver grid.

While the PDE itself is linear, the mapping that predicts data
$F(m) := m \mapsto P_r H(m)^{-1}q$, the so-called the forward modeling
operator, is known to be highly nonlinear and oscillatory in the case of
the high frequency Helmholtz equation \citep{sun2013waveform}, which
corresponds to propagating anywhere between 50-1000 wavelengths for
realistic models of interest. Without loss of generality, we will
consider the Helmholtz equation as our prototypical model in the sequel.
The level of formalism, however, will ultimately be related to parameter
estimation problems that make use of real-world data, which is
inherently band-limited. We will therefore not be overly concerned with
convergence as the mesh- or element-size tends to zero, as the
informative capability of our acquired data is only valid up until a
certain resolution dictated by the resolution of our measurement device.
We will focus solely on the discretized formulation of the problem from
hereon out.

We can compute relevant derivatives of the objective function with
straightforward, ableit cumbersome, matrix calculus. Consider the state
equation $u(m) = H(m)^{-1}q$. Using the chain rule, we can derive
straightforward expressions for the directional derivative
$Du(m)[\delta m]$ as
\begin{equation}
Du(m)[\delta m] = -H(m)^{-1} DH(m)[\delta m]u(m).
\label{dU}
\end{equation}
 Here $DH(m)[\delta m]$ is the directional derivative of the mapping
$m \mapsto H(m),$ which is assumed to be smooth. To emphasize the
dependence on the linear argument, we let $T$ denote the linear operator
defined by $T\delta m = DH(m)[\delta m]u(m)$, which outputs a vector in
model space. Note that $T = T(m,u(m))$, but we drop this dependence for
notational simplicity. We let $T^*$ denote the adjoint of the linear
mapping $\delta m \mapsto T\delta m.$ The associated adjoint mapping
of~\eqref{dU} is therefore
\begin{equation*}
Du(m)[\cdot]^* y = -T^* H(m)^{-H}y.
\end{equation*}
 The (forward) action of the Jacobian $J$ of the forward modelling
operator $F(m) = P_r u(m)$ is therefore given by
$J\delta m = P_r Du(m)[\delta m]$.

We also derive explicit expressions for the Jacobian adjoint,
Gauss-Newton Hessian, and full Hessian matrix-vector products, as
outlined in Table~\eqref{derivstuff}, although we leave the details for
Appendix (\#userfriendly). For even medium sized 2D problems, it is
computationally infeasible to store these matrices explicitly and
therefore we only have access to matrix-vector products. The number of
matrix-vector products for each quantity are outlined in
Table~\eqref{derivcosts} and are per-source and per-frequency. By
adhering to the principle of `the code should reflect the math', once we
have the relevant formula from Table~\eqref{derivstuff}, the resulting
implementation will be as simple as copying and pasting these formula in
to our code in a straightforward fashion, which results in little to no
computational overhead, as we shall see in the next section.

\begin{table*}
\centering
\begin{tabular}{ll}
\toprule\addlinespace
$u(m)$ & $H(m)^{-1}q$\tabularnewline
$Du(m)[\delta m]$ & $-H(m)^{-1}T\delta m$\tabularnewline
$Du(m)[\cdot]^*y$ & $-T^*H(m)^{-H} y $\tabularnewline
$F(m)$ & $P u(m)$\tabularnewline
$J\delta m := DF(m)[\delta m]$ & $P Du(m)[\delta m]$\tabularnewline
$T\delta m$ & $DH(m)[\delta m]u(m)$, e.g.,~\eqref{dH}\tabularnewline
$T^*z$ & PDE-dependent, e.g.,~\eqref{dHadj}\tabularnewline
$DT^*[\delta m,\delta u]z $ & PDE-dependent,
e.g.,~\eqref{DTadj}\tabularnewline
$V(m)$ & $-H(m)^{-H} P^T \nabla \phi$\tabularnewline
$DV(m)[\delta m]$ &
$H(m)^{-H} (-DH(m)[\delta m] V(m) - P^T \nabla^2 \phi(Pu)[PDu(m)[\delta m]])$\tabularnewline
$H_{GN}\delta m := J^HJ\delta m$ &
$T^* H(m)^{-H}P^T PH(m)^{-1}T\delta m$\tabularnewline
$\nabla f(m)$ & $T^* V(m)$\tabularnewline
$\nabla^2 f(m)[\delta m]$ &
$DT^*[\delta m,Du(m)[\delta m]] V(m) + T^* DV(m)[\delta m]$\tabularnewline
\bottomrule
\end{tabular}
\caption{Quantities of interest for PDE-constrained
optimization.}\label{derivstuff}
\end{table*}

\begin{table*}
\centering
\begin{tabular}{ll}
\toprule\addlinespace
Quantity & \# PDEs\tabularnewline
\midrule
$f(m)$ & 1\tabularnewline
$f(m),\nabla f(m)$ & 2\tabularnewline
$H_{GN}\delta m$ & 3\tabularnewline
$\nabla^2 f(m)[\delta m]$ & 4\tabularnewline
\bottomrule
\end{tabular}
\caption{Number of PDEs (per frequency/source) for each optimization
quantity of interest}\label{derivcosts}
\end{table*}

\section{From Inverse Problems to Software
Design}\label{from-inverse-problems-to-software-design}

Given the large number of summands in~\eqref{mainprob} and the high cost
of solving each PDE $u_{i,j} = H_j(m)^{-1}q_{i,j}$, there are a number
of techniques one can employ to reduce up per-iteration costs and to
increase convergence speed to a local minimum, given a fixed
computational budget. Stochastic gradient techniques
\citep{bottou2010large, bottou1998online} aim to reduce the per
iteration costs of optimization algorithms by treating the sum as an
expectation and approximate the average by a random subset. The batching
strategy aims to use small subsets of data in earlier iterations, making
significant progress towards the solution for lower cost. Only in later
iterations does the algorithm require a larger number of per-iteration
sources to ensure convergence \citep{friedlander2012hybrid}. In a
distributed parallel environment, the code should also employ batch
sizes that are commensurate with the available parallel resources. One
might also seek to solve the PDEs to a lower tolerance at the initial
stages, as in \citep{vanLeeuwen2014SISC3Dfds}, and increasing the
tolerance as iterations progress, an analogous notion to batching. By
exploiting curvature information by minimizing a quadratic model of
$f(m)$ using the Gauss-Newton method, one can further enhance
convergence speed of the outer algorithm.

In order to employ these high-level algorithmic techniques and
facilitate code reuse, our optimization method should be a black-box, in
the sense that it is completely oblivious to the underlying structure of
the inverse problem, calling a user-defined function that returns an
objective value, gradient, and Gauss-Newton Hessian operator. Our
framework should be flexible enough so that, for 2D problems, we can
afford to store the sparse matrix $H(m)$ and utilize the resulting
efficient sparse linear algebra tools for inverting this matrix, while
for large-scale 3D problems, we can only compute matrix-vector products
with $H(m)$ with coefficients constructed on-the-fly. Likewise,
inverting $H(m)$ should only employ Krylov methods that use these
primitives, such as FGMRES \citep{saad1993flexible}. These restrictions
are very realistic for the seismic inverse problem case, given the large
number of model and data points involved as well as the limited
available memory for each node in a distributed computational
environment. In the event that a new robust preconditioner developed for
the PDE system matrix, we should be able to easily swap out one
algorithm for another, without touching the outer optimization method.
Likewise, if researchers develop a much more efficient stencil for
discretizing the PDE, develop a new misfit objective
\citep{aravkin2012robust}, or add model-side constraints
\citep{peters2016cvp}, we would like to easily integrate such changes in
to the framework with minimal code changes. Our code should also expose
an interface to allow a user or algorithm to perform source/frequency
subsampling from arbitrarily chosen indices.

\begin{figure}
\centering
\includegraphics[width=1.000\hsize]{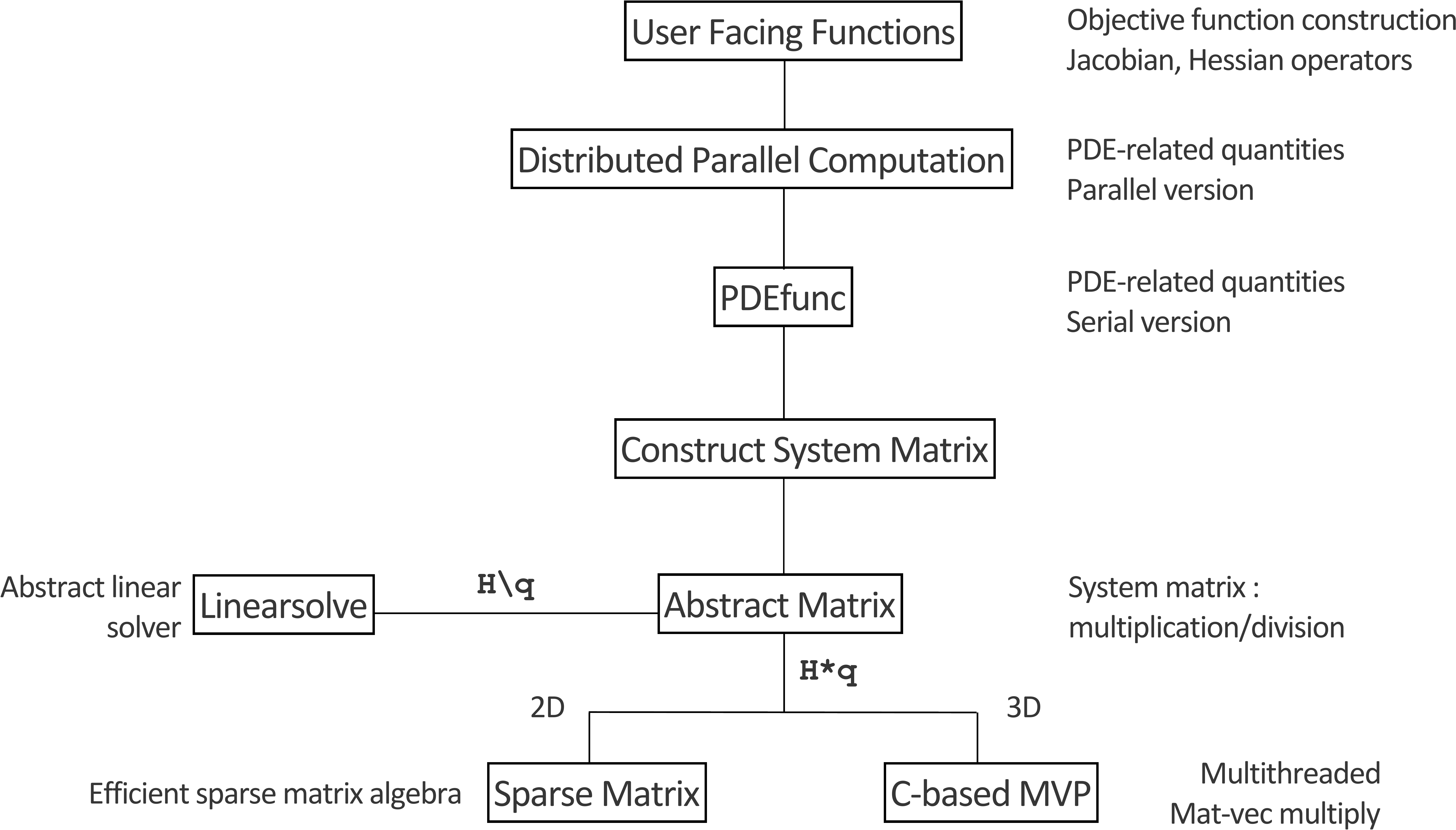}
\caption{Software Hierarchy.}\label{softorg}
\end{figure}

We decouple the various components of the inverse problem context by
using an appropriate software hierarchy, which manages complexity
level-by-level and will allow us to test components individually to
ensure their correctness and efficiency, as shown in
Figure~\ref{softorg}. Each level of the hierarchy is responsible for a
specific set of procedural requirements and defers the details of lower
level computations to lower levels in the hierarchy.

At the topmost level, our user-facing functions are responsible for
constructing a misfit function suitable for black-box optimization
routines such as LBFGS or Newton-type methods
\citep{liu1989limited, zhu1997algorithm, grippo1989truncated}. This
procedure consists of

\begin{itemize}
\itemsep1pt\parskip0pt\parsep0pt
\item
  handling subsampling of sources/frequencies for the distributed data
  volume
\item
  coarsening the model, if required (for low frequency 3D problems)
\item
  constructing the function interface that returns the objective,
  gradient, and requested Hessian at the current point.
\end{itemize}

For the Helmholtz case in particular, coarsening the model allows us to
keep the number of degrees of freedom to a minimum, in line with the
requirements of the discretization. In order to accommodate stochastic
optimization algorithsm, we provide an option for a batch mode
interface, which allows the objective function to take in as input both
a model vector and a set of source-frequency indices. This option allows
either the user or the outer stochastic algorithm to dynamically specify
which sources and frequencies should be computed at a given time. This
auxiliary information is passed along to lower layers of the hierarchy.
Additionally, we provide methods to construct the Jacobian,
Gauss-Newton, and Full Hessian as SPOT operators. For instance, when
writing $\text{Hess}*v$ as in a linear solver, Matlab implicitly calls
the lower level functions that actually solve the PDEs to compute this
product, all the while never forming the matrix explicitly. Information
hiding in this fashion allows us to use existing linear solver codes
written in Matlab (Conjugate Gradient, MINRES, etc.) to solve the
resulting systems. For the parameter inversion problem, the user can
choose to have either the Gauss-Newton or full Hessian operators
returned at the current point.

Lower down the hierarchy, we have our \texttt{PDEfunc} layer. This
function is responsible for computing the quantities of interest, i.e.,
the objective value, gradient, Hessian or Gauss-Newton Hessian
matrix-vector product, forward modelling operator, or linearized forward
modelling operator and its adjoint. At this stage in the hierarchy, we
are concerned with `assembling' the relevant quantities based on PDE
solutions in to their proper configurations, rather than how exactly to
obtain such solutions, i.e., we implement the formulas in
Table~\ref{derivstuff}. Here the PDE system matrix is a SPOT operator
that has methods for performing matrix-vector products and matrix-vector
divisions, which contains information about the particular stencil to
use as well as which linear solvers and preconditioners to call. When
dealing with 2D problems, this SPOT operator is merely a shallow wrapper
around a sparse matrix object and contains its sparse factorization,
which helps speed up solving PDEs with multiple right hand sides. In
order to discretize the delta source function, we use Kaiser-windowed
sinc interpolation \citep{hicks2002arbitrary}. At this level in the
hierarchy, our code is not sensitive to the discretization or even the
particular PDE we are solving, as we demonstrate in Section
(\#expoisson) with a finite volume discretization of the Poisson
equation. To illustrate how closely our code resembles the underlying
mathematics, we include a short snippet of code from our
\texttt{PDEfunc} below in Figure~\ref{pdefunc}.

\begin{lstlisting}[caption={Excerpt from the code of
\texttt{PDEfunc}.}, label=pdefunc, float=htbp]
% Set up interpolation operators
% Source grid -> Computational Grid
Ps = opInterp('sinc',model.xsrc,xt,model.ysrc,yt,model.zsrc,zt);
% Computational grid -> Receiver grid
Pr = opInterp('sinc',model.xrec,xt,model.yrec,yt,model.zrec,zt)';
% Sum along sources dimension
sum_srcs = @(x) to_phys*sum(real(x),2);
% Get Helmholtz operator, computational grid struct, its derivative
[Hk,comp_grid,T,DT_adj] = discrete_pde_system(v,model,freq(k),params);
U = H \ Q;
switch func
 case OBJ
    [phi,dphi] = misfit(Pr*U,getData(Dobs,data_idx),current_src_idx,freq_idx);
    f = f + phi;
    if nargout >= 2
        V = H' \ ( -Pr'* dphi);
        g = g + sum_srcs(T(U)'*V);
    end

 case FORW_MODEL
    output(:,data_idx) = Pr*U;

 case JACOB_FORW
    dm = to_comp*vec(input);
    dU = H\(-T(U)*dm);
    output(:,data_idx) = Pr*dU;

 case JACOB_ADJ
    V = H'\( -Pr'* input(:,data_idx) );
    output = output + sum_srcs(T(U)'*V);

 case HESS_GN
    dm = to_comp*vec(input);
    dU = H\(-T(U)*dm);
    dU = H'\(-Pr'*Pr*dU);
    output = output + sum_srcs(T(U)*dU);

 case HESS
    dm = to_comp*vec(input);
    [~,dphi,d2phi] = misfit(Pr*U,getData(Dobs,data_idx),current_src_idx,freq_idx);
    dU = H\(-T(U)*dm);
    V = H'\(-Pr'*dphi);
    dV = H'\(-T(V)*dm - Pr'* reshape(d2phi*vec(Pr*dU),nrec,size(U,2)));
    output = output + sum_srcs(DT_adj(U,dm,dU)*V + T(U)*dV);
\end{lstlisting}

\begin{figure}
\centering
\includegraphics[width=0.750\hsize]{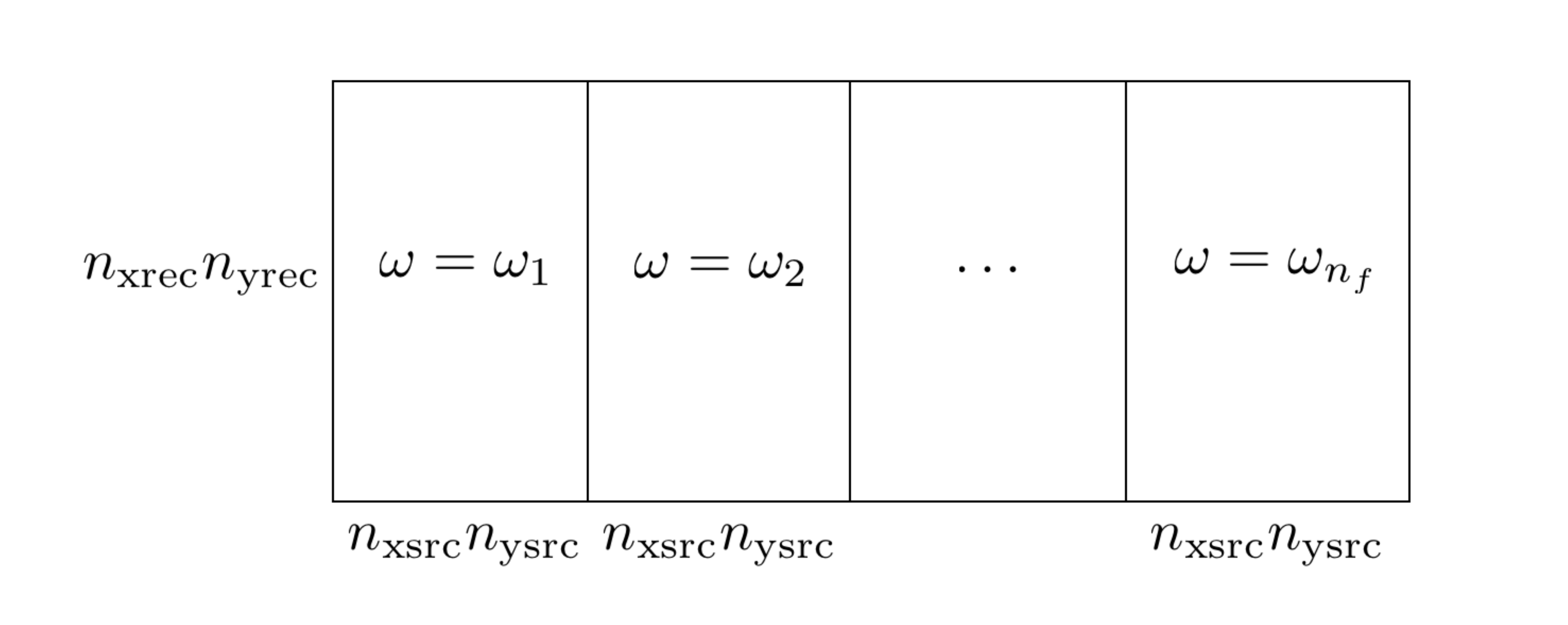}
\caption{Data distributed over the joint (source, frequency)
indices.}\label{datadist}
\end{figure}

In our parallel distribution layer, we compute the result of
\texttt{PDEfunc} in an embarrassingly parallel manner by distributing
over sources and frequencies and summing the results computed across
various Matlab workers. This distribution scheme uses Matlab's Parallel
Toolbox, which is capable of Single Program Multiple Data (SPMD)
computations that allow us to call \texttt{PDEfunc} identically on
different subsets of source/frequency indices. The data is distributed
as in Figure~\eqref{datadist}. The results of the local worker
computations are then summed together (for the objective, gradient,
adjoint-forward modelling, Hessian-vector products) or assembled in to a
distributed vector (forward modelling, linearized forward modelling).
Despite the ease of use in performing these parallel computations, the
parallelism of Matlab is \textbf{not} fault-tolerant, in that if a
single worker or process crashes at any point in computing a local value
with \texttt{PDEfunc}, the parent calling function aborts as well. In a
large-scale computing environment, this sort of behaviour is unreliable
and therefore we recommend swapping out this `always-on' approach with a
more resilient model such as a map reduce paradigm. One possible
implementation workaround is to measure the elapsed time of each worker,
assuming that the work is distributed evenly. In the event that a worker
times out, one can simply omit the results of that worker and sum
together the remaining function and gradient values. In some sense, one
can account for random machine failure or disconnection by considering
it another source of stochasticity in an outer-level stochastic
optimization algorithm.

Further down the hierarchy, we construct the system matrix for the PDE
in the following manner. As input, we require the current parameters
$m$, information about the geometry of the problem (current frequency,
grid points and spacing, options for number of PML points), as well as
performance and linear solver options (number of threads, which
solver/preconditioner to use, etc.). This function also extends the
medium parameters to the PML extended domain, if required. The resulting
outputs are a SPOT operator of the Helmholtz matrix, which has suitable
routines for performing matrix-vector multiplications and divisions, a
struct detailing the geometry of the pml-extended problem, and the
mappings $T$ and $DT^*[\delta m]$ from Table~\eqref{derivstuff}. It is
in this function that we also construct the user-specified
preconditioner for inverting the Helmholtz (or other system) matrix.

The actual operator that is returned by this method is a SPOT operator
that performs matrix-vector products and matrix-vector divisions with
the underlying matrix, which may take a variety of forms. In the 2D
regime, we can afford to explicitly construct the sparse matrix and we
utilize the efficient sparse linear algebra routines in Matlab for
solving the resulting linear system. In this case, we implement the 9pt
optimal discretization of \citep{chen2013optimal}, which fixes the
problems associated to the 9pt discretization of \citep{jo1996optimal},
and use the sparse LU decomposition built in to Matlab for inverting the
system. These factors are computed at the initialization of the system
matrix and are reused across multiple sources. In the 3D regime, we
implement a stencil-based matrix-vector product (i.e., one with
coefficients constructed on the fly) written in C++, using the compact
27-pt stencil of \citep{operto2007fd} along with its adjoint and
derivative. The stencil-based approach allows us to avoid having to
keep, in this case, 27 additional vectors of the size of the
PML-extended model in memory. This implementation is multi-threaded
along the $z-$axis using OpenMP and is the only `low-level' component in
this software framework, geared towards high performance. Since this
primitive operation is used throughout the inverse problem framework, in
particular for the iterative matrix-vector division, any performance
improvements made to this code will propagate throughout the entire
codebase. Likewise, if a `better' discretization of the PDE becomes
available, it can be easily integrated in to the software framework by
swapping out the existing function at this stage, without modifying the
rest of the codebase. When it is constructed, this SPOT operator also
contains all of the auxiliary information necessary for performing
matrix-vector products (either the matrix itself, for 2D, or the
PML-extended model vector and geometry information, for 3D) as well as
for matrix-vector divisions (options for the linear solver,
preconditioner function handle, etc.).

We allow the user access to multiplication, division, and other matrix
operations like Jacobi and Kaczmarz sweeps through a unified interface,
irrespective of whether the underlying matrix is represented explicitly
or implicitly via function calls. For the explicit matrix case, we have
implemented such basic operations in Matlab. When the matrix is
represented implicitly, these operations are presented by a standard
function handle or a `FuncObj' object, the latter of which mirrors the
Functor paradigm in C++. In the Matlab case, the `FuncObj' stores a
reference to a function handle, as well as a list of arguments. These
arguments can be partially specialized upon construction, whereby only
some of the arguments are specified at first. Later on when they are
invoked, the remainder of their arguments are passed along to the
function. In a sense, they implement the same functionality of anonymous
functions in Matlab without the variable references to the surrounding
workspace, which can be costly when passing around vectors of size $N^3$
and without the variable scoping issues inherent in Matlab. We use this
construction to specify the various functions that implement the
specific operations for the specific stencils. Through this delegation
pattern, we present a unified interface to the PDE system matrix
irrespective of the underlying stencil or even PDE in question.

When writing \texttt{u\phantom{\ }=\phantom{\ }H\textbackslash{}q} for
this abstract matrix object $H$, Matlab calls the `linsolve' function,
which delegates the task of solving the linear system to a
user-specified solver. This function sets up all of the necessary
preamble for solving the linear system with a particular method and
preconditioner. Certain methods such as the row-based Carp-CG method,
introduced in \citep{carpcg} and applied to seismic problems in
\citep{vanLeeuwen2014SISC3Dfds}, require initial setup, which is
performed here. This construction allows us to easily reuse the idea of
`solving a linear system with a specific method' in setting up the
multi-level preconditioner of the next section, which reduces the
overall code complexity since the multigrid smoothers themselves can be
described in this way.

In order to manage the myriad of options available for computing with
these functions, we distinguish between two classes of options and use
two separate objects to manage them. `LinSolveOpts' is responsible for
containing information about solving a given linear system, i.e., which
solver to use, how many inner/outer iterations to perform, relative
residual tolerance, and preconditioner. `PDEopts' contains all of the
other information that is pertinent to \texttt{PDEfunc} (e.g., how many
fields to compute at a time, type of interpolation operators to use for
sources/receivers, etc.) as well as propagating the options available
for the PDE discretization (e.g., stencil to use, number of PML points
to extend the model, etc.).

\subsection{Extensions}\label{softwareextensions}

This framework is flexible enough for us to integrate additional models
beyond the standard adjoint-state problem. We outline two such
extensions below.

\subsubsection{Penalty Method}\label{penalty-method}

Given the highly oscillatory nature of the forward modelling operator,
due to the presence of the inverse Helmholtz matrix $H(m)^{-1}q$, one
can also consider relaxing the exact constraint $H(m)u(m)=q$ in to an
unconstrained and penalized form of the problem. This is the so-called
Waveform Reconstruction Inversion approach \citep{leeuwen2014EAGEntf},
which results in the following problem, for a least-squares data-misfit
penalty,
\begin{equation}
\min_m = \dfrac{1}{2}\|Pu(m)-d\|_2^2 + \dfrac{\lambda^2}{2} \|H(m)u(m) - q\|_2^2
\label{wri}
\end{equation}
 where $u(m)$ solves the least-squares system
\begin{equation}
u(m) = \argmin_{u} \dfrac{1}{2} \|Pu - d\|_2^2 + \dfrac{\lambda^2}{2} \|H(m)u-q\|_2^2.
\label{wrifieldeq}
\end{equation}
 The notion of variable projection \citep{aravkin2012estimating}
underlines this method, whereby the field $u$ is projected out of the
problem by solving~\eqref{wrifieldeq} for each fixed $m$. We perform the
same derivations for problem~\eqref{wri} in Appendix (\#wrideriv). Given
the close relationship between the two methods, the penalty method
formulation is integrated into the same function as our FWI code, with a
simple flag to change between the two computational modes. The full
expressions for the gradient, Hessian, and Gauss-Newton Hessian of WRI
are outlined in Appendix B.

\subsubsection{2.5D}\label{d}

For a 3D velocity model that is invariant with respect to one dimension,
i.e., $v(x,y,z) = h(x,z) \text{ for all } y$, so
$m(x,y,z) = \omega^2 g(x,z) $ for $g(x,z) = \dfrac{1}{h^2(x,z)}$, we can
take a Fourier transform in the y-coordinate of~\eqref{helmholtz} to
obtain
\begin{equation*}
(\partial_x^2 + \partial_z^2 + \omega^2 g(x,z) - k^2_{y})\hat{u}(x,k_y,z) = S(\omega)\delta(x-x_s)\delta(z-z_s)e^{-ik_y y_s}.
\end{equation*}
 This so-called 2.5D modeling/inversion allows us to mimic the physical
behaviour of 3D wavefield propagation (e.g., $1/r$ amplitude decay vs
$1/r^{1/2}$ decay in 2D, point sources instead of line sources, etc.)
without the full computational burden of solving the 3D Helmholtz
equation \citep{song1995frequency}. We solve a series of 2D problems
instead, as follows.

Multiplying both sides by $e^{ik_y y_s}$ and setting
$\tilde{u}_{k_y}(x,z) = e^{ik_yy_s} \hat{u}(x,k_y,z)$, we have that, for
each fixed $k_y$, $\tilde{u}_{k_y}$ is the solution of
$H(k_y)\tilde{u}_{k_y}=S(\omega)\delta(x-x_s)\delta(z-z_s)$ where
$H(k_y) = (\partial_x^2 + \partial_z^2 + \omega^2 g(x,z) - k^2_{y})$.

We can recover $u(x,y,z)$ by writing
\begin{equation*}
\begin{aligned}
u(x,y,z) &= \dfrac{1}{2\pi}\int_{-\infty}^{\infty} \hat{u}_{k_y}(x,z) e^{ik_y y} dk_y \\
 &= \dfrac{1}{2\pi}\int_{-\infty}^{\infty} \tilde{u}_{k_y}(x,z) e^{ik_y (y-y_s)} dk_y \\
 &= \dfrac{1}{\pi}\int_0^{\infty} \tilde{u}_{k_y}(x,z) \cos(k_y (y-y_s)) dk_y \\
 &= \dfrac{1}{\pi}\int_0^{k_{nyq}} \tilde{u}_{k_y}(x,z) \cos(k_y (y-y_s)) dk_y.
\end{aligned}
\end{equation*}
 Here the third line follows from the symmetry between $k_y$ and $-k_y$
and the fourth line restricts the integral to the Nyquist frequency
given by the sampling in $y$, namely $k_{nyq} = \frac{\pi}{\Delta y}$
\citep{song1995frequency}. One can further restrict the range of
frequencies to $[0,p\cdot k_c]$ where
$k_c = \frac{\omega}{\min_{x,z}v(x,z)}$ is the so-called critical
frequency and $p \ge 1, p \approx 1.$ In this case, waves corresponding
to a frequency much higher than that of $k_c$ do not contribute
significantly to the solution. By evaluating this integral with, say, a
Gauss-Legendre quadrature, the resulting wavefield can be expressed as
\begin{equation*}
u(x,y,z) = \sum_{i=1}^{N} w_i \tilde{u}_{k^i_y}(x,z),
\end{equation*}
 which is a sum of 2D wavefields. This translates in to operations such
as computing the Jacobian or Hessian having the same sum structure and
allows us to incorporate 2.5D modeling and inversion easily in to the
resulting software framework.

\section{Multi-level Recursive Preconditioner for the Helmholtz
Equation}\label{multi-level-recursive-preconditioner-for-the-helmholtz-equation}

Owing to the PML layer and indefiniteness of the underlying PDE system
for sufficiently high frequencies, the Helmholtz system matrix is
complex-valued, non-Hermitian, indefinite, and therefore challenging to
solve using Krylov methods without adequate preconditioning. There have
been a variety of ideas proposed to precondition the Helmholtz system,
including multigrid methods \citep{stolk2014multigrid}, methods
inverting the shifted Laplacian system
\citep{riyanti2007parallel, plessix2007helmholtz, erlangga2004class},
sweeping preconditioners
\citep{engquist2011sweeping, liu2015recursive, poulson2013parallel, liu2016additive},
domain decomposition methods
\citep{stolk2013rapidly, stolk2016improved, boubendir2012quasi, gander2013domain},
and Kaczmarz sweeps \citep{van2012preconditioning}, among others. These
methods have varying degrees of applicability in this framework. Some
methods such as the sweeping preconditioners rely on having explicit
representations of the Helmholtz matrix and keeping track of dense LU
factors on multiple subdomains. Their memory requirements are quite
steep as a result and they require a significant amount of bookkeeping
to program correctly, in addition to their large setup time, which is
prohibitive in inversion algorithms where the velocity is being updated.
Many of these existing methods also make stringent demands of the number
of points per wavelength $N_{\text{ppw}}$ needed to succeed, in the
realm of $10$ to $15$, which results in prohibitively large system
matrices for the purposes of inversion. As the standard 7pt
discretization requires a high $N_{\text{ppw}}$ in order to adequately
resolve the solutions of the phases \citep{operto2007fd}, this results
in a large computational burden as the system size increases.

Our aim in this section is to develop a preconditioner that is scalable,
in that it uses the multi-threading resources on a given computational
node, easy to program, matrix-free, without requiring access to the
entries of the system matrix, and leads to a reasonably low number of
outer Krylov iterations. It is more important to have a larger number of
sources distributed to multiple nodes rather than using multiple nodes
solving a single problem when working in a computational environment
where there is a nonzero probability of node failure.

We follow the development of the preconditioners in
\citep{calandra2013twogrid, lago2015EAGEtrg}, which uses a multigrid
approach to preconditioning the Helmholtz equation. The preconditioner
in \citep{calandra2013twogrid} approximates a solution
to~\eqref{mainprob} with a two-level preconditioner. Specifically, we
start with a standard multigrid V-cycle \citep{briggs2000multigrid} as
described in Figure~\eqref{alg:vcycle}. The smoothing operator aims to
reduce the amplitude of the low frequency components of the solution
error, while the restriction and prolongation operators. For an
extensive overview of the multigrid method, we refer the reader to
\citep{briggs2000multigrid}. We use linear interpolation as the
prolongation operator and its adjoint as for restriction, although other
choices are possible \citep{de1990matrix}.

\begin{scholmdAlgorithm*}
~~V-cycle~to~solve~$H_f x_f = b_f$~on~the~fine~scale

~~Input:~Current~estimate~of~the~fine-scale~solution~$x_f$\\\hspace*{0.333em}\hspace*{0.333em}Smooth~the~current~solution~using~a~particular~smoother~(Jacobi,~CG,~etc.)~to~produce~$\tilde{x}_f$\\\hspace*{0.333em}\hspace*{0.333em}Compute~the~residual~$r_f = b_f - A_f \tilde{x}_f$\\\hspace*{0.333em}\hspace*{0.333em}Restrict~the~residual,~right~hand~side~to~the~coarse~level~$r_c = Rr_f$,~$b_c = Rb_f$\\\hspace*{0.333em}\hspace*{0.333em}Approximately~solve~$H_c x_c = r_c$\\\hspace*{0.333em}\hspace*{0.333em}Interpolate~the~coarse-scale~solution~to~the~fine-grid,~add~it~back~to~$\tilde{x}_f$,~$\tilde{x}_f \leftarrow x_f + Px_c$\\\hspace*{0.333em}\hspace*{0.333em}Smooth~the~current~solution~using~a~particular~smoother~(Jacobi,~CG,~etc.)~to~produce~$\tilde{x}_f$\\\hspace*{0.333em}\hspace*{0.333em}Output:~$\tilde{x}_f$
\caption{Standard multigrid V-cycle}\label{alg:vcycle}
\end{scholmdAlgorithm*}

In the original work, the coarse-scale problem is solved with GMRES
preconditioned by the approximate inverse of the shifted Laplacian
system, which is applied via another V-cycle procedure. In this case, we
note that the coarse-scale problem is merely an instance of the original
problem and since we apply the V-cycle in~\eqref{alg:vcycle} to
precondition the original problem, we can recursively apply the same
method to precondition the coarse-scale problem as well. This process
cannot be iterated beyond two-levels typically because of the minimum
grid points per wavelength sampling requirements for wave phenomena
\citep{cohen2013higher}.

Unfortunately as-is, the method of \citep{calandra2013twogrid} was only
designed with the standard, 7-pt stencil in mind, rather than the more
robust 27-pt stencil of \citep{operto2007fd}. The work of
\citep{lago2015EAGEtrg} demonstrates that a straightforward application
of the previous preconditioner to the new stencil fails to converge. The
authors attempt to extend these ideas to this new stencil by replacing
the Jacobi iterations with the Carp-CG algorithm \citep{carpcg}, which
acts as a smoother in its own right. For realistically sized problems,
this method performs poorly as the Carp-CG method is inherently
sequential and attempts to parallelize it result in degraded convergence
performance \citep{carpcg}, in particular when implemented in a
stencil-based environment. As such, we propose to reuse the existing
fast matrix-vector kernels we have developed thus far and set our
smoother to be GMRES \citep{saad1986gmres} with an identity
preconditioner. Our coarse level solver is FGMRES
\citep{saad1993flexible}, which allows us to use our nonlinear,
iteration-varying preconditioner. Compared to our reference
stencil-based Carp sweep implementation in C, a single-threaded
matrix-vector multiplication is 30 times faster. In the context of
preconditioning linear systems, this means that unless the convergence
rate of the Carp sweeps are $30/k_s$ faster than the GMRES-based
smoothers, we should stick to the faster kernel for our problems.
Furthermore, we observed experimentally in testing that using a shifted
Laplacian preconditioner, as in
\citep{calandra2013twogrid, lago2015EAGEtrg} on the second level caused
an increase in the number of outer iterations, slows down convergence.
As such, we have replaced preconditioning the shifted Laplacian system
by preconditioning the Helmholtz itself, solved with FGMRES, which
results in much faster convergence.

\begin{figure}
\centering
\includegraphics[width=1.000\hsize]{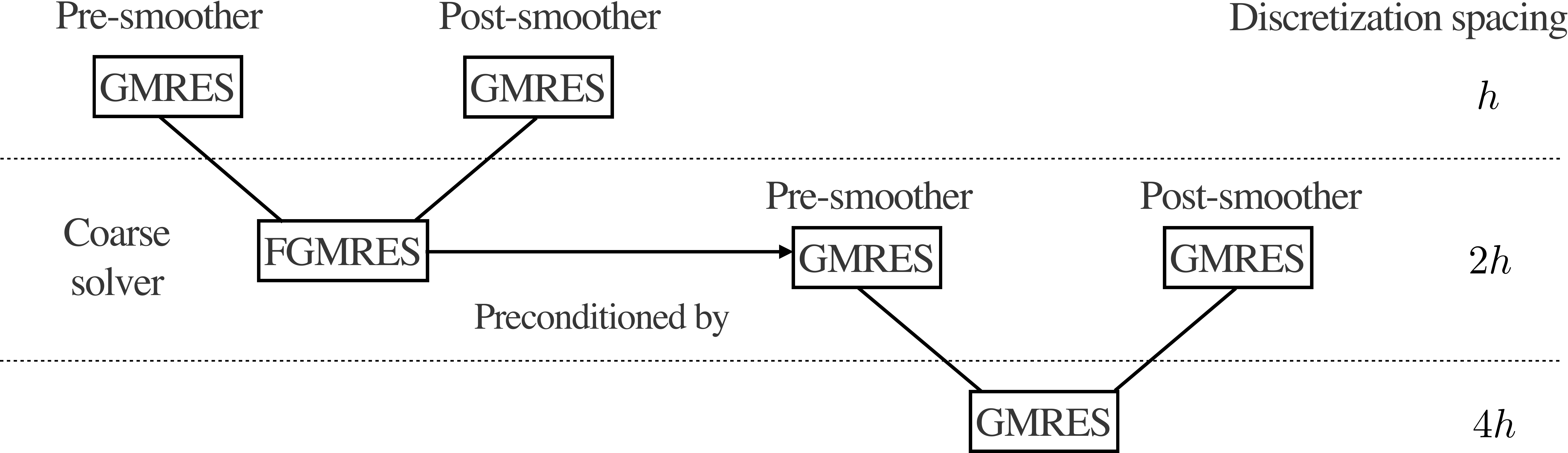}
\caption{ML-GMRES preconditioner. The coarse-level problem (relative to
the finest grid spacing) is preconditioned recursively with the same
method as the fine-scale problem.}\label{mlgmres}
\end{figure}

A diagram of the full algorithm, which we denote $ML-GMRES$, is depicted
in Figure~\eqref{mlgmres}. This algorithm is matrix-free, in that we do
not have to construct any of the intermediate matrices explicitly and
instead merely compute matrix-vector products, which will allow us to
apply this method to large systems.

We experimentally study the convergence behaviour of this preconditioner
on a 3D constant velocity model and leave a theoretical study of
ML-GMRES for future work. By fixing the preconditioner memory vector to
be $(k_{s,o},k_{s,i},k_{c,o},k_{c,i}) = (3,5,3,5)$, we can study the
performance of the preconditioner as we vary the number of wavelengths
in each direction, denoted $n_{\lambda}$, as well as the number of
points per wavelength, denoted $n_{\text{ppw}}$. For a fixed domain, as
the former quantity increases, the frequency increases while as the
latter quantity increases, the grid sampling decreases. As an aside, we
prefer to parametrize our problems in this manner as these quantities
are independent of the scaling of the domain, velocity, and frequency,
which are often obscured in preconditioner examples in research. These
two quantities, $n_{\lambda}$ and $n_{\text{ppw}}$, on the other hand,
are explicitly given parameters and comparable across experiments. We
append our model with a number of PML points equal to one wavelength on
each side. Using 5 threads per matrix-vector multiply, we use FGMRES
with 5 inner iterations as our outer solver, solve the system to a
relative residual of $10^{-6}$. The results are displayed in
Table~\eqref{table:precond_performance}.

\begin{table*}
\centering
\begin{tabular}{llll}
\toprule\addlinespace
$n_{\lambda} / n_{\text{ppw}}$ & 6 & 8 & 10\tabularnewline
\midrule
5 & 2 ($43^3$, 4.6) & 2 ($57^3$, 6.9) & 2 ($71^3$, 10.8)\tabularnewline
10 & 3 ($73^3$, 28.9) & 2 ($97^3$, 38.8) & 2 ($121^3$,
76.8)\tabularnewline
25 & 8 ($161^3$, 809) & 3 ($217^3$, 615) & 3 ($271^3$,
1009.8)\tabularnewline
40 & 11 ($253^3$, 4545) & 3 ($337^3$, 2795) & 3 ($421^3$,
4747)\tabularnewline
50 & 15 ($311^3$, 11789) & 3 ($417^3$, 5741) & 3 ($521^3$,
10373)\tabularnewline
\bottomrule
\end{tabular}
\caption{Preconditioner performance as a function of varying points per
wavelength and number of wavelengths. Values are number of outer FGMRES
iterations. In parenthesis are displayed the number of grid points
(including the PML) and overall computational time (in seconds),
respectively.}\label{table:precond_performance}
\end{table*}

We upper bound the memory usage of ML-GMRES as follows. We note that the
FGMRES($k_o,k_i$), GMRES($k_o,k_i$) solvers, with $k_o$ outer iterations
and $k_i$ inner iterations, store $2k_i+6$ vectors and $k_i+5$ vectors,
respectively. In solving the discretized $H(m)u = q$ with $N = n^3$
complex-valued unknowns, we require $(2k_i+6)N$ memory for the outer
FGMRES solver, $2N$ for additional vectors in the V-cycle, in addition
to $(k_{s,i}+5)N$ memory for the level-1 smoother, $(2k_{c,i}+6)(N/8)$
memory for the level-2 outer solver, $(k_{s,i}+5)(N/8)$ memory for the
level-2 smoother, and $(k_{c,i}+5)(N/64)$ memory for the level-3 solver.
The total peak memory of the entire solver is therefore
\begin{equation*}
(2k_i+6)N + \max( (k_{s,i}+5)N, (2k_{c,i} + 6)(N/8) + \max((k_{s,i}+5)N/8 , (2k_{c,i}+6)N/64))
\end{equation*}
 Using the same memory settings as before, our preconditioner requires
at most $26$ vectors to be stored. Although the memory requirements of
our preconditioner are seemingly large, they are approximately the same
cost as storing the entire 27-pt stencil coefficients of $H(m)$
explicitly and much smaller than the LU factors in (Table 1,
\citep{operto2007fd}). The memory requirements of this preconditioner
can of course be reduced by reducing $k$, $k_s$ or $k_c$, at the expense
of an increased computational time per wavefield solve. To compute the
real world memory usage of this preconditioner, we consider a constant
model problem with $10$ wavelengths and a varying number of points per
wavelength so that the total number of points, including the PML region,
is fixed. In short, we are not changing the effective difficulty of the
problem, but merely increase the oversampling factor to ensure that the
convergence remains the same with increasing model size. The results in
Table~\eqref{table:precmem} indicate that ML-GMRES is performing
slightly better than expected from a memory point of view and the
computational time scaling is as expected.

\begin{table*}
\centering
\begin{tabular}{llll}
\toprule\addlinespace
Grid Size & Time(s) & Peak memory (GB) & Number of vectors of size
$N$\tabularnewline
\midrule
$128^3$ & 46 & 0.61 & $20N$\tabularnewline
$256^3$ & 213 & 5.8 & $23N$\tabularnewline
$512^3$ & 1899 & 48.3 & $24N$\tabularnewline
\bottomrule
\end{tabular}
\caption{Memory usage for a constant-velocity problem as the number of
points per wavelength increases}\label{table:precmem}
\end{table*}

Despite the strong empirical performance of this preconditioner, we note
that performance tends to degrade as $n_{\text{ppw}} < 8$, even though
the 27-pt compact stencil is rated for $n_{\text{ppw}} = 4$ in
\citep{operto2007fd}. Intuitively this makes sense as on the coarsest
grid $n_{\text{ppw}} < 2$, which is under the Nyquist limit, and leads
to stagnating convergence. In our experience, this discrepancy did not
disappear when using the shifted Laplacian preconditioner on the
coarsest level. It remains to be seen how multigrid methods for the
Helmholtz equation with $n_{\text{ppw}} < 8$ will be developed in future
research.

\section{Numerical Examples}\label{pdenumericalexamples}

\subsection{Validation}\label{validation}

\begin{figure*}
\centering
\includegraphics[width=0.500\hsize]{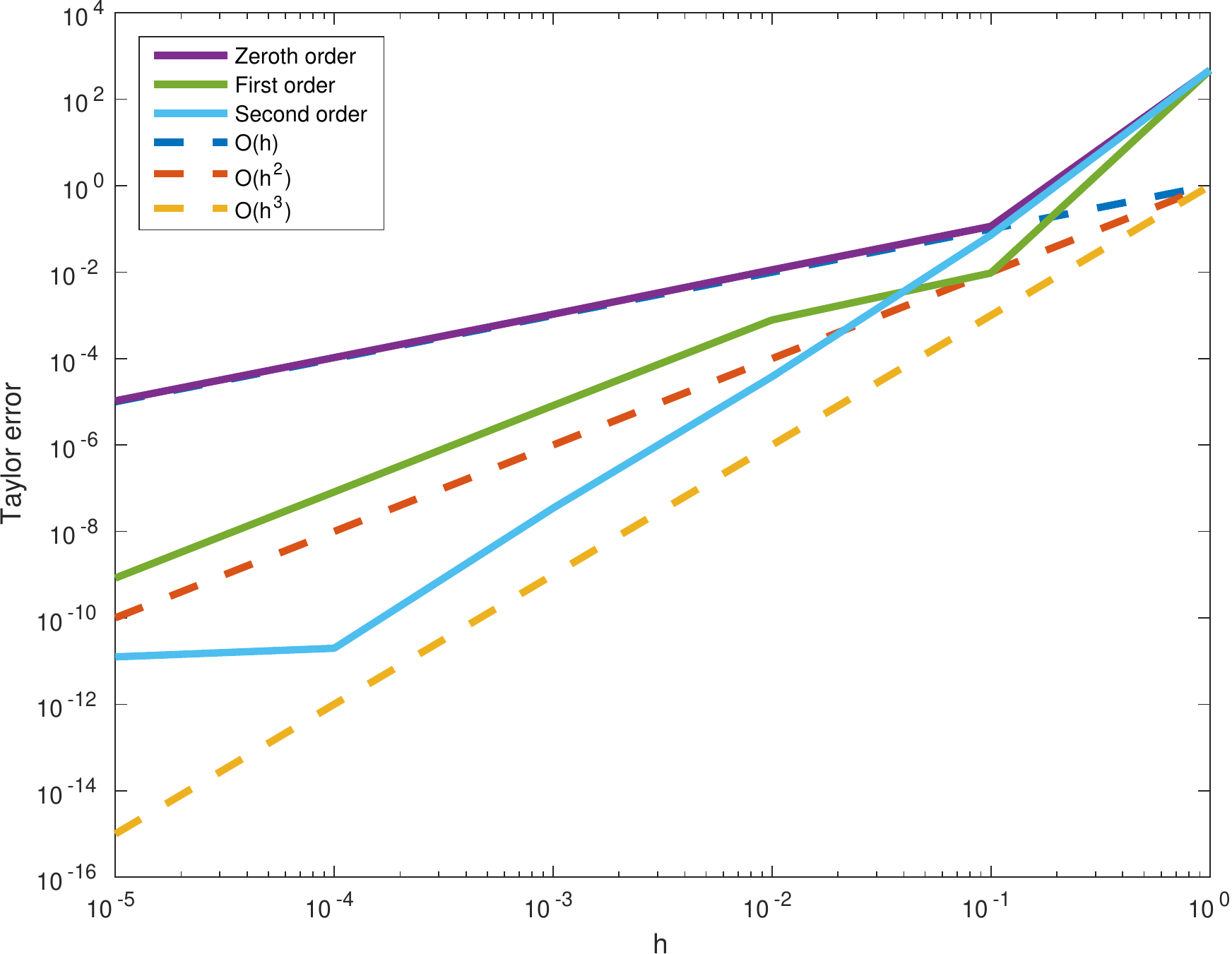}
\caption{Numerical Taylor error for a 3D reference
model.}\label{taylorerr}
\end{figure*}

To verify the validity of this implementation, specifically that the
code as implemented reflects the underlying mathematics, we ensure that
the following tests pass. The Taylor error test stipulates that, for a
sufficiently smooth multivariate function $f(m)$ and an arbitrary
perturbation $\delta m$, we have that
\begin{equation*}
\begin{aligned}
&f(m+h\delta m) - f(m) &=&\; O(h)\\
&f(m+h\delta m) - f(m) - h \langle \nabla f(m), \delta m \rangle &=& \; O(h^2)\\
&f(m+h\delta m) - f(m) - h \langle \nabla f(m), \delta m \rangle - \dfrac{h^2}{2} \langle \delta m, \nabla^2 f(m)\delta m\rangle &=&\; O(h^3).
\end{aligned}
\end{equation*}
 We verify this behaviour numerically for the least-squares misfit
$f(m)$ and for a constant 3D velocity model $m$ and a random
perturbation $\delta m$ by solving for our fields to a very high
precision, in this case up to a relative residual of $10^{-10}$. As
shown in Figure~\eqref{taylorerr}, our numerical codes indeed pass this
test, up until the point where $h$ becomes so small that the numerical
error in the field solutions dominates the other sources of error.

The adjoint test requires us to verify that, for the functions
implementing the forward and adjoint matrix-vector products of a linear
operator $A$, we have, numerically,
\begin{equation*}
\langle Ax, y \rangle = \langle x, A^H y \rangle,
\end{equation*}
 for all vectors $x,y$ of appropriate length. We suffice for testing
this equality for randomly generated vectors $x$ and $y$, made complex
if necessary. Owing to the presence of the PML extension operator for
the Helmholtz equation, we set $x$ and $y$, if necessary, to be
zero-filled in the PML-extension domain and along the boundary of the
original domain. We display the results in
Table~\eqref{table:adjointtest}.

\begin{table*}
\centering
\begin{tabular}{llll}
\toprule\addlinespace
& $\langle Ax, y\rangle $ & $\langle x, A^Hy\rangle$ & Relative
difference\tabularnewline
\midrule
Helmholtz & $6.5755 - 5.2209 i \cdot 10^{0}$ &
$6.5755 - 5.2209 i \cdot 10^{0}$ &
$2.9004 \cdot 10^{-15}$\tabularnewline
Jacobian & $1.0748 \cdot 10^{-1}$ & $1.0748 \cdot 10^{-1}$ &
$1.9973 \cdot 10^{-9}$\tabularnewline
Hessian & $-1.6465 \cdot 10^{-2}$ & $-1.646 \cdot 10^{-2}$ &
$1.0478 \cdot 10^{-10}$\tabularnewline
\bottomrule
\end{tabular}
\caption{Adjoint test results for a single instance of randomly
generated vectors $x$, $y$, truncated to four digits for spacing
reasons. The linear systems involved are solved to the tolerance of
$10^{-10}$.}\label{table:adjointtest}
\end{table*}

\begin{figure}
\centering
\captionsetup[subfigure]{labelformat=empty}
\subfloat[Analytic
solution]{\includegraphics[width=0.250\hsize]{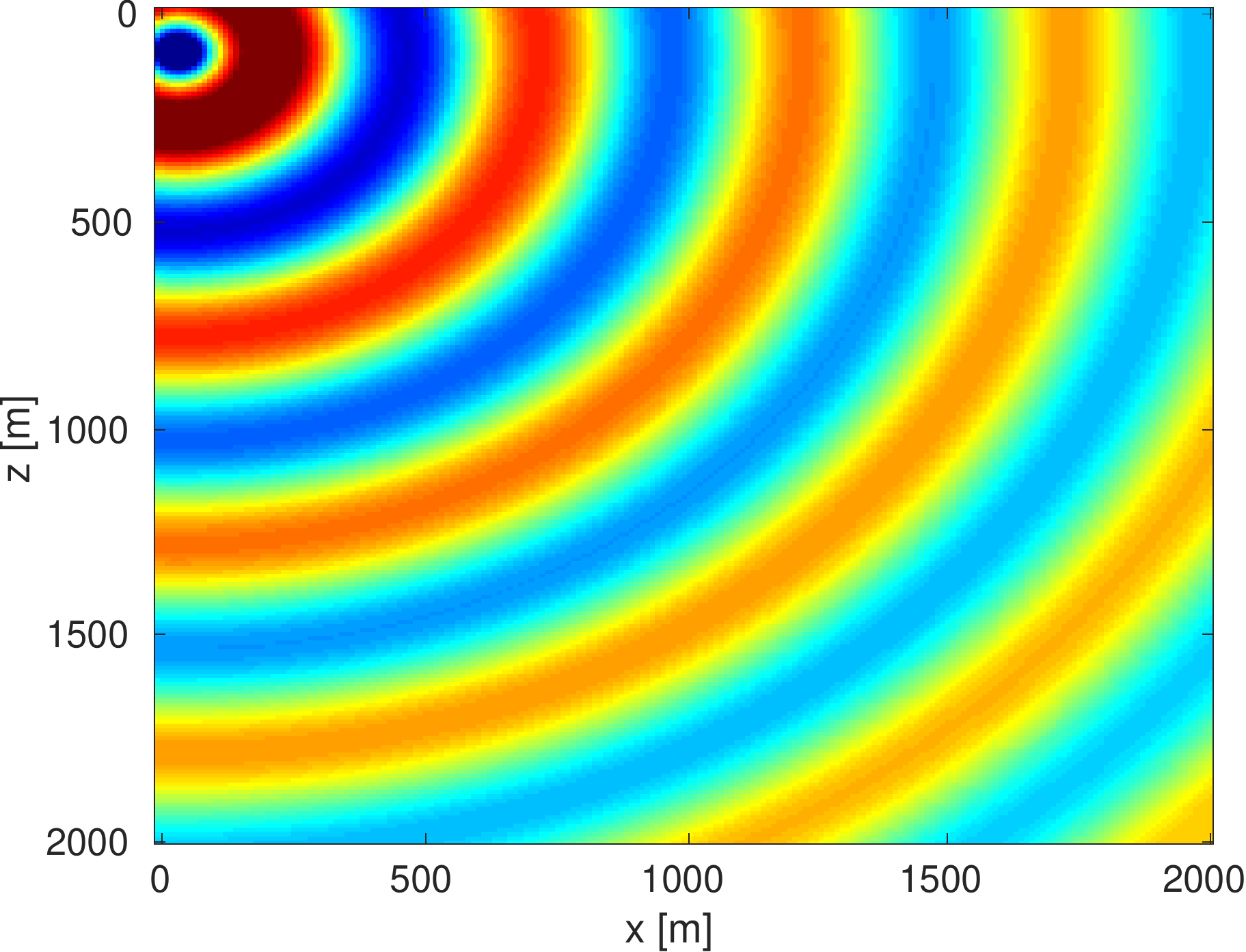}}
\subfloat[Computed
solution]{\includegraphics[width=0.250\hsize]{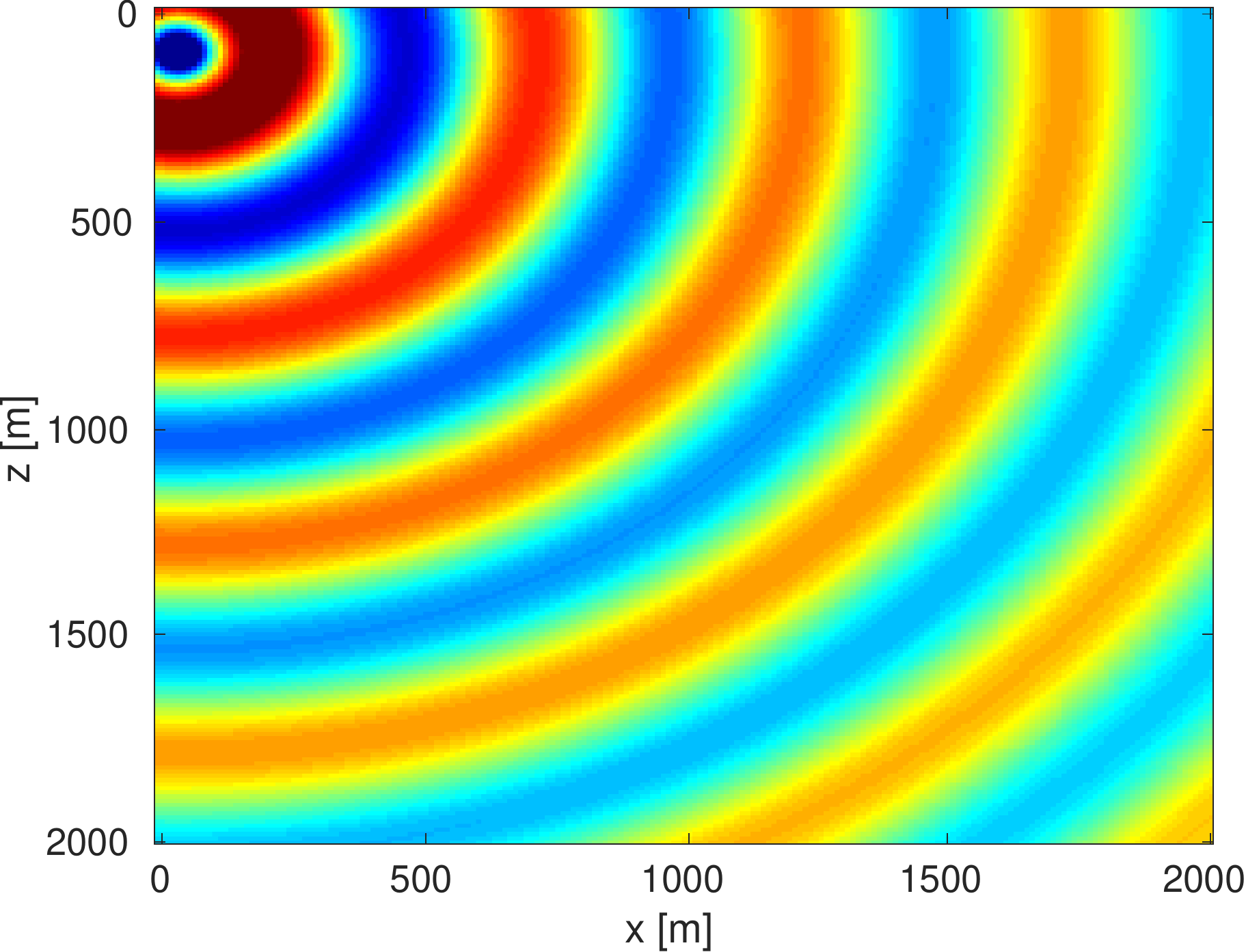}}
\subfloat[Difference
x100]{\includegraphics[width=0.250\hsize]{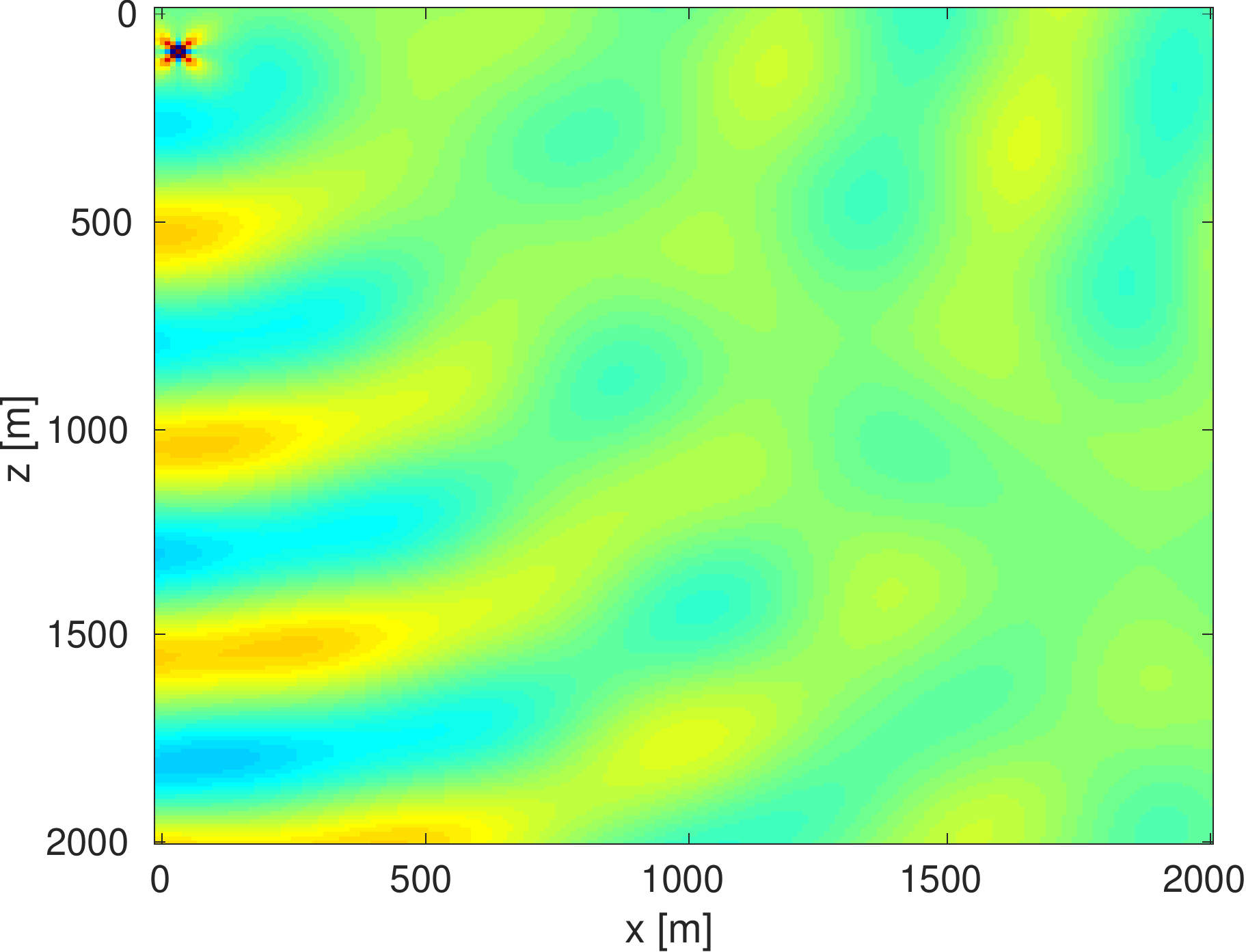}}
\\
\subfloat[]{\includegraphics[width=0.250\hsize]{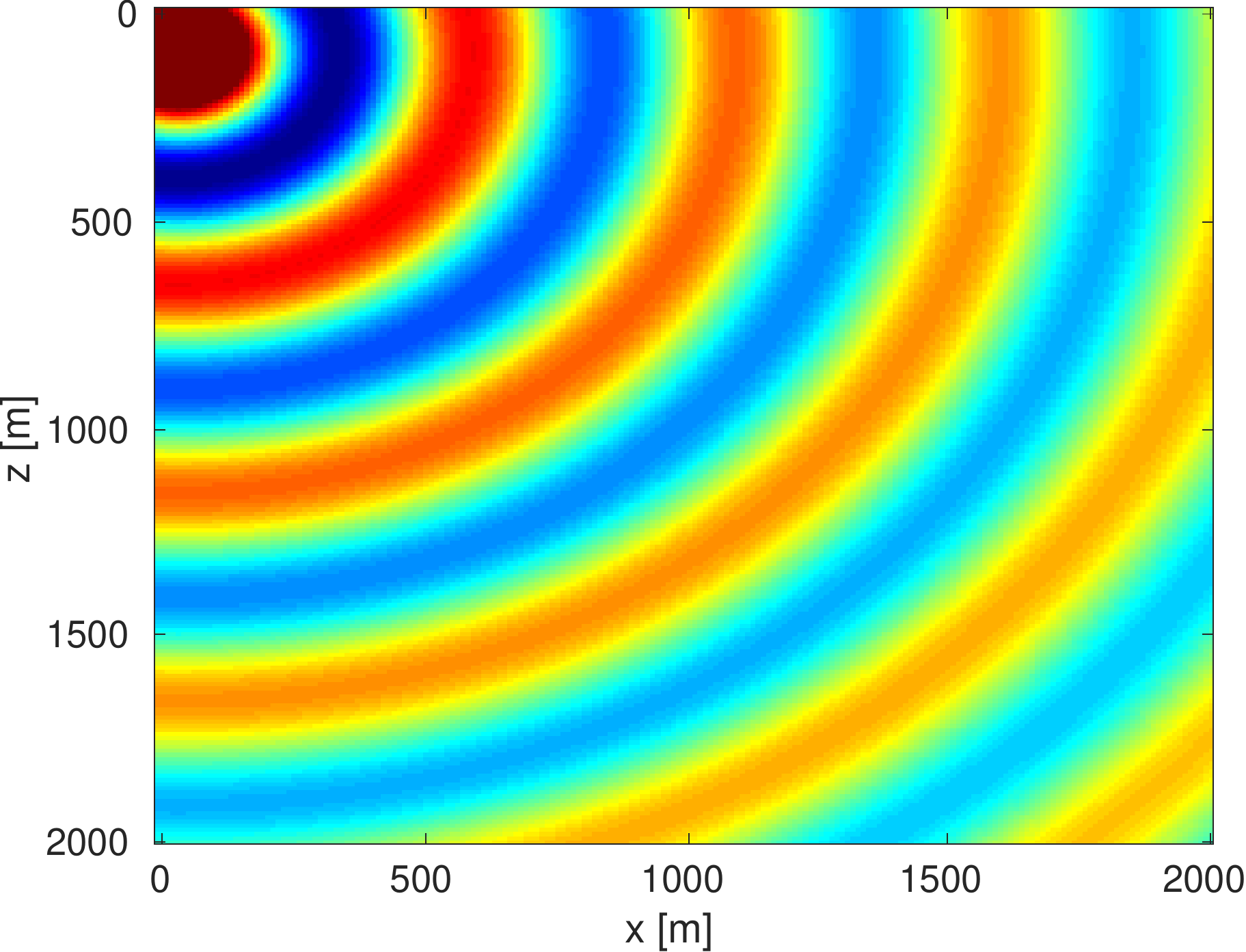}}
\subfloat[]{\includegraphics[width=0.250\hsize]{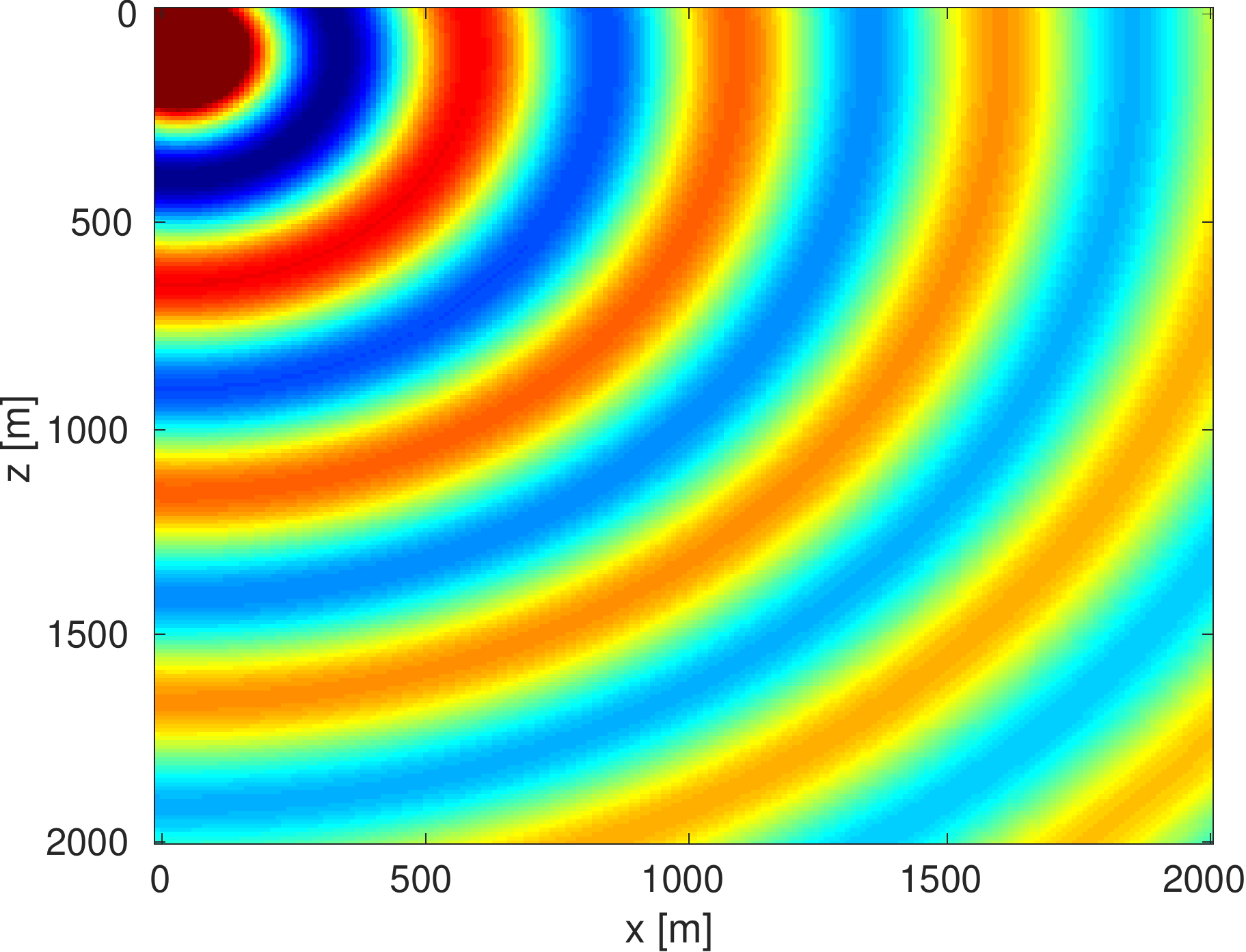}}
\subfloat[]{\includegraphics[width=0.250\hsize]{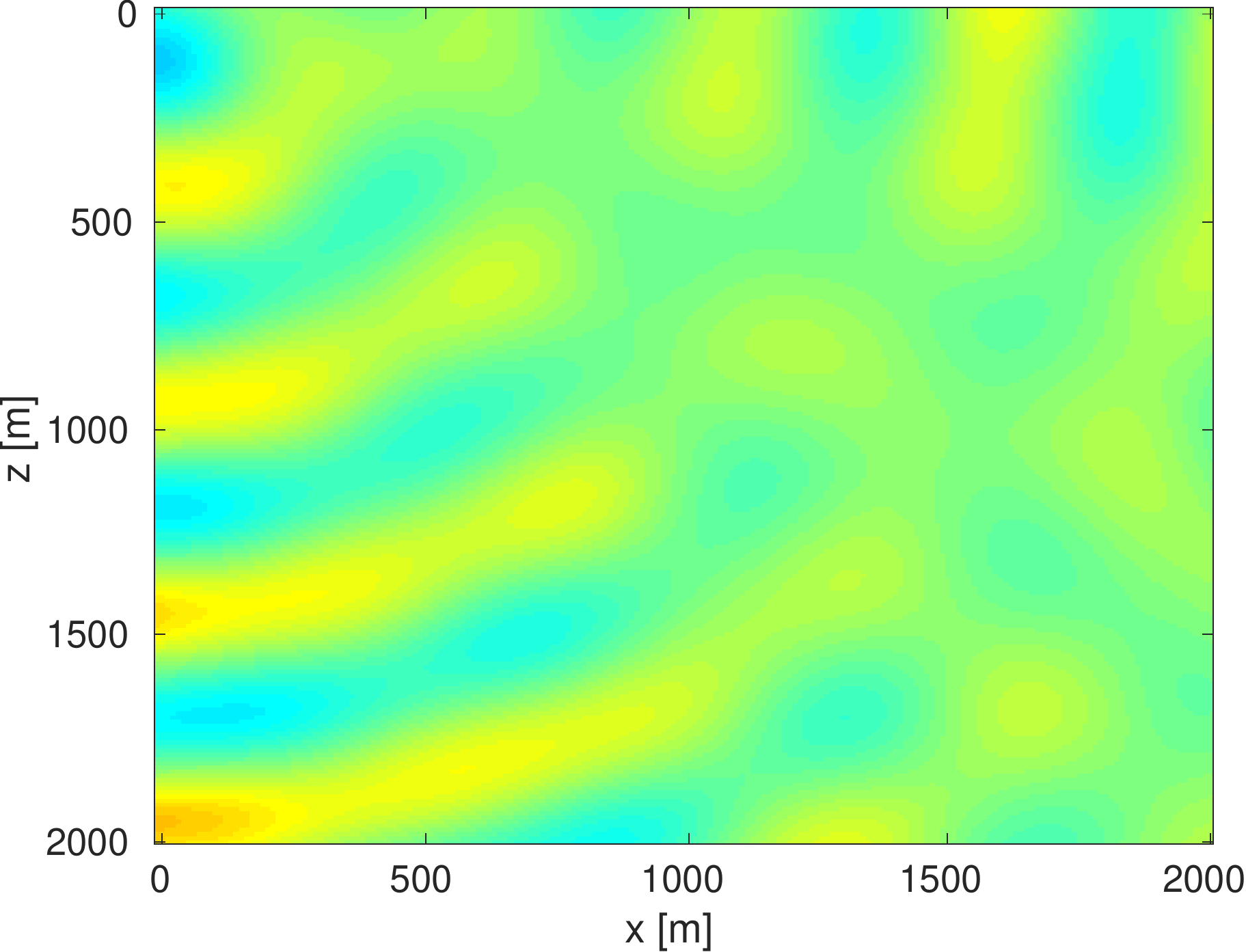}}
\caption{Analytic and numerical solutions for the 2D Helmholtz equation
for a single source. Difference is displayed on a colorbar 100x smaller
than the solutions. Top row is the real part, bottom row is the
imaginary part.}\label{analytic2D}
\end{figure}

\begin{figure}
\centering
\captionsetup[subfigure]{labelformat=empty}
\subfloat[Analytic
solution]{\includegraphics[width=0.250\hsize]{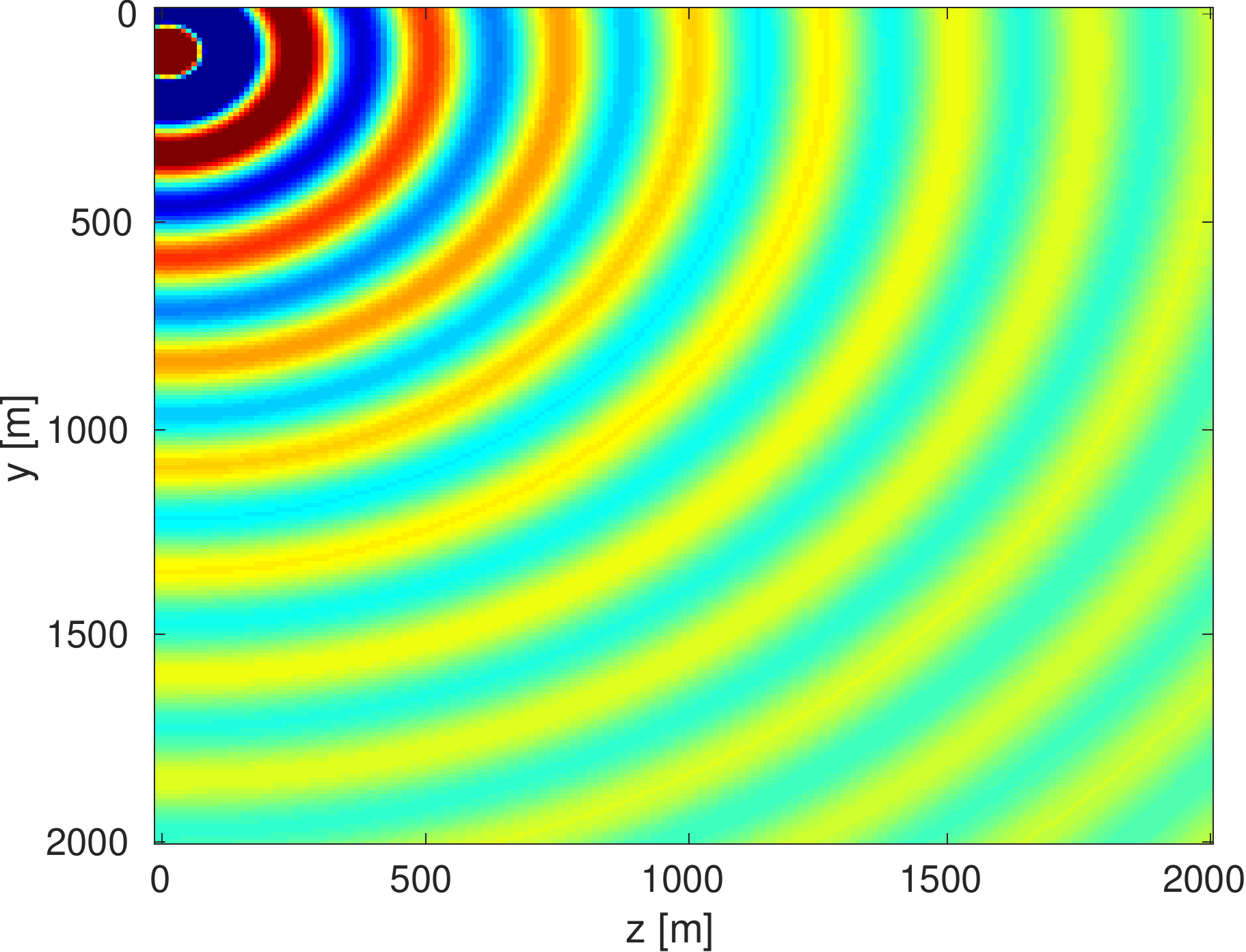}}
\subfloat[Computed
solution]{\includegraphics[width=0.250\hsize]{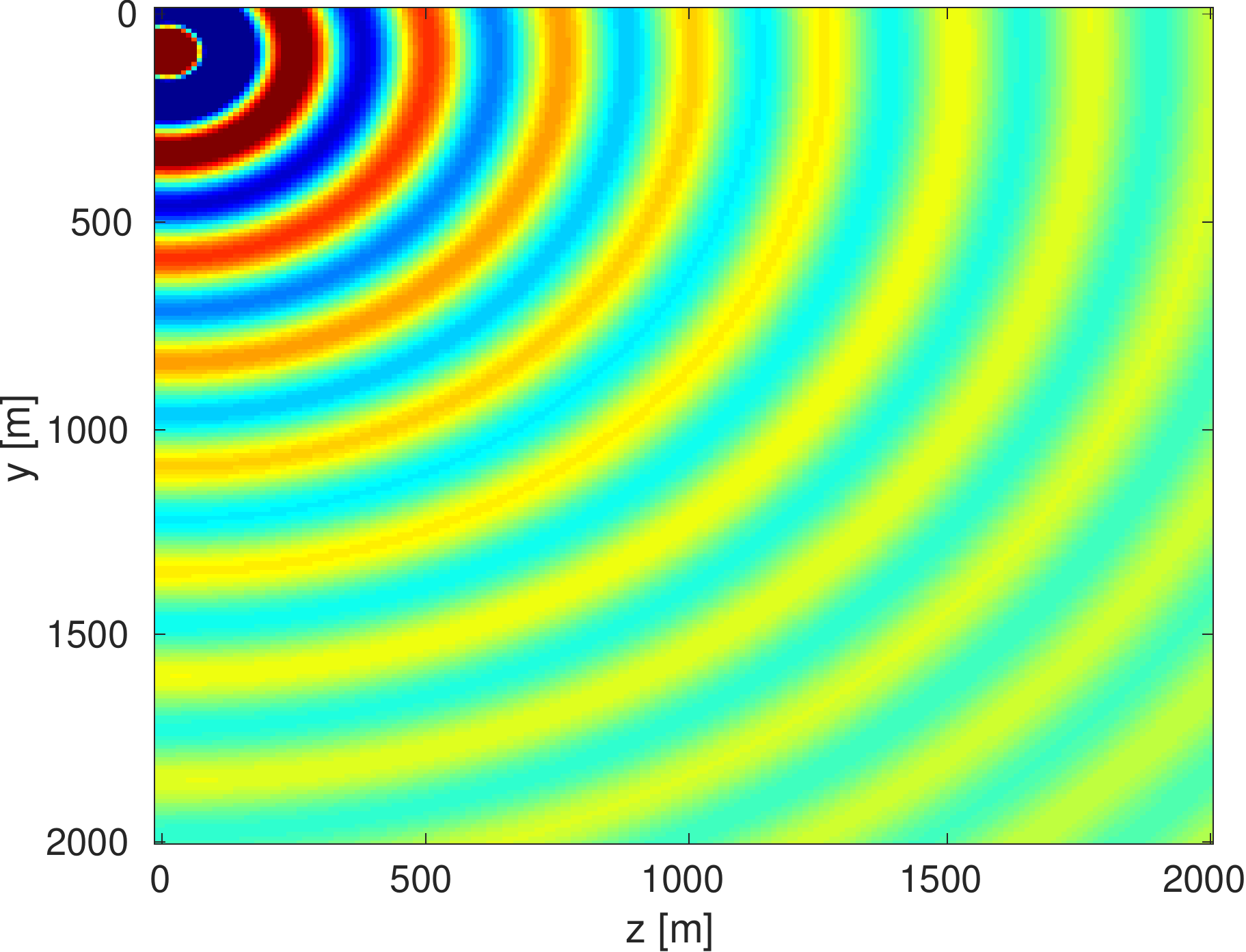}}
\subfloat[Difference
x10]{\includegraphics[width=0.250\hsize]{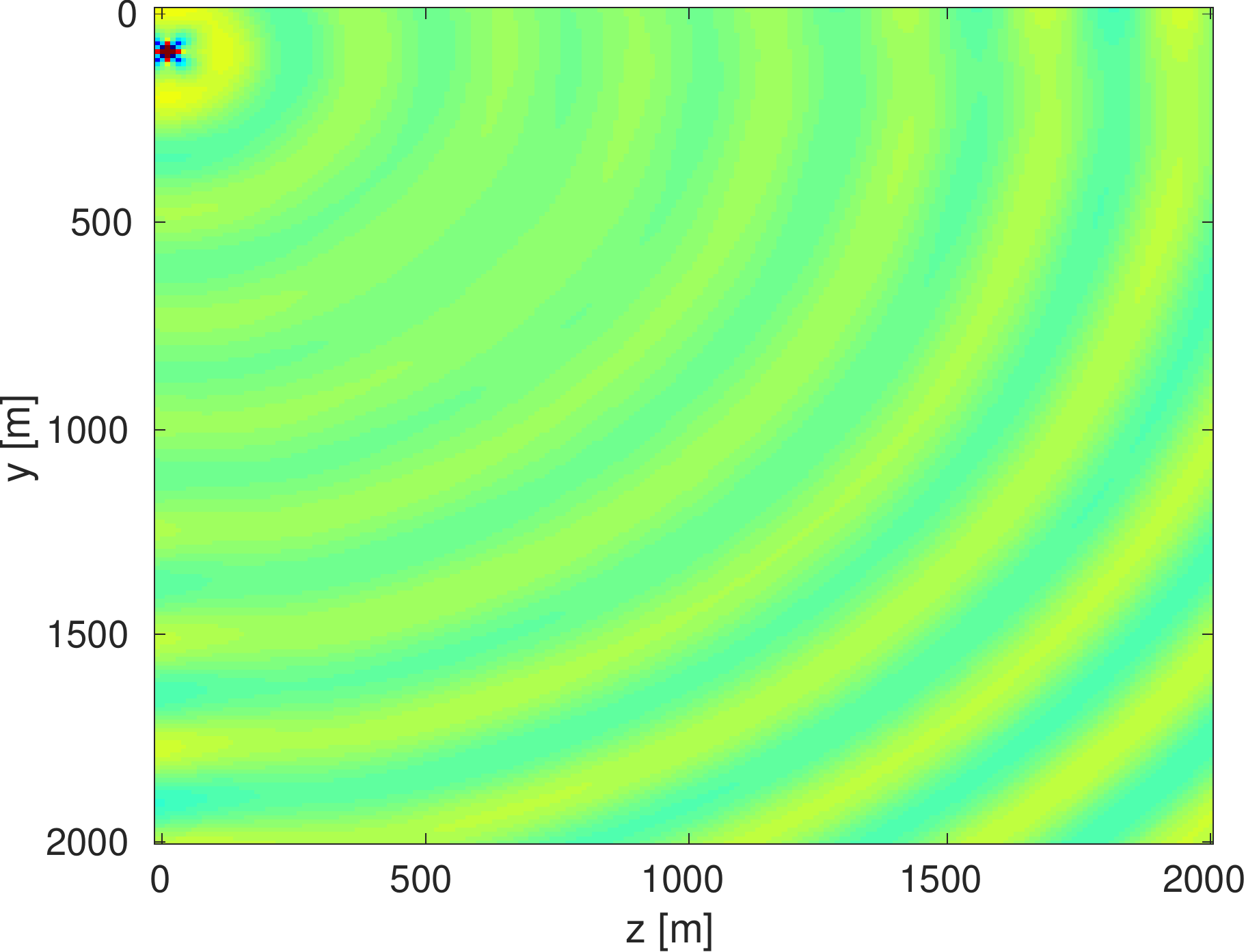}}
\\
\subfloat[]{\includegraphics[width=0.250\hsize]{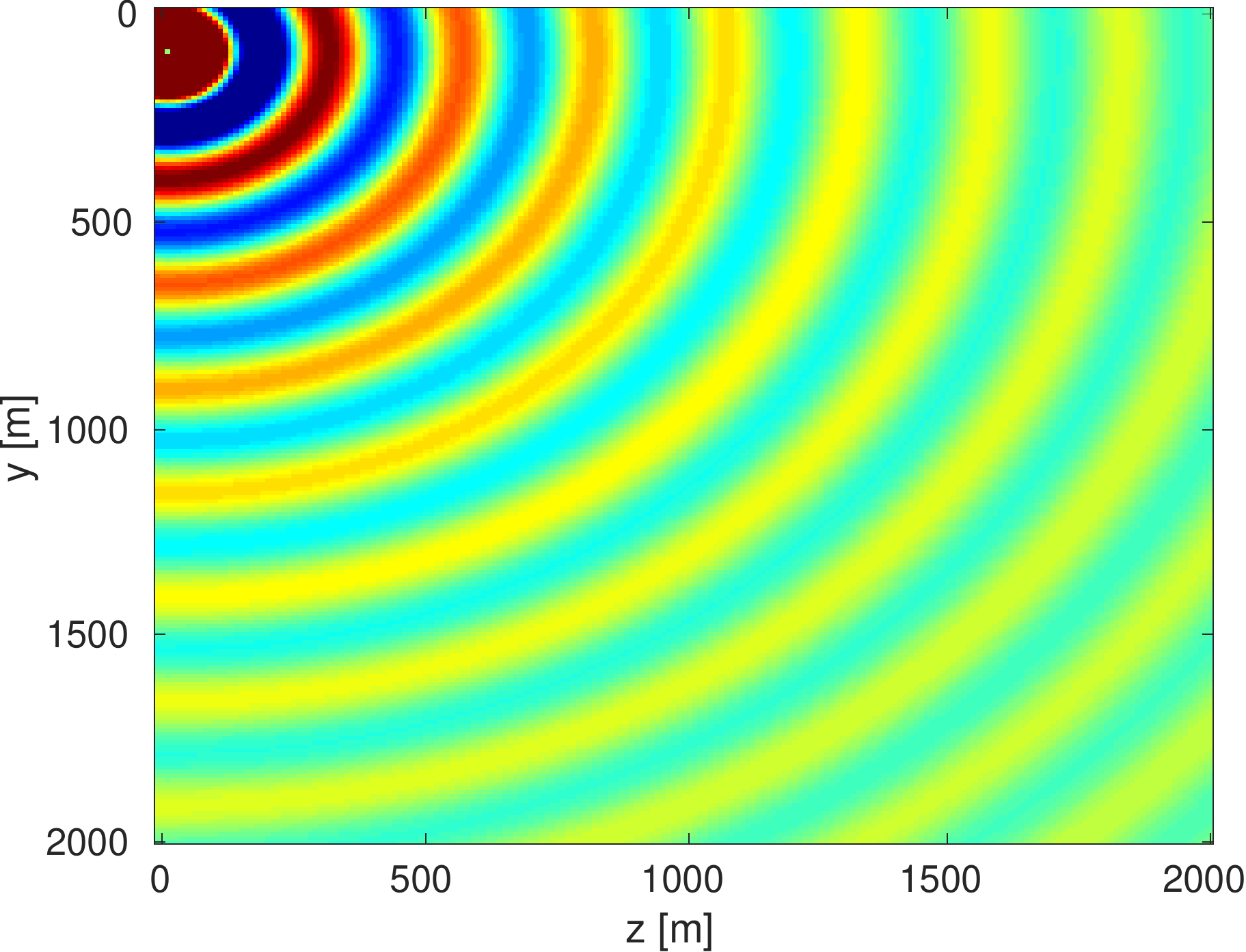}}
\subfloat[]{\includegraphics[width=0.250\hsize]{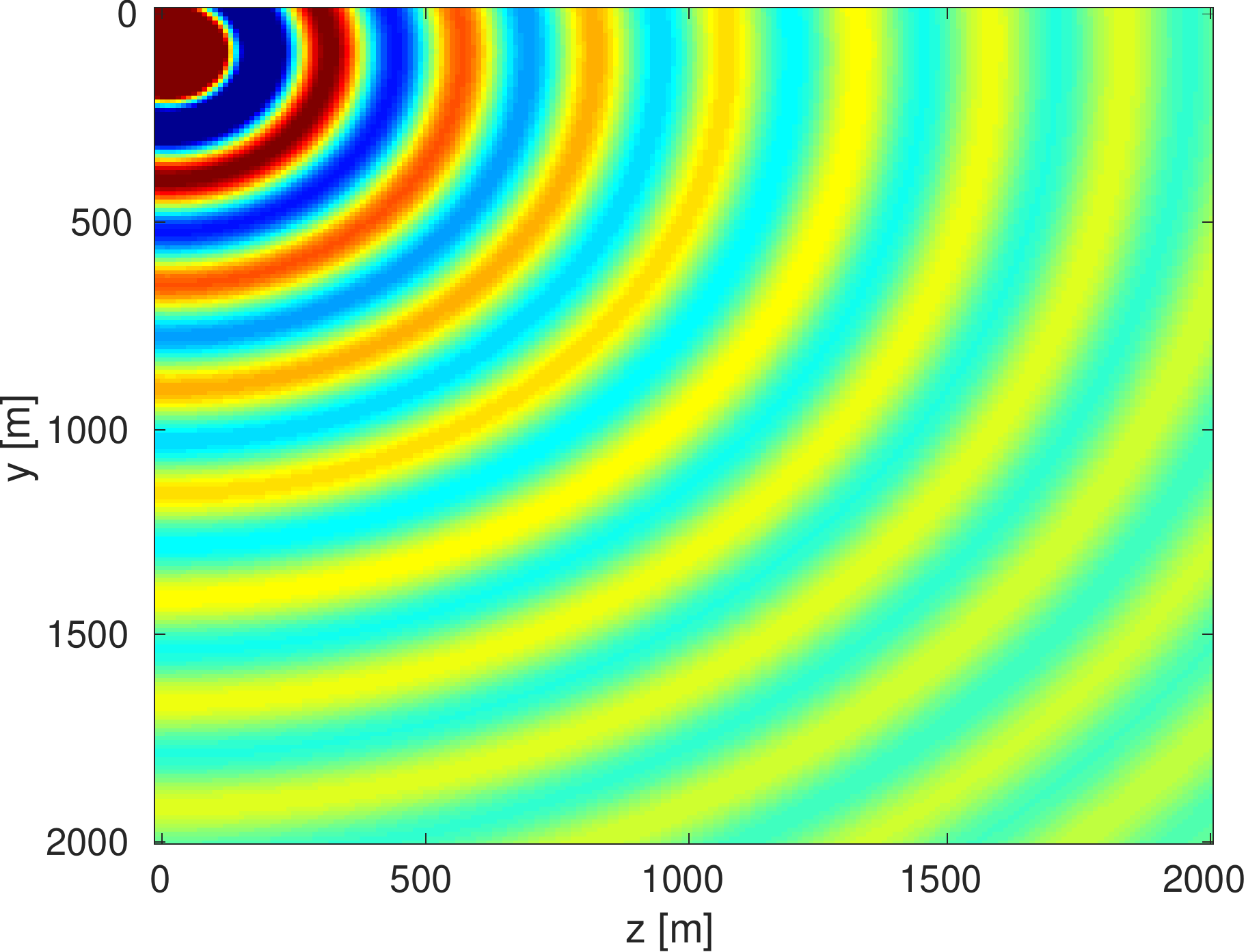}}
\subfloat[]{\includegraphics[width=0.250\hsize]{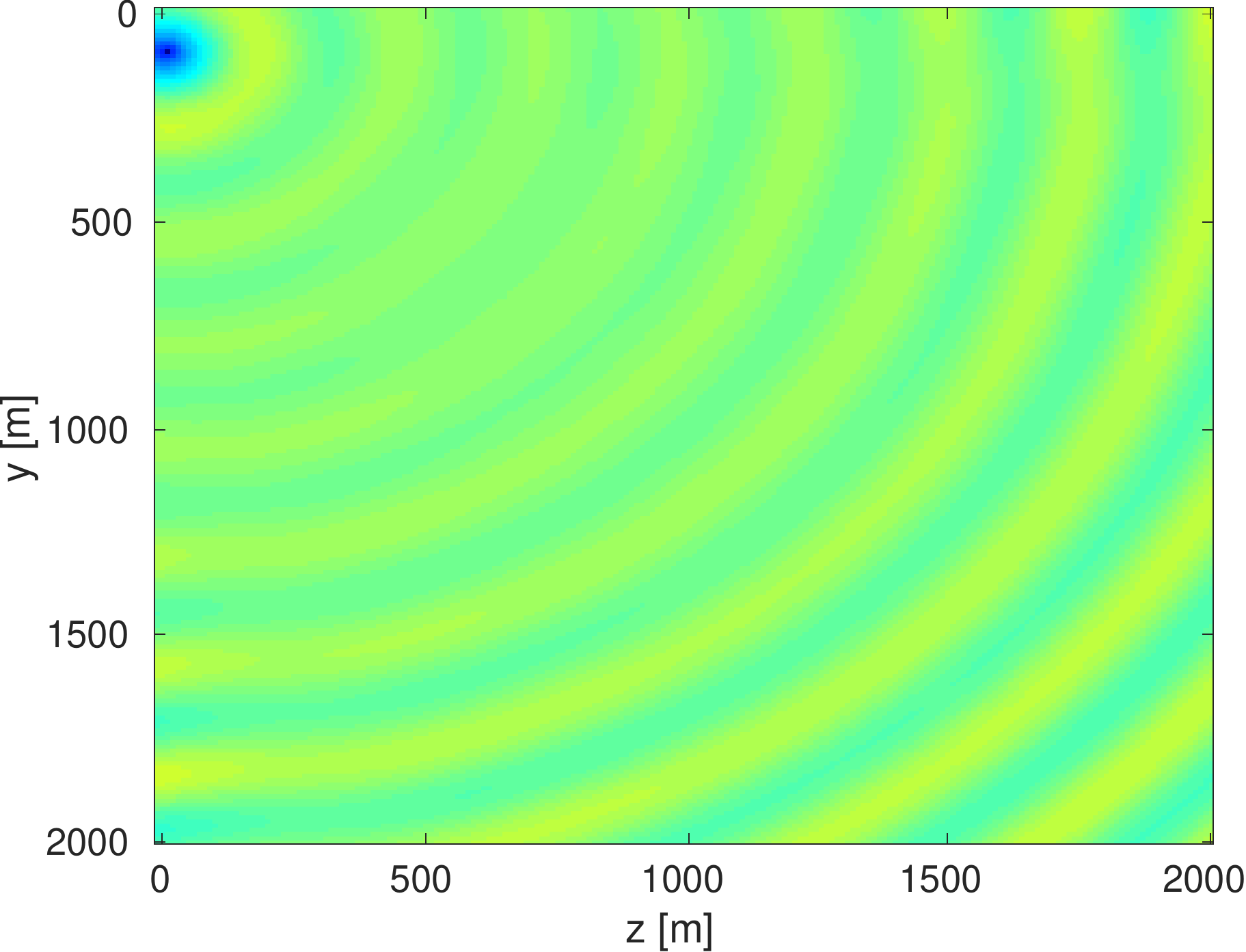}}
\caption{Analytic and numerical solutions for the 3D Helmholtz equation
(depicted as a 2D slice) for a single source. Difference is displayed on
a colorbar 10x smaller than the solutions. Top row is the real part,
bottom row is the imaginary part.}\label{analytic3D}
\end{figure}

\begin{figure}
\centering
\captionsetup[subfigure]{labelformat=empty}
\subfloat[Analytic]{\includegraphics[width=0.200\hsize]{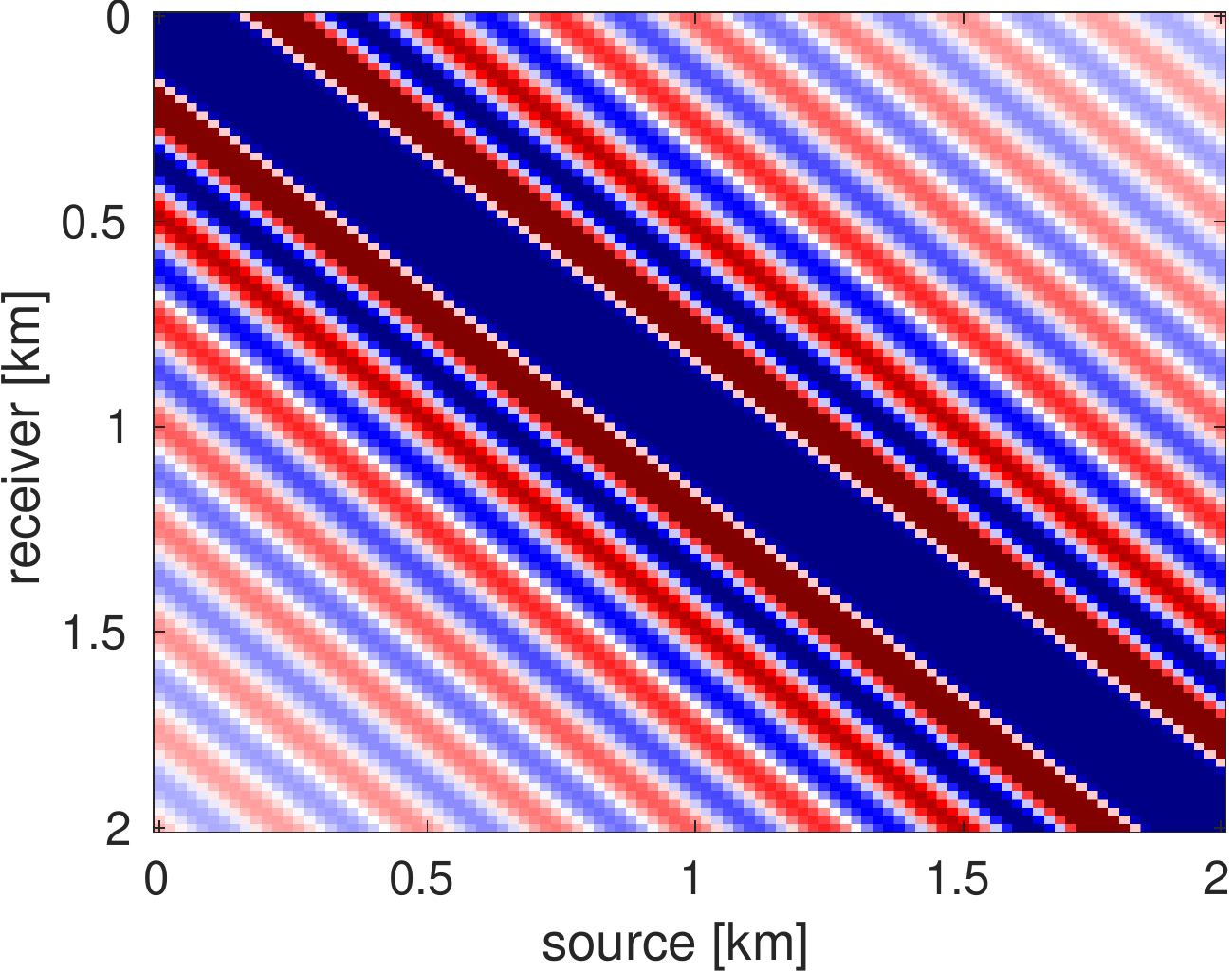}}
\subfloat[2.5D]{\includegraphics[width=0.200\hsize]{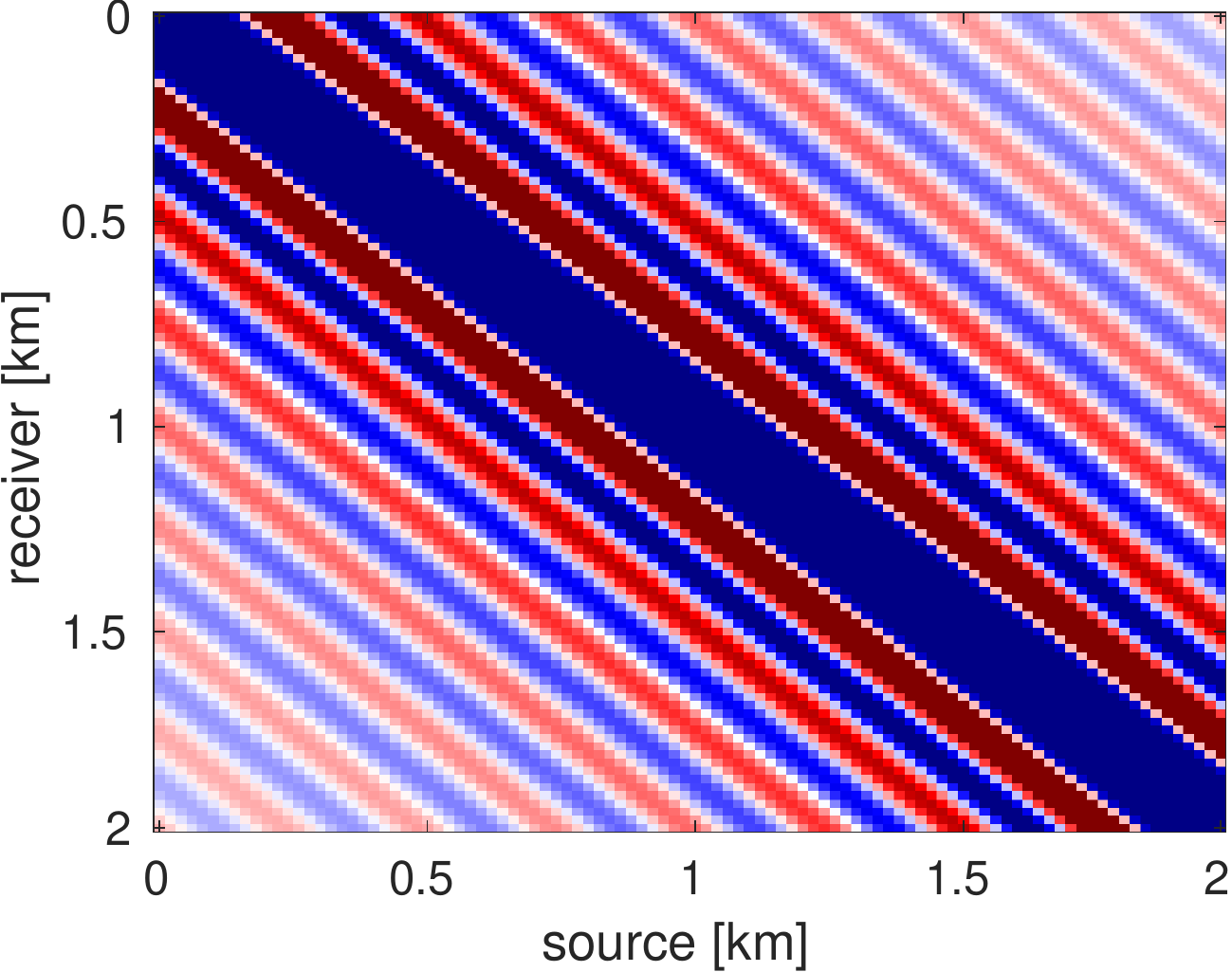}}
\subfloat[2.5D
difference]{\includegraphics[width=0.200\hsize]{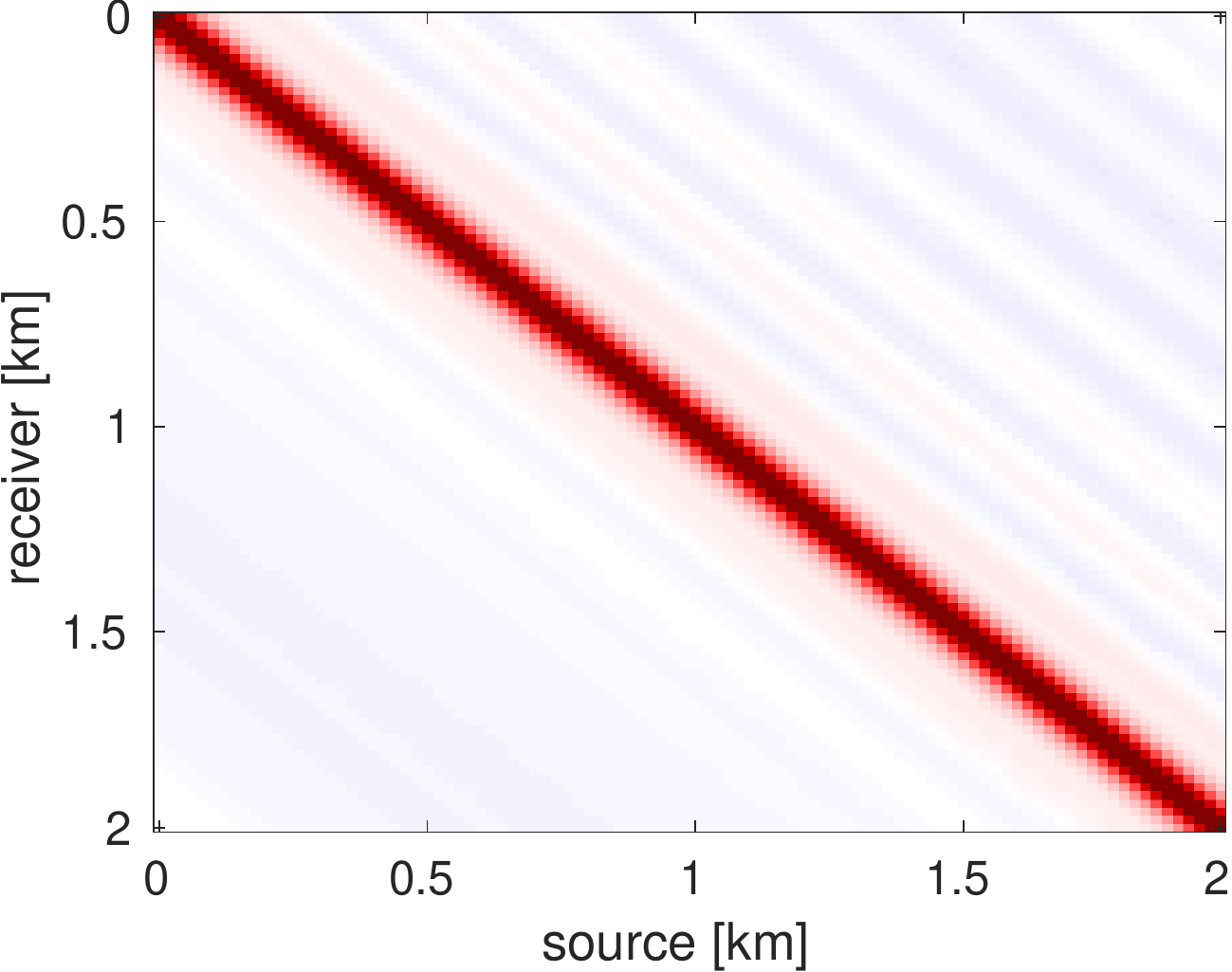}}
\subfloat[3D]{\includegraphics[width=0.200\hsize]{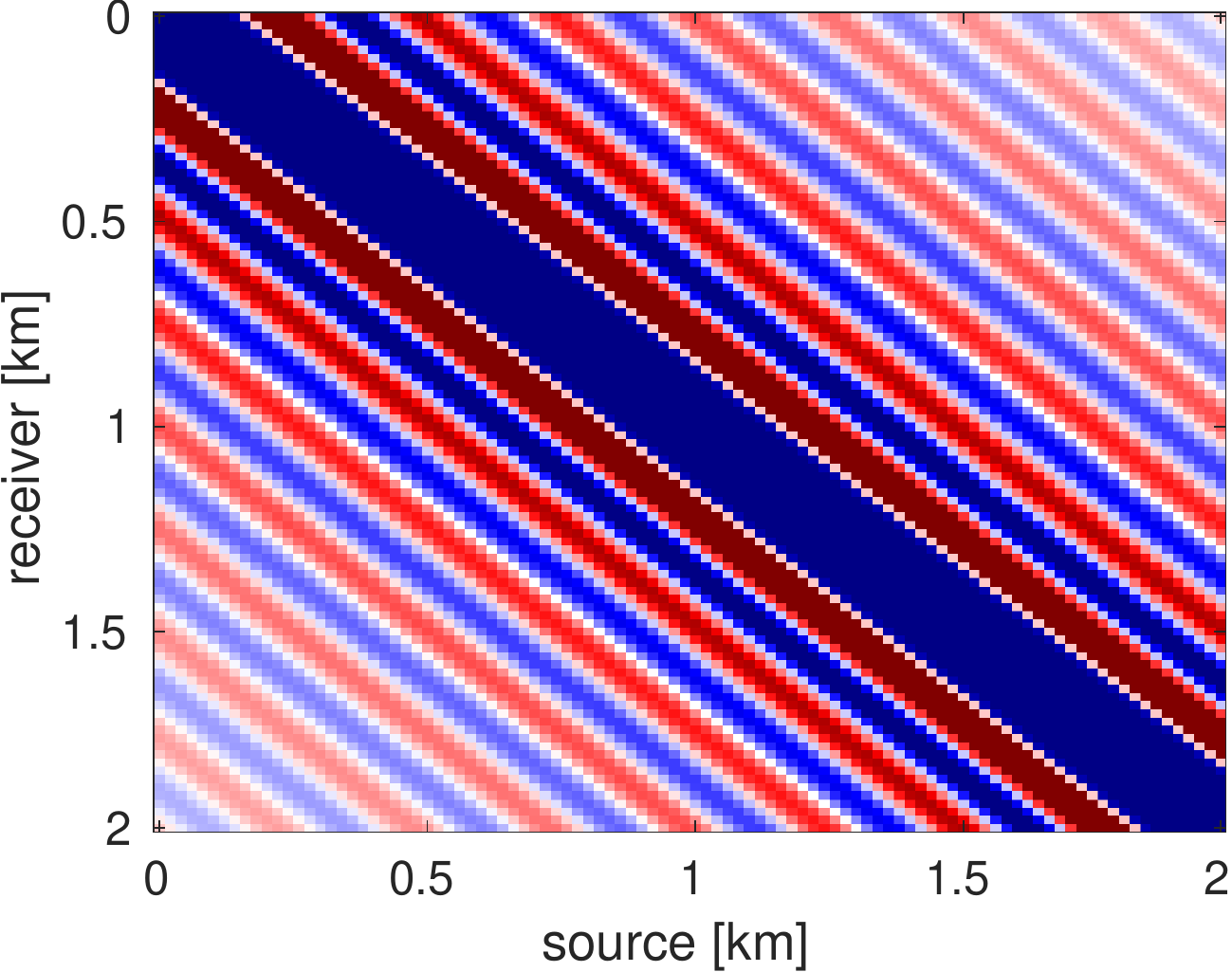}}
\subfloat[3D
difference]{\includegraphics[width=0.200\hsize]{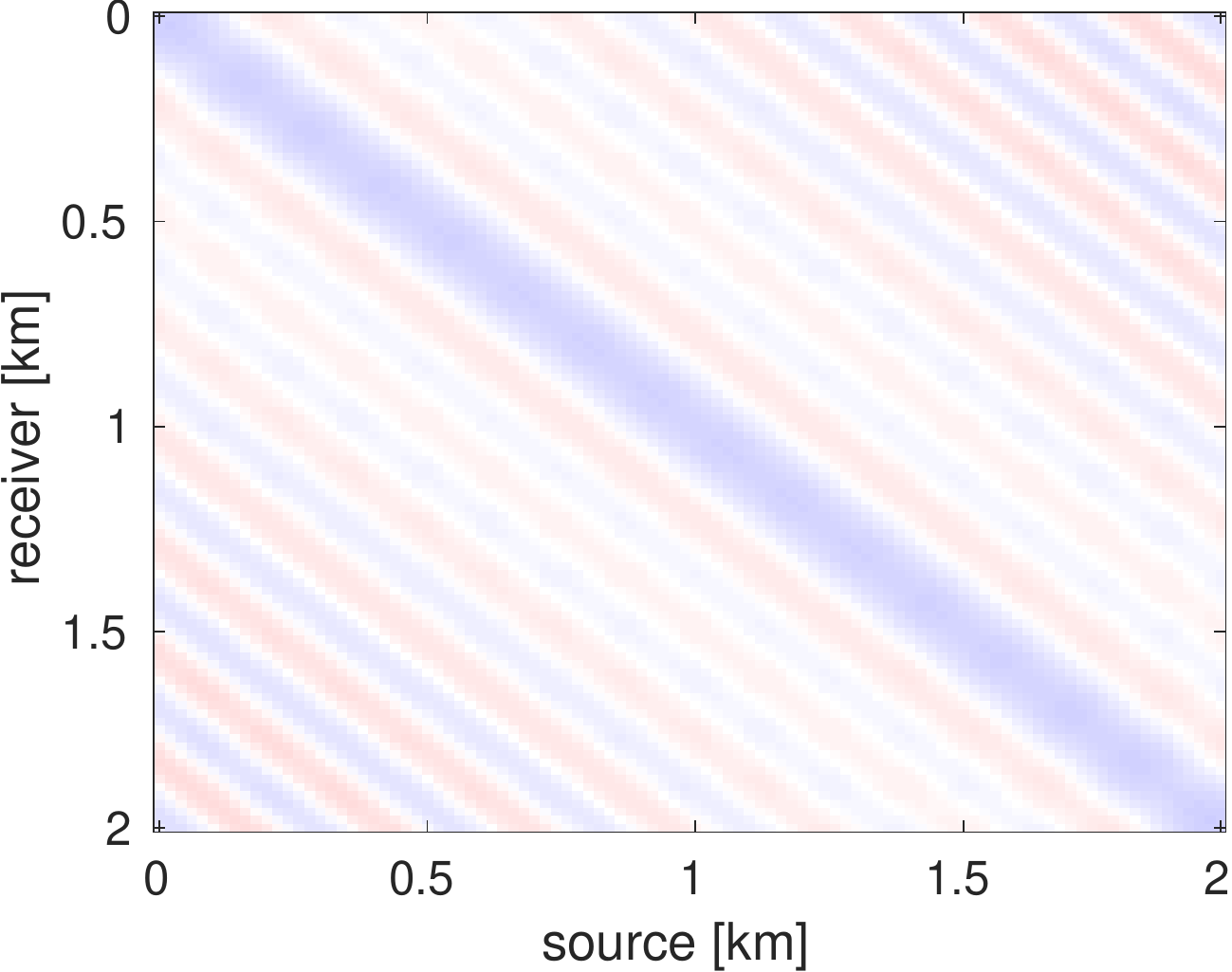}}
\\
\subfloat[Analytic]{\includegraphics[width=0.200\hsize]{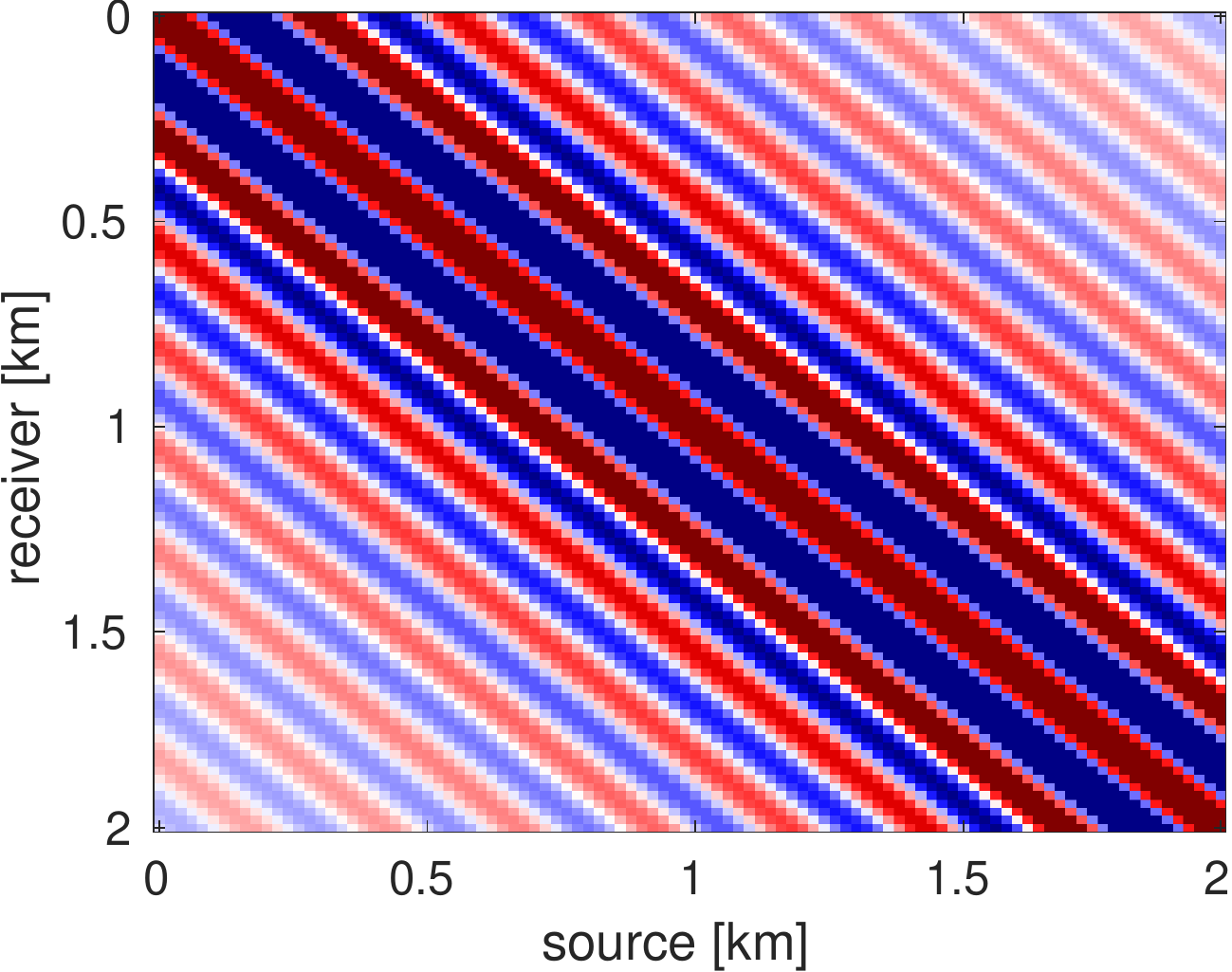}}
\subfloat[2.5D]{\includegraphics[width=0.200\hsize]{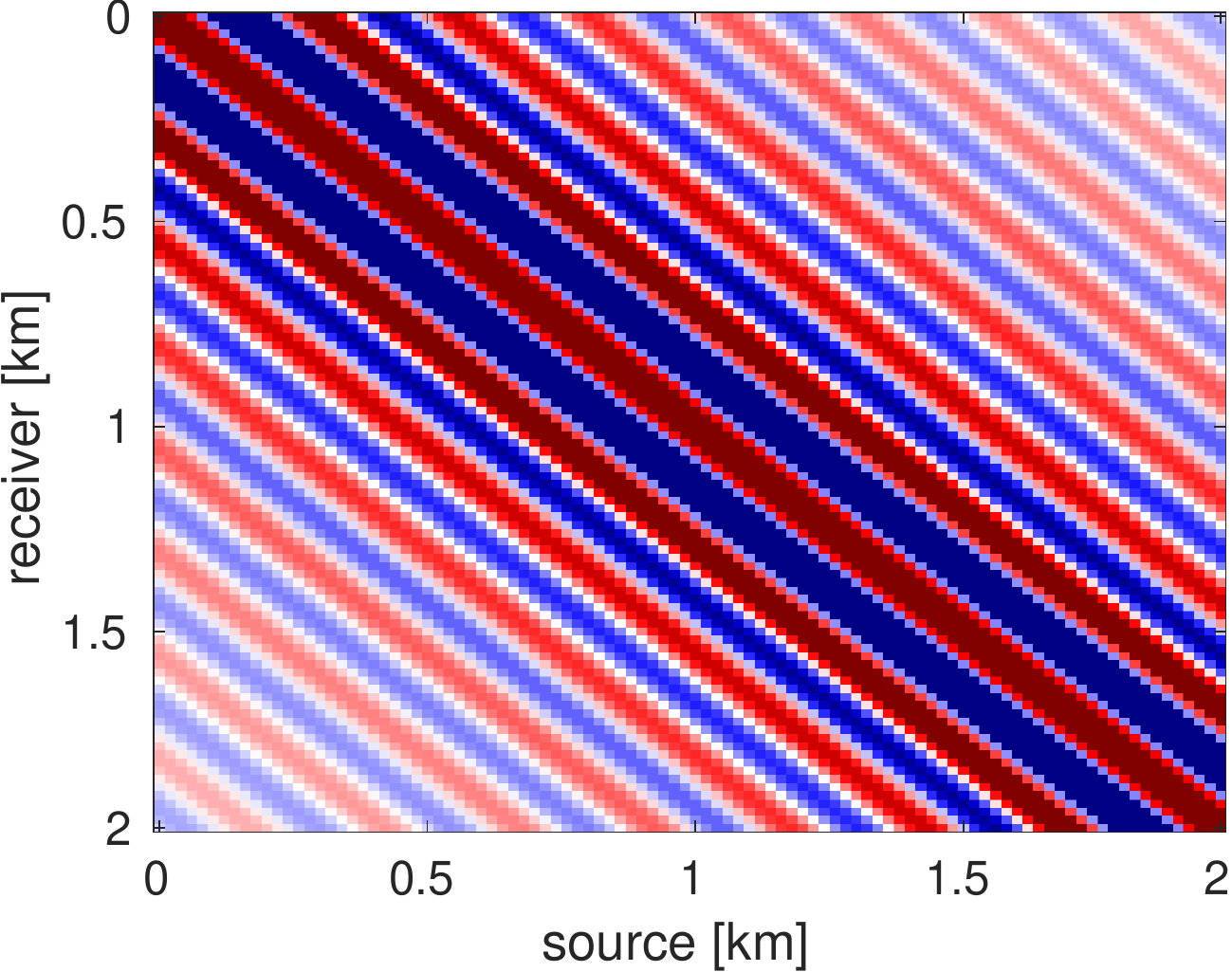}}
\subfloat[2.5D
difference]{\includegraphics[width=0.200\hsize]{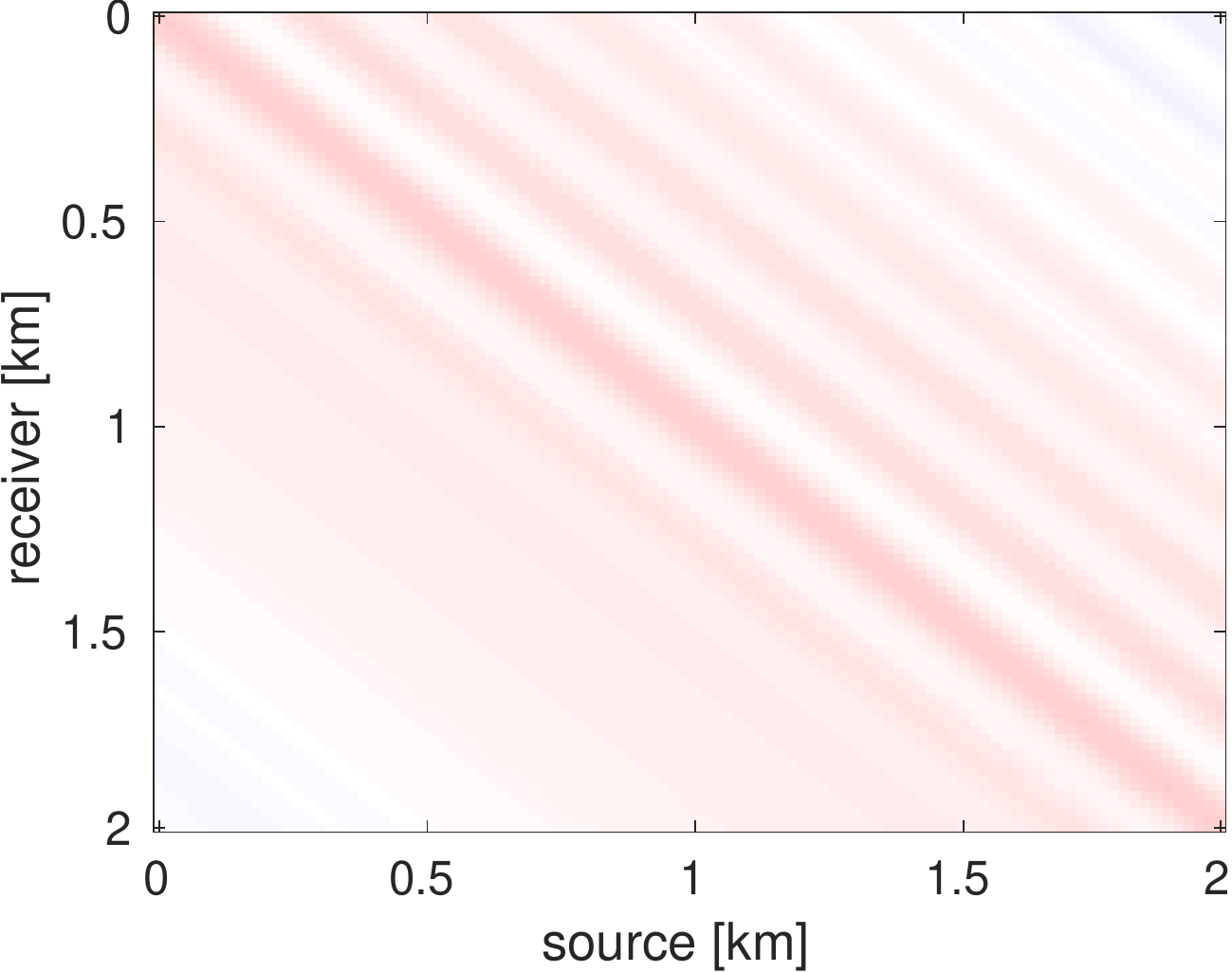}}
\subfloat[3D]{\includegraphics[width=0.200\hsize]{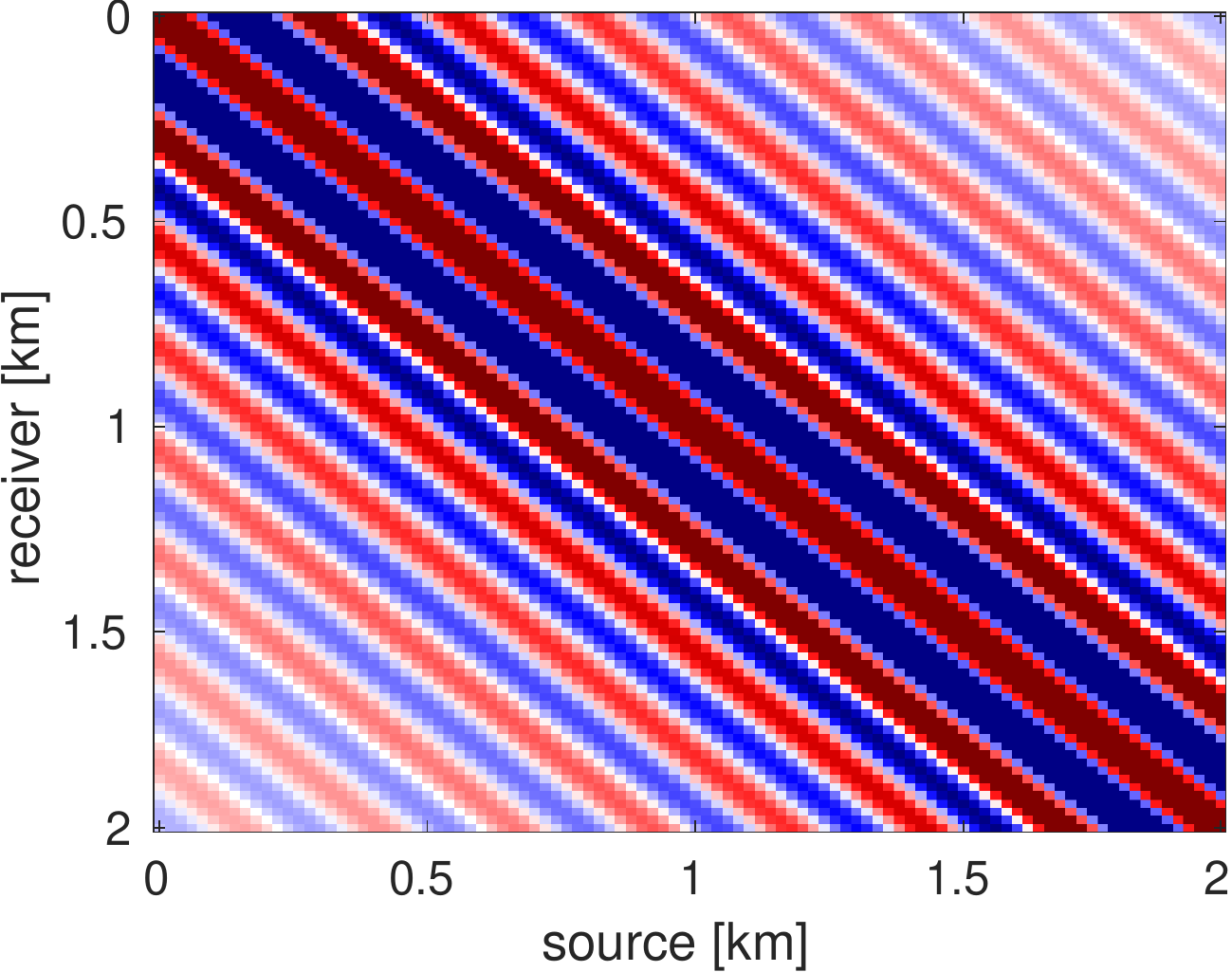}}
\subfloat[3D
difference]{\includegraphics[width=0.200\hsize]{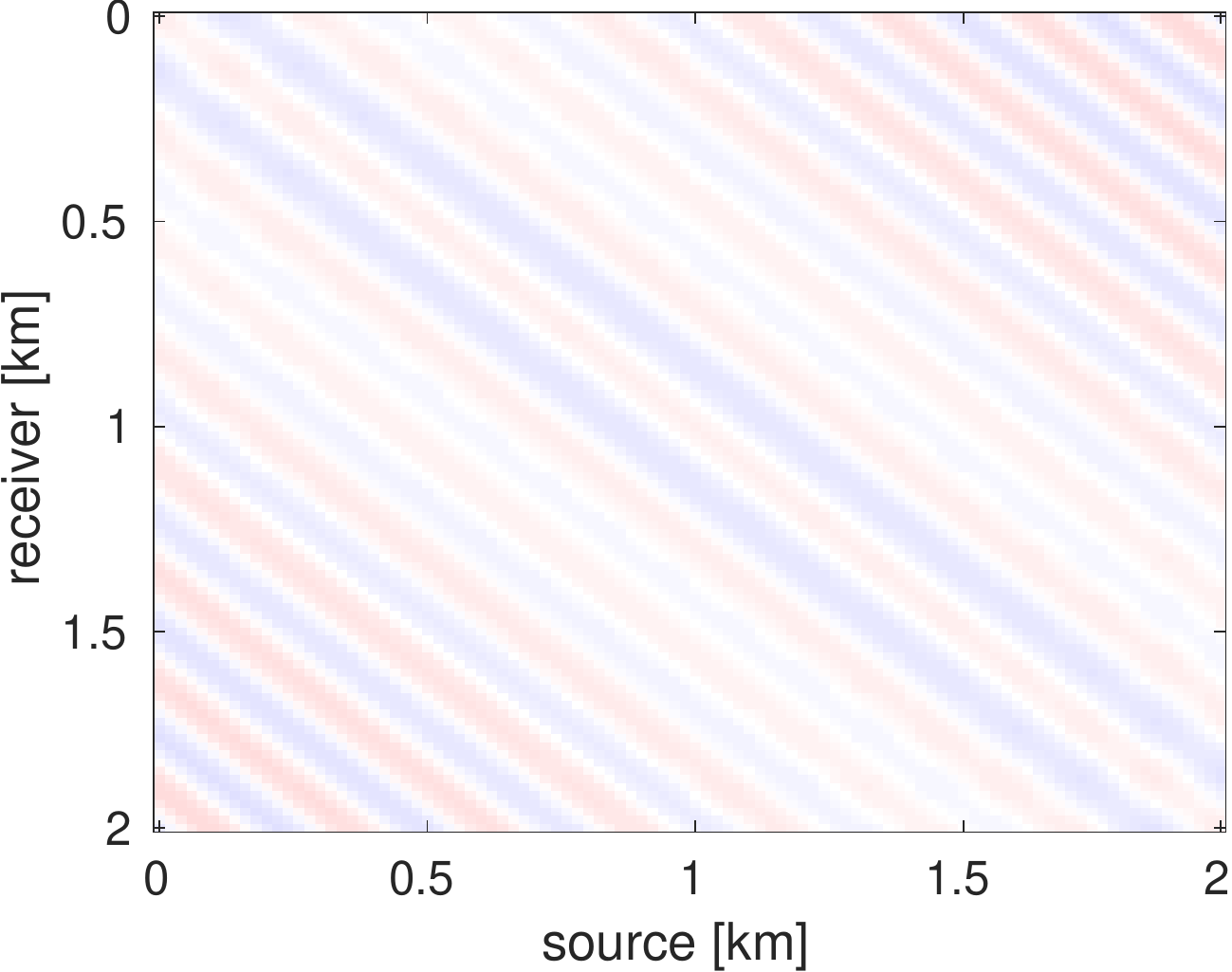}}
\caption{Analytic and numerical solutions for the 2.5D Helmholtz system
for a generated data volume with 100 sources, 100 receivers, and 100
y-wavenumbers. The 2.5D data took 136s to generate and the 3D data took
8200s, both on a single machine with no data parallelization. Top row:
real part, bottom row: imaginary part.}\label{analytic25D}
\end{figure}

We also compare the computed solutions in a homogeneous medium to the
corresponding analytic solutions in 2D and 3D, i.e.,
\begin{equation*}
\begin{aligned}
   G(x,x_s) &= -\dfrac{i}{4}H_0(\kappa \|x-x_s\|_2) & \quad \text{ in 2D }\\
   G(x,x_s) &= \dfrac{e^{i \kappa \|x-x_s\|_2}}{4\pi \|x-x_s\|_2} & \quad \text{ in 3D }
\end{aligned}
\end{equation*}
 where $\kappa = \dfrac{2\pi f}{v_0}$ is the wavenumber, $f$ is the
frequency in Hz, $v_0$ is the velocity in $m/s$, and $H_0$ is the Bessel
function of the third kind (the Hankel function). Figures
(\ref{analytic2D}, \ref{analytic3D}, \ref{analytic25D}) show the
analytic and numerical results of computing solutions with the 2D, 3D,
and 2.5D kernels, respectively. Here we see that the overall phase
dispersion is low for the computed solutions, although we do incur a
somewhat larger error around the source region as expected. The
inclusion of the PML also prevents any visible artificial reflections
from entering the solutions, as we can see from the (magnified) error
plots.

\subsection{Full Waveform Inversion}\label{fwiexample}

\begin{figure}
\centering
\captionsetup[subfigure]{labelformat=empty}
\subfloat[]{\includegraphics[width=0.330\hsize]{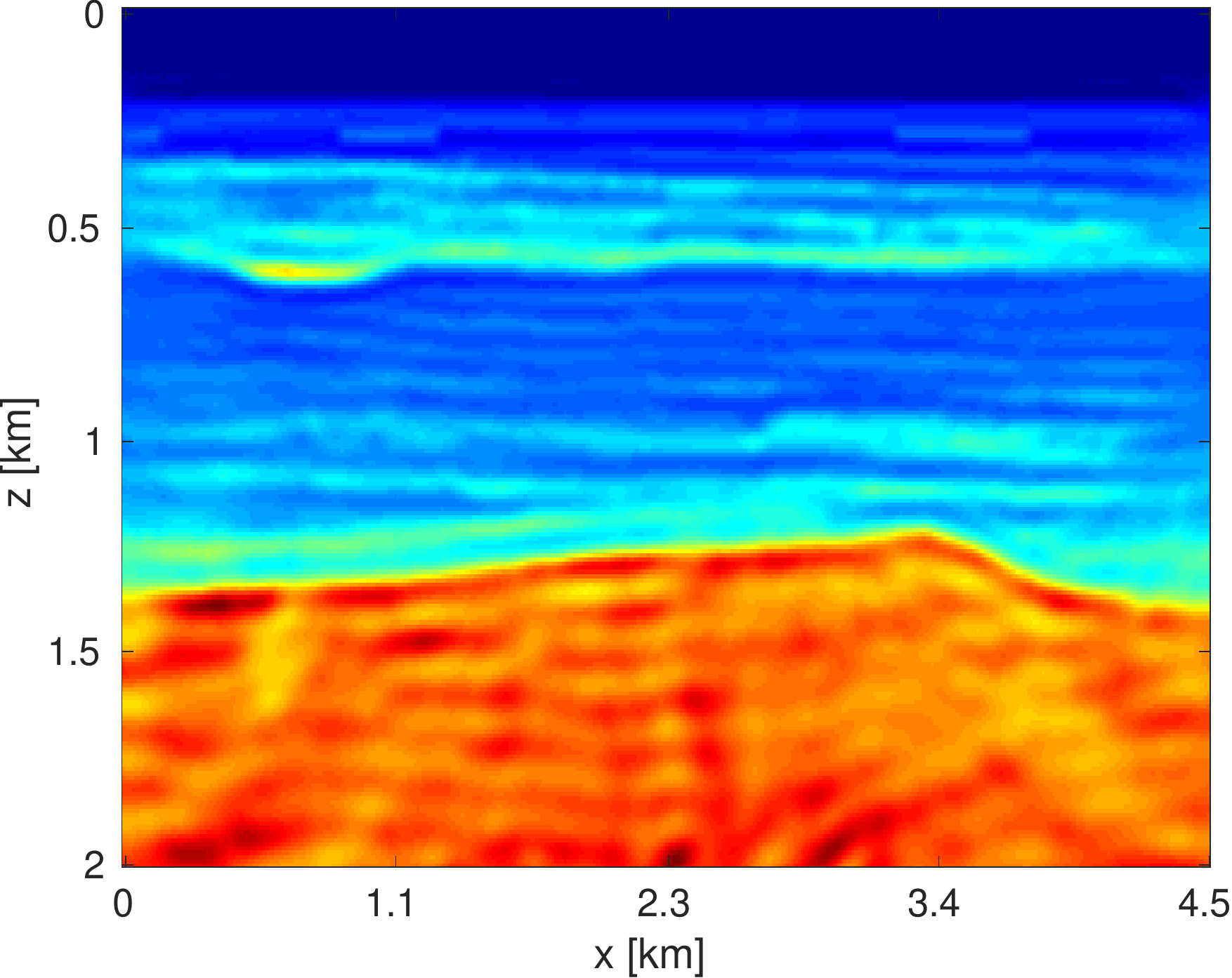}}
\subfloat[]{\includegraphics[width=0.330\hsize]{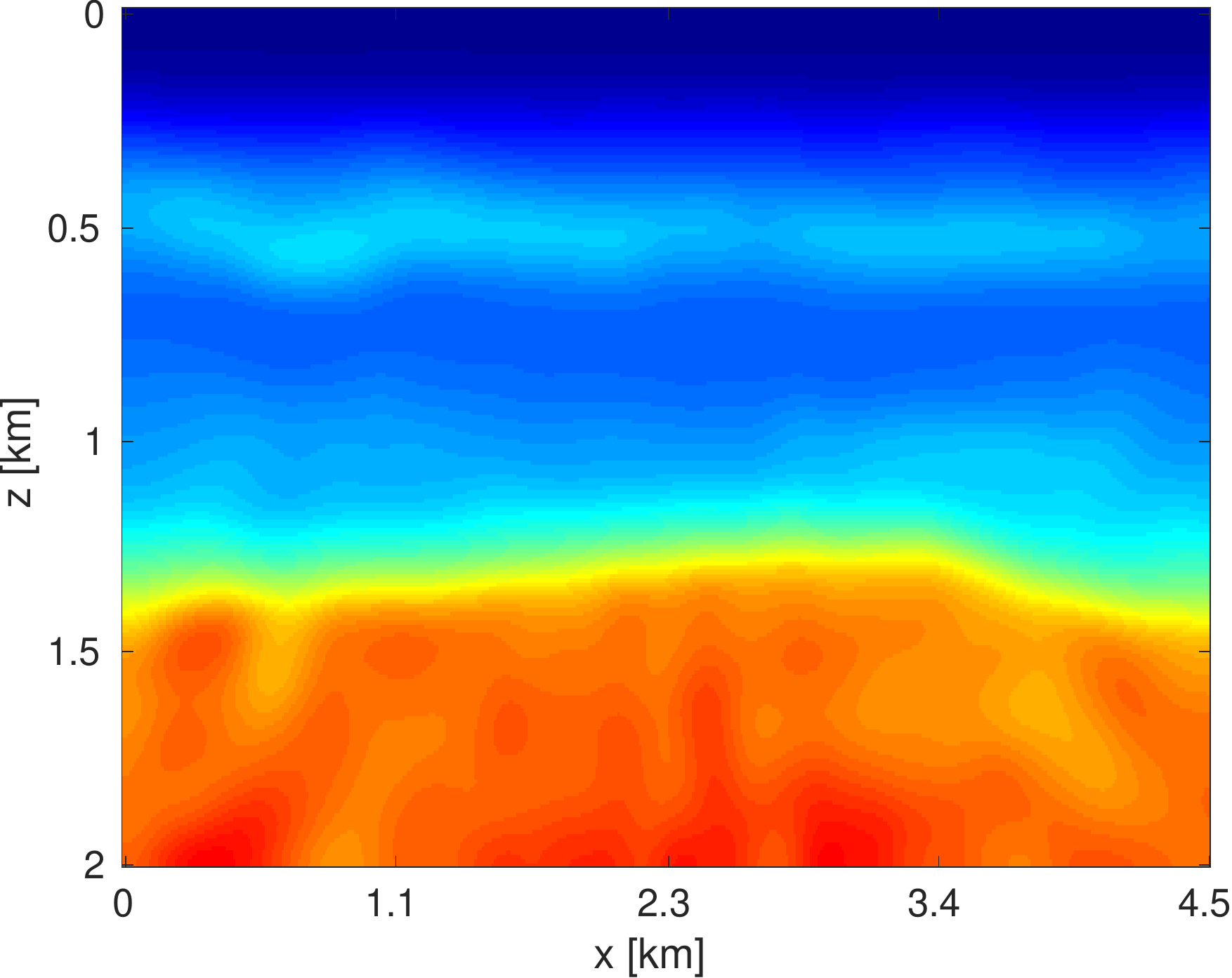}}
\subfloat[]{\includegraphics[width=0.330\hsize]{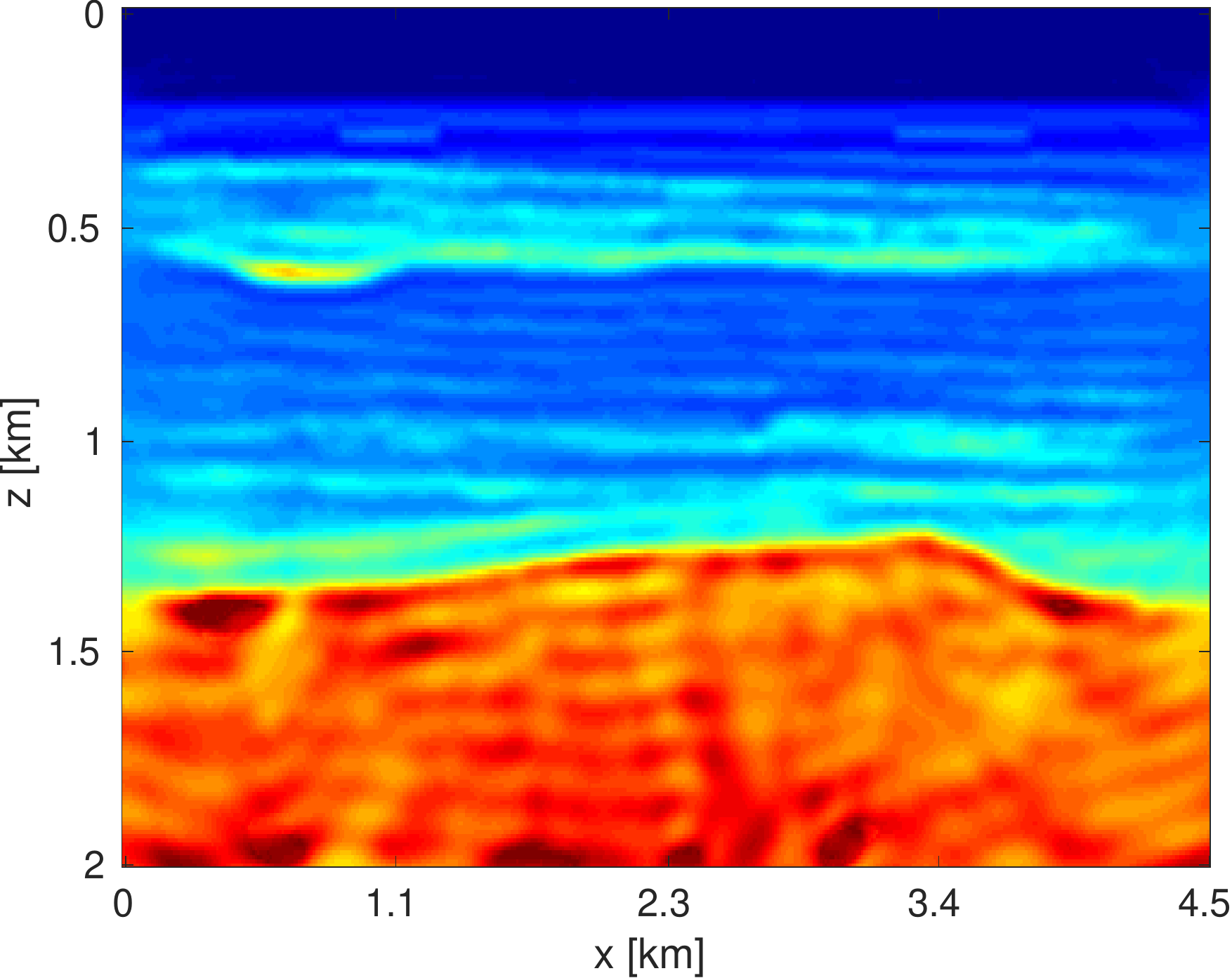}}
\caption{True (left) and initial (middle) and inverted (right)
models.}\label{ex_fwi2d_compass}
\end{figure}

To demonstrate the effectiveness of our software environment, we perform
a simple 2D FWI experiment on a 2D slice of the 3D BG Compass model. We
generate data on this 2km x 4.5km model (discretized on a 10m grid) from
3Hz to 18Hz (in 1Hz increments) using our Helmholtz modeling kernel.
Employing a frequency continuation strategy allows us to mitigate
convergence issues associated with local minima
\citep{virieux2009overview}. That is to say, we partition the entire
frequency spectrum ${3, 4, \dots, 18}$ in to overlapping subsets, select
a subset at which to invert invert the model, and use the resulting
model estimate as a warm-start for the next frequency band. In our case,
we use frequency bands of size $4$ with an overlap of $2$ between bands.
At each stage of the algorithm, we invert the model using $20$
iterations of a box-constrained LBFGS algorithm from
\citep{schmidt2009optimizing}. An excerpt from the full script that
produces this example is shown in Listing~\eqref{exfwi2dcode}. The
results of this algorithm are shown in Figure~\eqref{ex_fwi2d_compass}.
As we are using a band with a low starting frequency (3Hz in this case),
FWI is expected to perform well in this case, which we see that it does.
Although an idealized experimental setup, our framework allows for the
possibility of testing out algorithms that make use of higher starting
frequencies, or use frequency extrapolation as in
\citep{wang2016frequency, li2016}, with minimal code changes.

\begin{lstlisting}[caption={Excerpt from the script that produces this
example}, label=exfwi2dcode, float=htbp]
% Set the initial model (a smoothed version of the true model)
mest = m0;
% Loop over subsets of frequencies
for j=1:size(freq_partition,1)
    % Extract the current frequencies at this batch
    fbatch = freq_partition(j,:);
    % Select only sources at this frequency batch
    srcfreqmask = false(nsrc,nfreq);
    srcfreqmask(:,fbatch) = true;
    params.srcfreqmask = srcfreqmask;
    % Construct objective function for these frequencies
    obj = misfit_setup(mest,Q,Dobs,model,params);
    % Call the box constrained LBFGS method
    mest = minConf_TMP(obj,mest,mlo,mhi,opts);
end
\end{lstlisting}

\subsection{Sparsity Promoting Seismic Imaging}\label{exsparseimaging}

\begin{figure}
\centering
\captionsetup[subfigure]{labelformat=empty}
\subfloat[True
image]{\includegraphics[width=0.330\hsize]{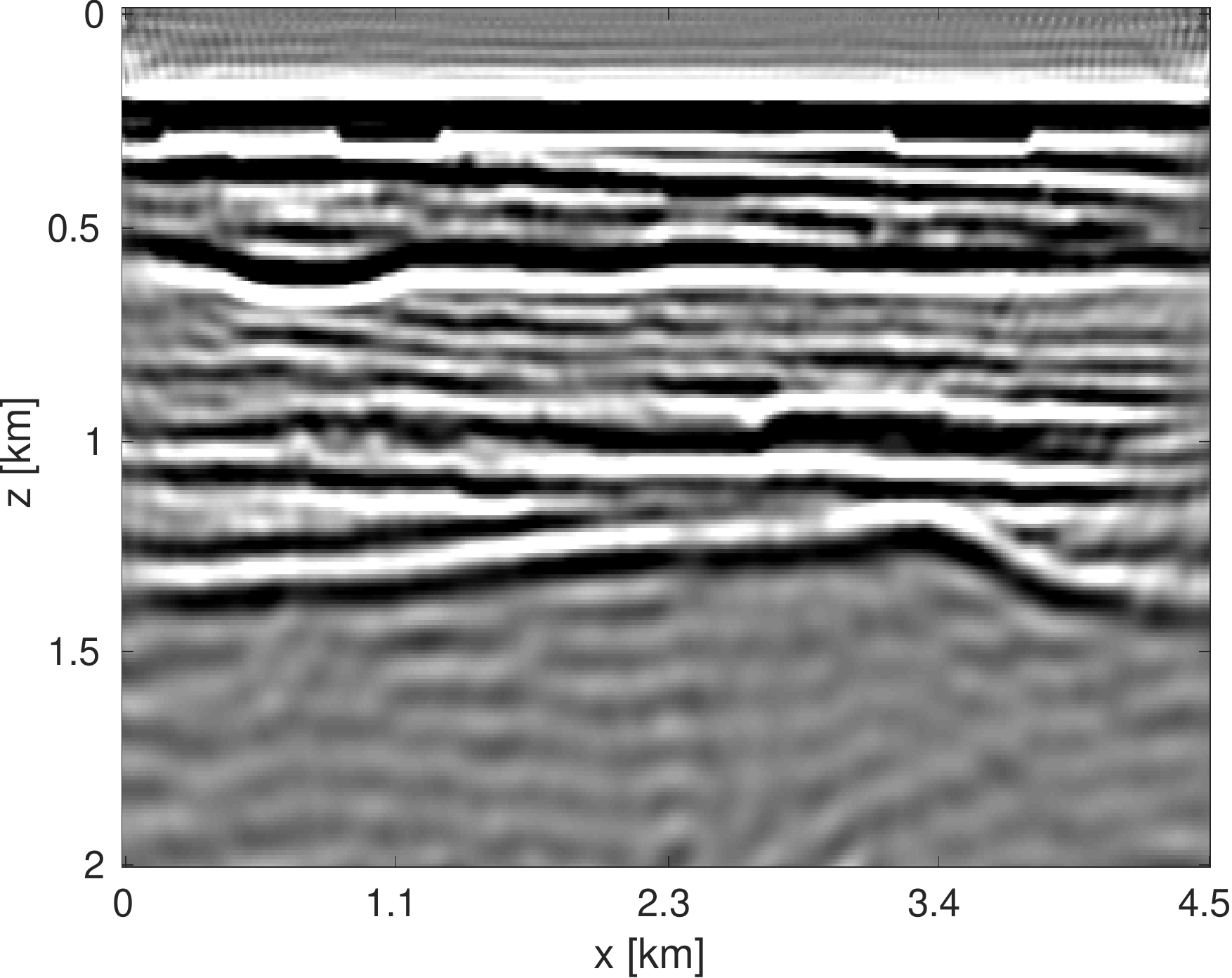}}
\subfloat[Full data
inversion]{\includegraphics[width=0.330\hsize]{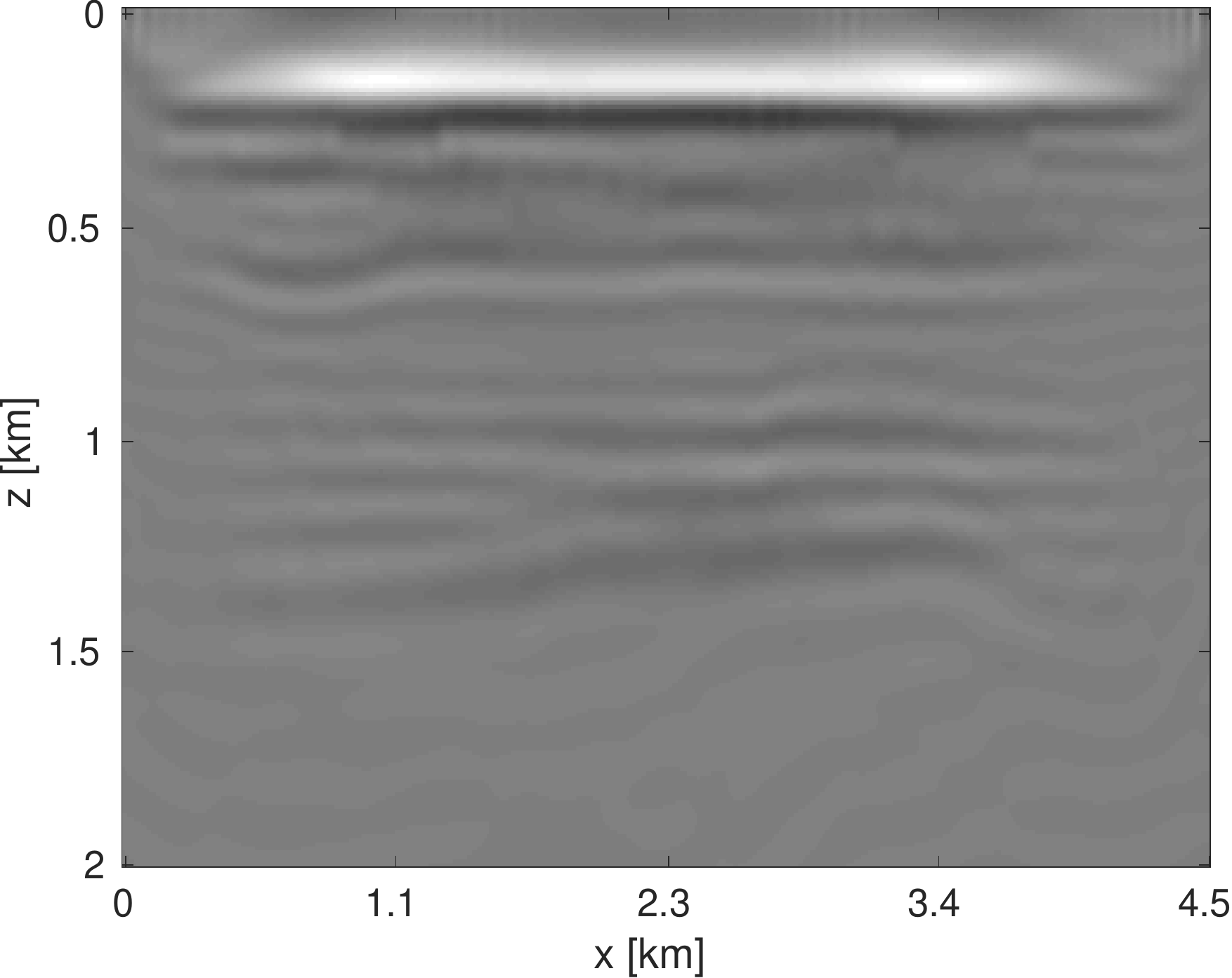}}
\subfloat[Linearized Bregman
inversion]{\includegraphics[width=0.330\hsize]{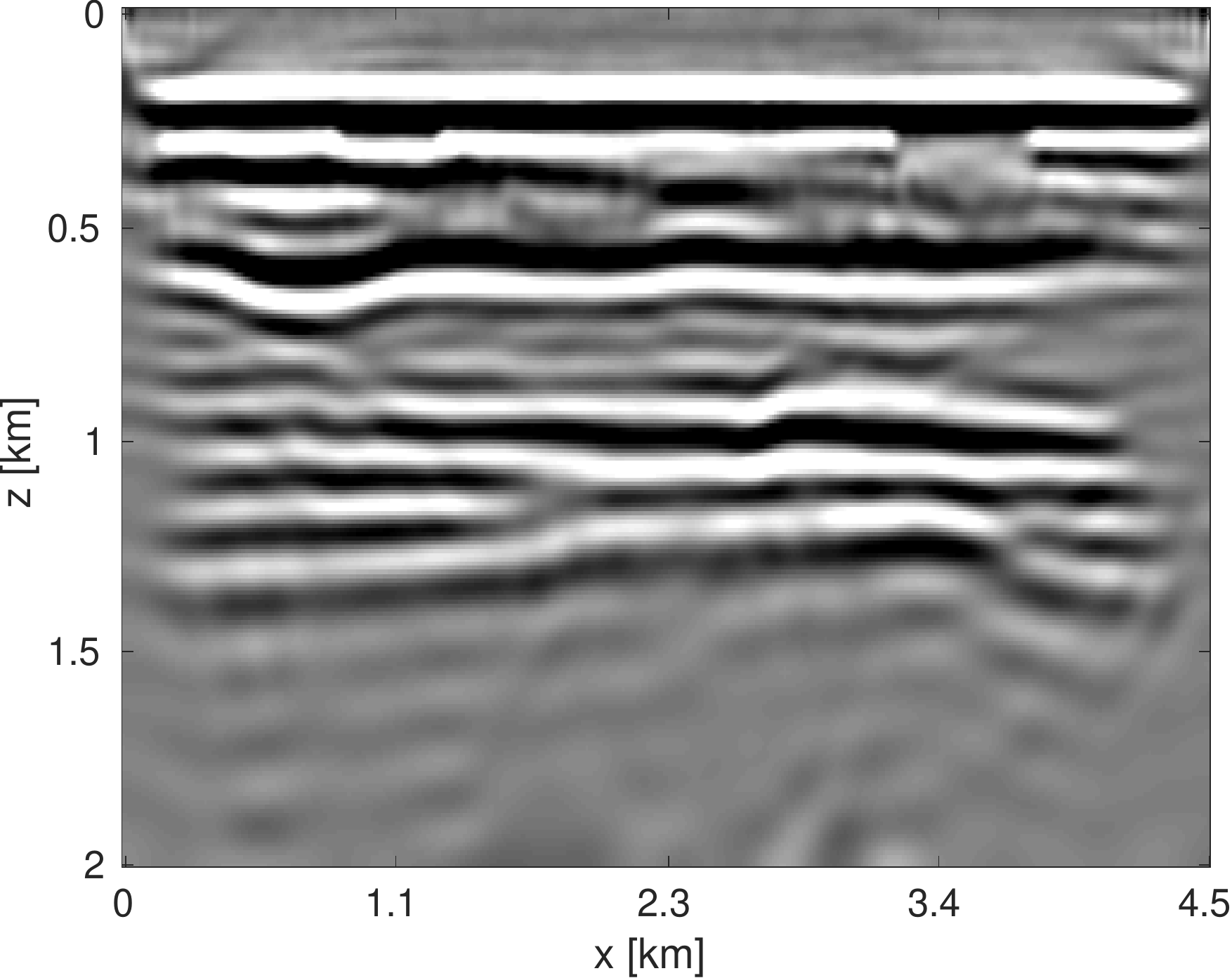}}
\caption{Sparse seismic imaging - full data least-squares inversion
versus linearized Bregman with randomized subsampling}\label{ex_imaging}
\end{figure}

The seismic imaging problem aims to reconstruct a high resolution
reflectivity map $\delta m$ of the subsurface, given some smooth
background model $m_0$, by inverting the (overdetermined) Jacobian
system
\begin{equation*}
J(m_0) \delta m \approx \delta D.
\end{equation*}
 In this simple example, $\delta D$ is the image of the true
perturbation under the Jacobian. Attempting to tackle the least-squares
system directly
\begin{equation*}
\min_{\delta m} \sum_{i_s \in S} \sum_{i_f \in F} \|J_{i_s,i_f} \delta m - \delta D_{i_s,i_f}\|_2^2,
\end{equation*}
 where $S$ indexes the sources and $F$ indexes the frequencies, is
computationally daunting due to the large number of sources and
frequencies used. One straightforward approach is to randomly subsample
sources and frequencies, i.e., choose $S' \subset S$ and $F' \subset F$
and solve
\begin{equation*}
\min_{\delta m} \sum_{i_s \in S'} \sum_{i_f \in F'} \|J_{i_s,i_f} \delta m - \delta D_{i_s,i_f}\|_2^2,
\end{equation*}
 but the Jacobian can become rank deficient in this case, despite it
being full rank normally. One solution to this problem is to use sparse
regularization coupled with randomized subsampling in order to force the
iterates to head towards the true perturbation while still reducing the
per-iteration costs. There have been a number of instances of
incorporating sparsity of a seismic image in the Curvelet domain
\citep{candes2000curvelets, candes2004new}, in particular
\citep{herrmann2008sparsity, herrmann2012efficient, tu2015fast}. We use
the Linearized Bregman method
\citep{osher2011fast, cai2009linearized, cai2009convergence}, which
solves
\begin{equation*}
\begin{aligned}
\min_{x} & \|x\|_1 + \lambda \|x\|_2^2 \\
\text{s.t. } & Ax = b.
\end{aligned}
\end{equation*}
 Coupled with random subsampling as in \citep{herrmann2015fast}, the
iterations are shown in Algorithm~\eqref{linbreg}, a variant of which is
used in \citep{chai2016linearized}. Here
$S_{\lambda}(x) = \text{sign}(x)\max(0,|x|-\lambda)$ is the
componentwise soft-thresholding operator. In
Listing~\eqref{eximagingcode}, the reader will note the close adherence
of our code to Algorithm~\eqref{linbreg}, aside from some minor
bookkeeping code and pre- and post-multiplying by the curvelet transform
to ensure the sparsity of the signal. In this example, we place 300
equispaced sources and 400 receivers at the top of the model and
generate data for 40 frequencies from 3-12 Hz. At each iteration of the
algorithm, we randomly select 30 sources and 10 frequencies
(corresponding to the 10 parallel workers we use) and set the number of
iterations so that we perform 10 effective passes through the entire
data. Compared to the image estimate obtained from an equivalent (in
terms of number of PDEs solved) method solving the least-squares problem
with the LSMR \citep{fong2011lsmr} method, the randomly subsampled
method has made significantly more progress towards the true solution,
as shown in Figure~\eqref{ex_imaging}.

\begin{scholmdAlgorithm}
~~~for~$k = 1, 2, \dots, T$\\\hspace*{0.333em}\hspace*{0.333em}\hspace*{0.333em}\hspace*{0.333em}\hspace*{0.333em}\hspace*{0.333em}Draw~a~random~subset~of~indices~$I_k$\\\hspace*{0.333em}\hspace*{0.333em}\hspace*{0.333em}\hspace*{0.333em}\hspace*{0.333em}\hspace*{0.333em}$z_{k+1} \gets z_k - t_k A_{I_k}^T (b_{I_k} - A_{I_k}x_k)$\\\hspace*{0.333em}\hspace*{0.333em}\hspace*{0.333em}\hspace*{0.333em}\hspace*{0.333em}\hspace*{0.333em}$x_{k+1} \gets S_{\lambda}(x_k)$
\caption{Linearized Bregman with per-iteration randomized
subsampling}\label{linbreg}
\end{scholmdAlgorithm}

\begin{lstlisting}[caption={Excerpt from the code that produces this
example}, label=eximagingcode, float=htbp]
for k=1:T
  % Draw a random subset of sources and frequencies
  Is = rand_subset(nsrc,round(0.2*nsrc));
  If = rand_subset(nfreq,parpool_size());
  % Mask the objective to the sources/frequencies drawn
  srcfreqmask = false(nsrc,nfreq);
  srcfreqmask(Is,If) = true;
  params.srcfreqmask = srcfreqmask;
  % Construct the subsampled Jacobian operator + data
  A = oppDF(m0,Q,model,params);
  b = distributed_subsample_data(b_full,Is,If);
  % Linearized Bregman algorithm
  r = A*x-b;
  ATr = A'*r;
  t = norm(r)^2/norm(ATr)^2;
  z = z - t*ATr;
  x = C'*softThreshold(C*z,lambda);
end
\end{lstlisting}

\subsection{Electrical Conductivity Inversion}\label{expoisson}

To demonstrate the modular and flexible nature of our software
framework, we replace the finite difference Helmholtz PDE with a simple
finite volume discretization of the variable coefficient Poisson
equation
\begin{equation*}
\nabla(\sigma \cdot \nabla u) = q
\end{equation*}
 where $\sigma$ is the spatially-varying conductivity coefficient. We
discretize the field on the vertices of a regular rectangular mesh and
$\sigma$ at the cell centers, which results in the system matrix,
denoted $H(\sigma)$, written abstractly as
\begin{equation*}
H(\sigma) = \sum_{i=1}^{5} A_i \text{diag}(B_i \sigma)
\end{equation*}
 for constant matrices $A_i, B_i$. This form allows us to easily derive
expressions for $T : \delta \sigma \mapsto DH(\sigma)[\delta m]u$ and
$T^*$ as
\begin{equation*}
\begin{aligned}
T\delta \sigma &= \sum_{i=1}^{5} A_i \text{diag}(B_i \delta \sigma)u \\
T^* z &= \sum_{i=1}^{5} B_i^H \text{diag}(\overline{u}) A_i^H z.
\end{aligned}
\end{equation*}
 With these expressions in hand, we merely can slot the finite volume
discretization and the corresponding directional derivative functions in
to our framework without modifying any other code.

\begin{figure}
\centering
\captionsetup[subfigure]{labelformat=empty}
\subfloat[True
Model]{\includegraphics[width=0.330\hsize]{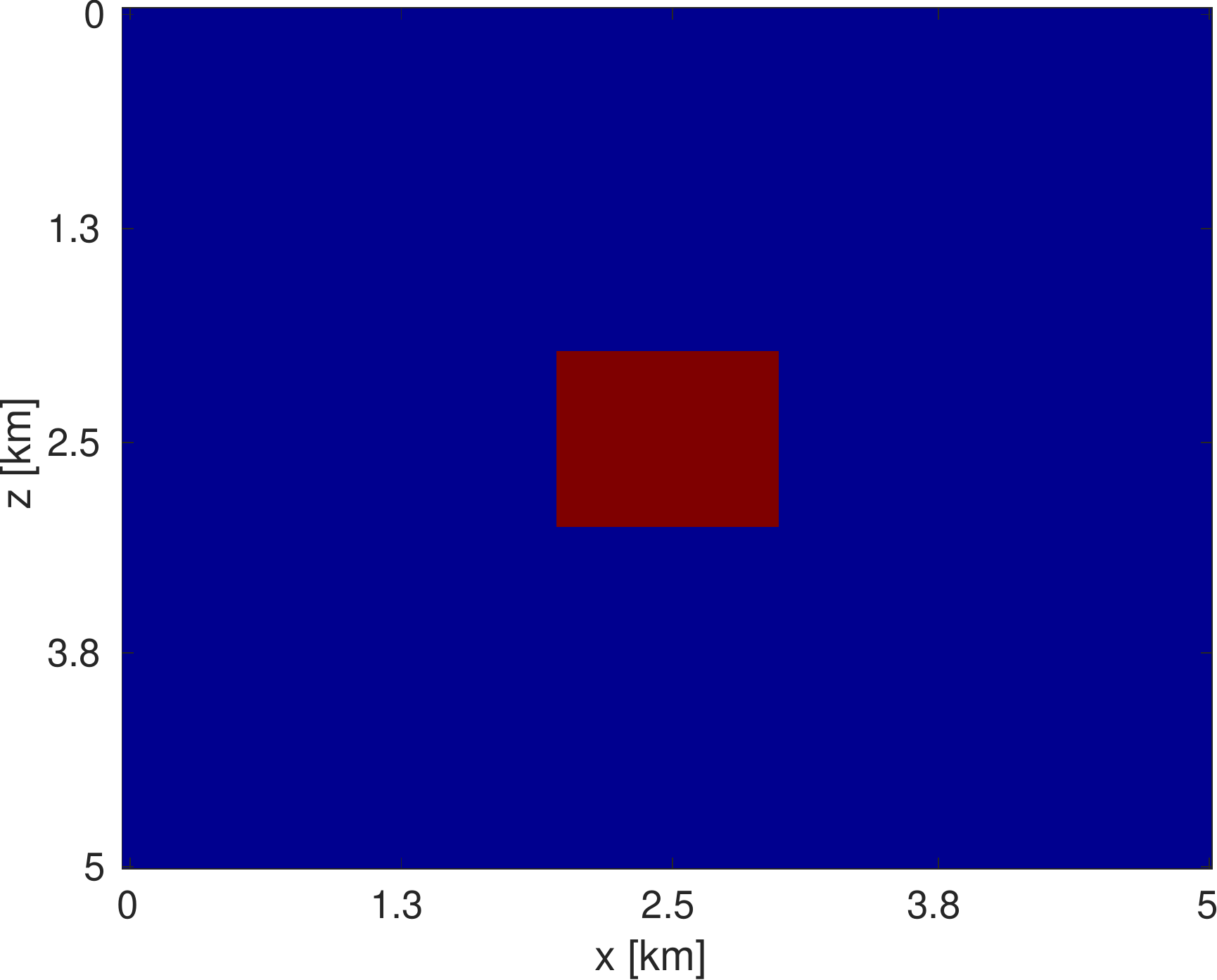}}
\subfloat[Initial
Model]{\includegraphics[width=0.330\hsize]{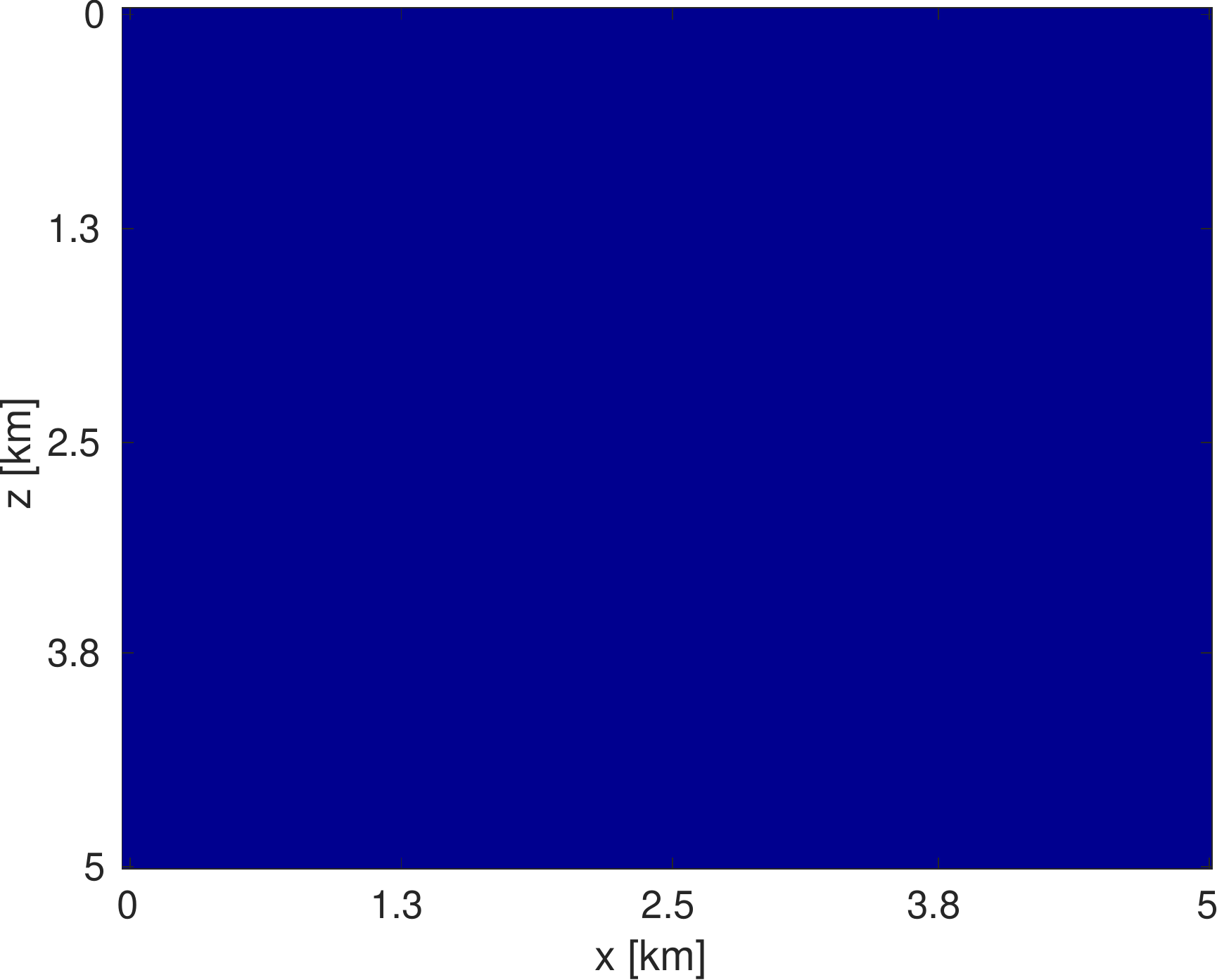}}
\subfloat[Inverted
Model]{\includegraphics[width=0.330\hsize]{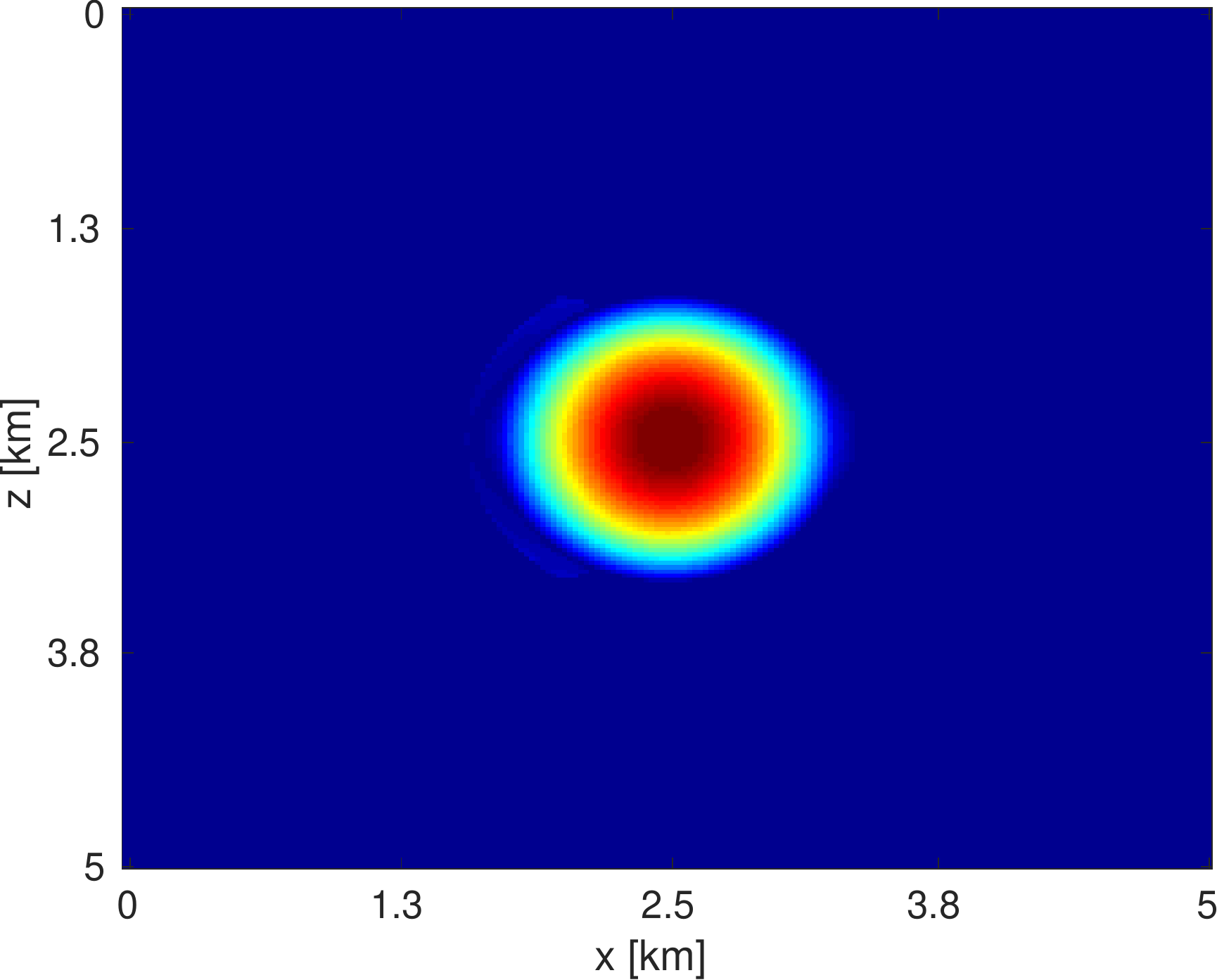}}
\caption{Inversion results when changing the PDE model from the
Helmholtz to the Poisson equation}\label{ex_poisson}
\end{figure}

Consider a simple constant conductivity model containing a square
anomaly with a 20\% difference compared to the background, encompassing
a region that is 5km x 5km with a grid spacing of 10m, as depicted in
Figure~\eqref{ex_poisson}. We place 100 equally spaced sources and
receivers at depths $z = 400m$ and $z = 1600m$, respectively. The
pointwise constraints we use are the true model for $z < 400m$ and
$z > 1600m$ and, for the region in between, we set
$\sigma_{\text{min}} = \min_{x} \sigma(x)$ and
$\sigma_{\text{max}} = 2\max_{x} \sigma(x)$. Our initial model is a
constant model with the correct background conductivity, shown in
Figure~\eqref{ex_poisson}. Given a current estimate of the conductivity
$\sigma_k$, we minimize a quadratic model of the objective subject to
bound constraints, i.e.,
\begin{equation*}
\begin{aligned}
\sigma_{k+1} = \argmin_{\sigma} & \langle \sigma-\sigma_k, g_k \rangle + \dfrac{1}{2} \langle \sigma-\sigma_k, H_k(\sigma-\sigma_k) \rangle \\
\text{s.t. } &\; \sigma_{\text{min}} \le \sigma \le \sigma_{\text{max}}
\end{aligned}
\end{equation*}
 where $g_k,H_k$ are the gradient and Gauss-Newton Hessian,
respectively. We solve $5$ of these subproblems, using $5$
objective/gradient evaluations to solve each subproblem using the bound
constrained LBFGS method of \citep{schmidt2009optimizing}. As we do not
impose any constraints on the model itself and the PDE itself is
smoothing, we are able to recover a very smooth version of the true
model, shown in Figure~\eqref{ex_poisson}, with the attendant code shown
in Listing~\eqref{expoissoncode}. This discretization and optimization
setup is by no means the optimal method to invert such conductivity
models but we merely outline how straightforward it is to incorporate
different PDEs in to our framework. Techniques such as total variation
regularization \citep{abubaker2001total} can be incorporated in to this
framework by merely modifying our objective function.

\begin{lstlisting}[caption={Excerpt from the script that produces this
example}, label=expoissoncode, float=htbp]
obj = misfit_setup(sigma0,Q,D,model,params);
sigma = sigma0;
for i=1:5
  % Evaluate objective
  [f,g,h] = obj(sigma);
  % Construct quadratic model
  q = @(x) quadratic_model(g,h,x,sigma);
  % Minimize quadratic model, subject to pointwise constraints
  sigma = minConf_TMP(q,sigma,sigma_min,sigma_max,opts);
end
\end{lstlisting}

\subsection{Stochastic Full Waveform
Inversion}\label{stochastic-full-waveform-inversion}

Our software design makes it relatively straightforward to apply the
same inversion algorithm to both a 2D and 3D problem in turn, while
changing very little in the code itself. We consider the following
algorithm for inversion, which will allow us to handle the large number
of sources and number of model parameters, as well as the necessity of
pointwise bound constraints \citep{herrmann2016SEGWScvp} on the
intermediate models. Our problem has the form
\begin{equation*}
\begin{aligned}
\min_{m} & \dfrac{1}{N_s} \sum_{i=1}^{N_s} f_i(m) \\
\text{s.t. } & \; m \in C
\end{aligned}
\end{equation*}
 where $m$ is our model parameter (velocity or slowness),
$f_i(m) = \frac{1}{2}\|P_r H(m)^{-1}q_i - d_i\|_2^2$ is the
least-squares misfit for source $i$, and $C$ is our convex constraint
set, which is $C = \{ m : m_{LB} \le m \le m_{UB} \}$ in this case. When
we have $p$ parallel processes and $N_s \gg p$ sources, in order to
efficiently make progress toward the solution of the above problem, we
stochastically subsample the objective and approximately minimize the
resulting problem, i.e., at the $k$th iteration we solve
\begin{equation*}
\begin{aligned}
m_{k} = \argmin_{m} & \dfrac{1}{|I_k|} \sum_{i \in I_k} f_i(m) \\
\text{s.t. } &\; m \in C
\end{aligned}
\end{equation*}
 for $I_k \subset \{1, \dots, N_s\}$ drawn uniformly at random and
$|I_k| = p \ll N_s$. We use the approximate solution $m_k$ as a warm
start for the next problem and repeat this process a total of $T$ times.
Given that our basic unit of work in these problems is computing the
solution of a PDE, we limit the number of iterations for each subproblem
solution so that each subproblem can be evaluated at a constant multiple
of evaluating the objective and gradient with the full data, i.e.,
$r_{\text{sub}} \lceil \frac{N_s}{p} \rceil$ iterations for a constant
$r_{\text{sub}}$. If an iteration stagnates, i.e., if the line search
fails, we increase the size of $|I_k|$ by a fixed amount. This algorithm
is similar in spirit to the one proposed in
\citep{vanLeeuwen2014SISC3Dfds}.

\subsubsection{2D Model --- BG Compass}\label{d-model-bg-compass}

\begin{figure}
\centering
\captionsetup[subfigure]{labelformat=empty}
\subfloat[]{\includegraphics[width=0.330\hsize]{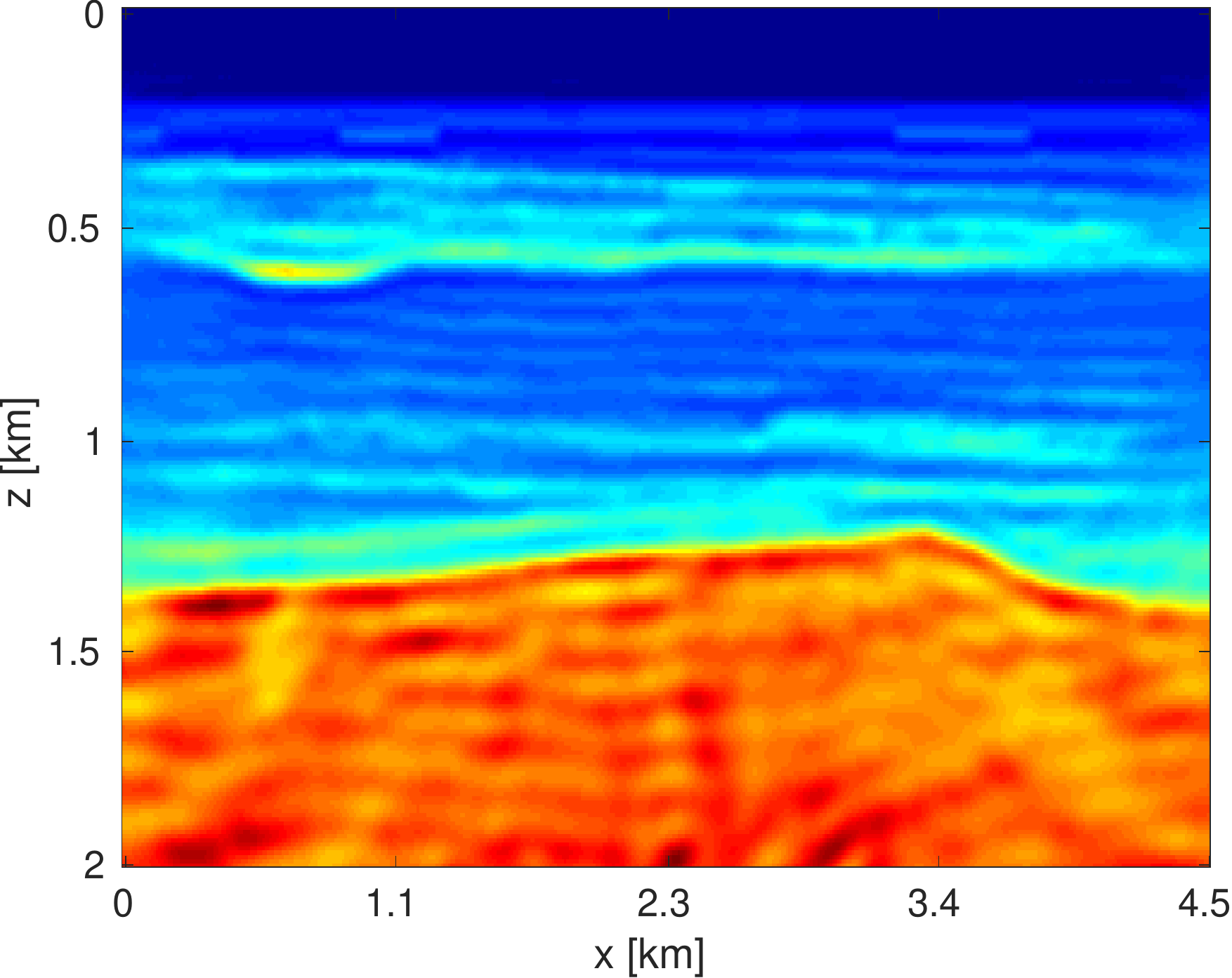}}
\subfloat[]{\includegraphics[width=0.330\hsize]{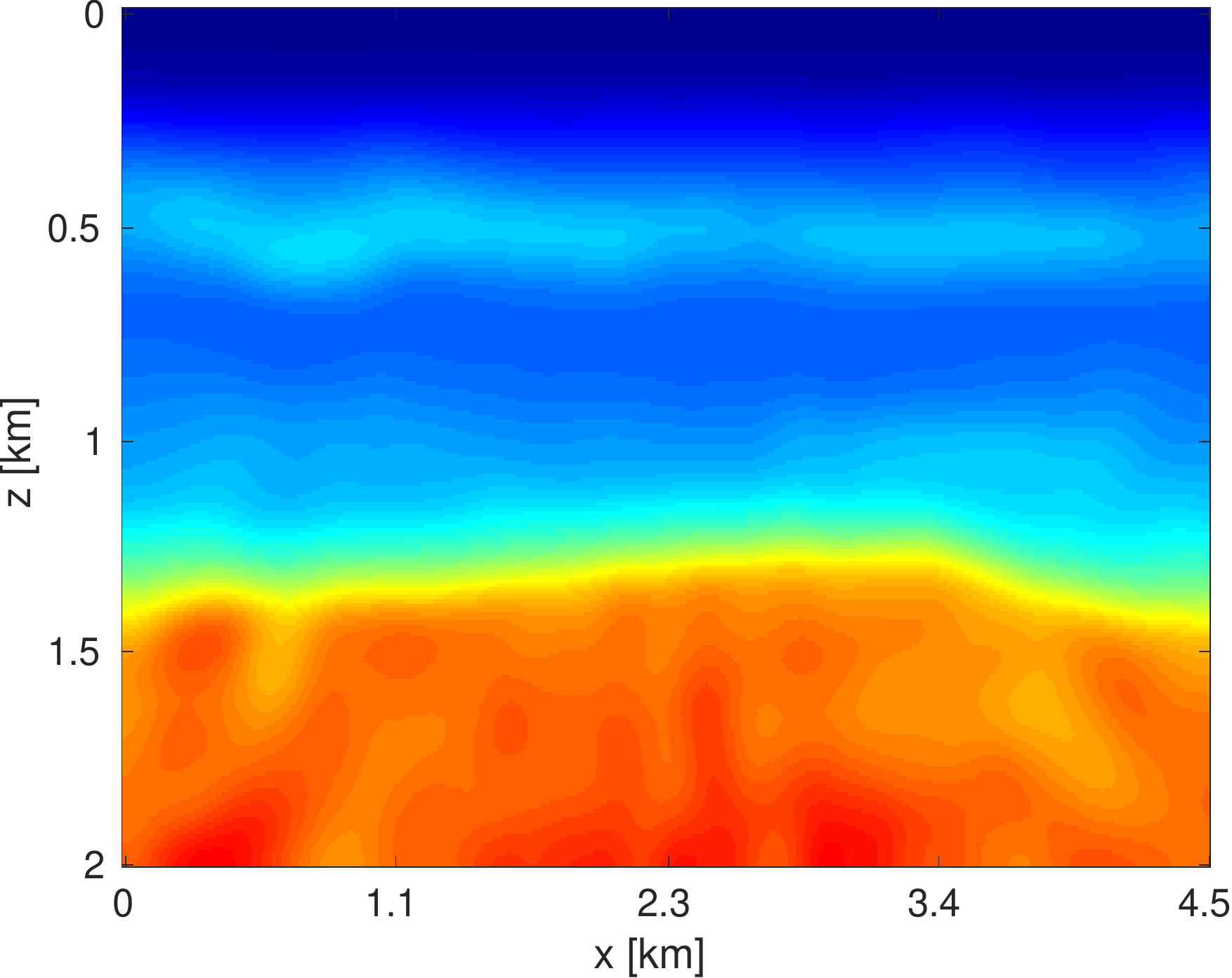}}
\subfloat[]{\includegraphics[width=0.330\hsize]{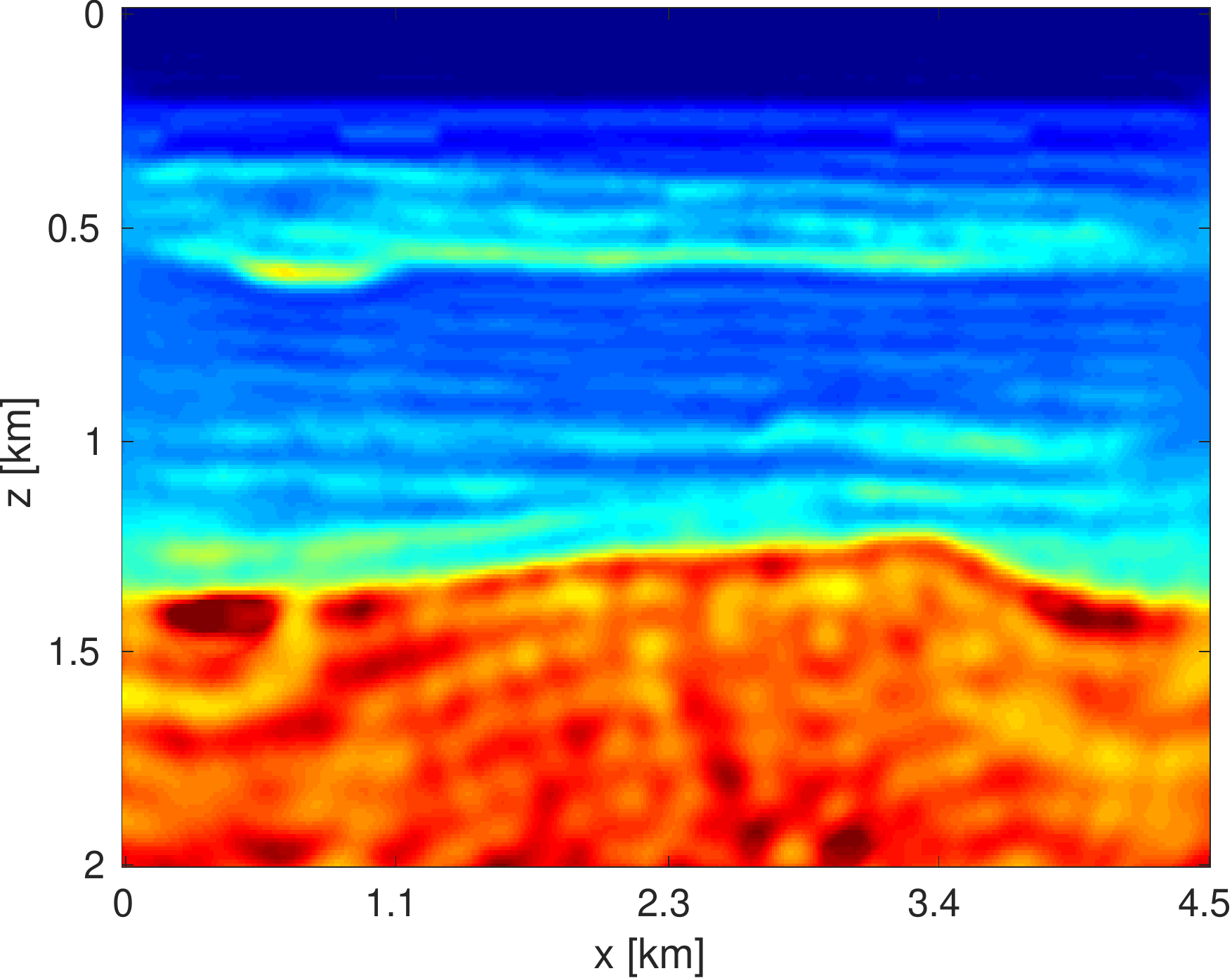}}
\caption{True (left) and initial (middle) and inverted (right)
models}\label{ex_fwi2d_compass_randsub}
\end{figure}

\begin{figure}
\centering
\includegraphics[width=0.500\hsize]{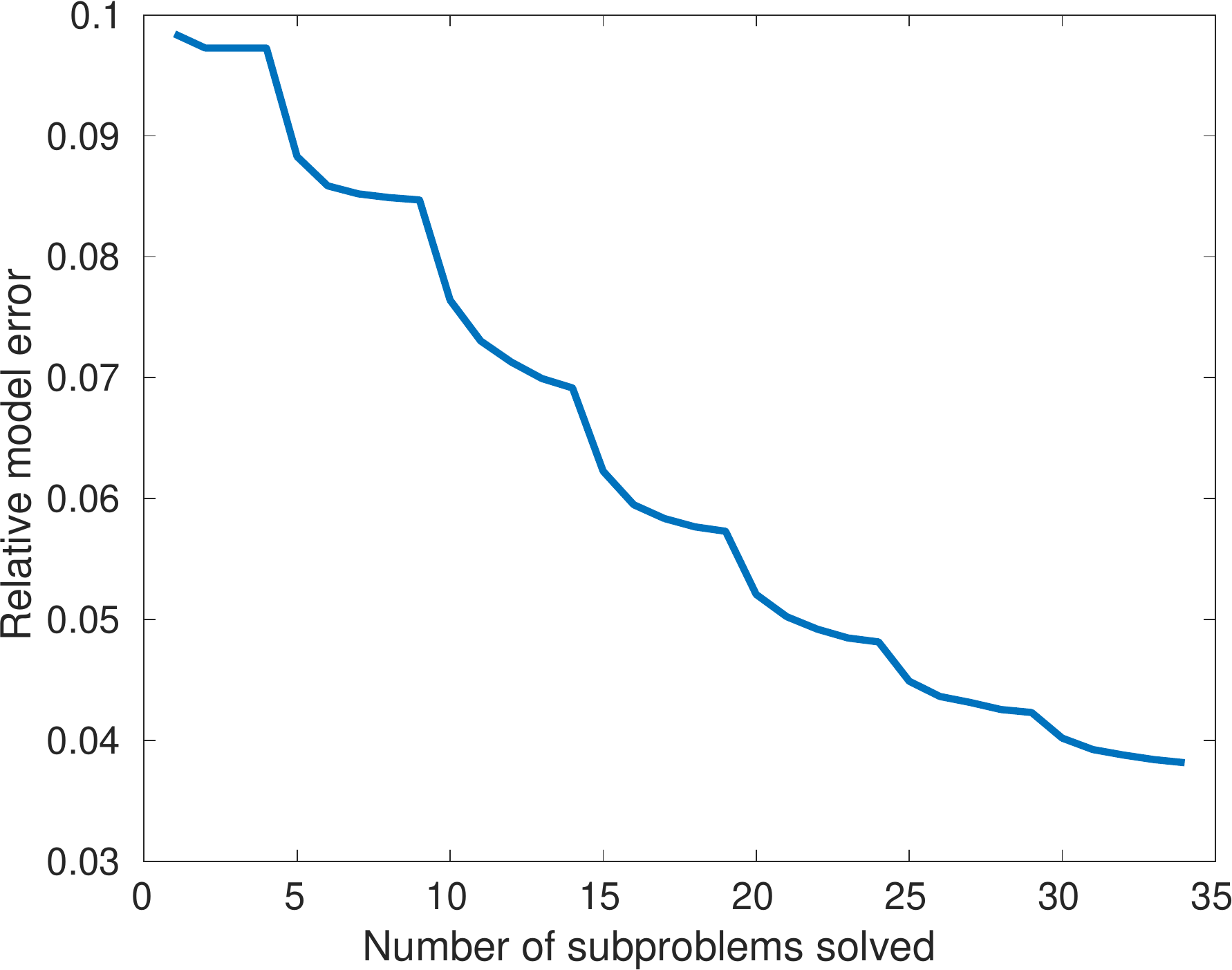}
\caption{Relative model error as a function of the number of randomized
subproblems solved.}\label{ex_fwi2d_compass_randsub_model_err}
\end{figure}

We apply the above algorithm to the BG Compass model, with the same
geometry and source/receiver configuration as outlined in Section
(\#fwiexample). We limit the total number of passes over the entire data
to be equal to $50\%$ of those used in the previous example, with $10$
re-randomization steps and $r_{\text{sub}} = 2$. Our results are shown
in Figure~\eqref{ex_fwi2d_compass_randsub}. Despite the smaller number
of overall PDEs solved, the algorithm converges to a solution that is
qualitatively hard to distinguish from Figure~\eqref{ex_fwi2d_compass}.
The overall model error as a function of number of subproblems solved is
depicted Figure~\eqref{ex_fwi2d_compass_randsub_model_err}. The model
error stagnates as the number of iterations in a given frequency batch
rises and continues to decrease when a new frequency batch is drawn.

\subsubsection{3D Model - Overthrust}\label{d-model---overthrust}

\begin{figure}
\centering
\captionsetup[subfigure]{labelformat=empty}
\subfloat[x=12500m]{\includegraphics[width=0.330\hsize]{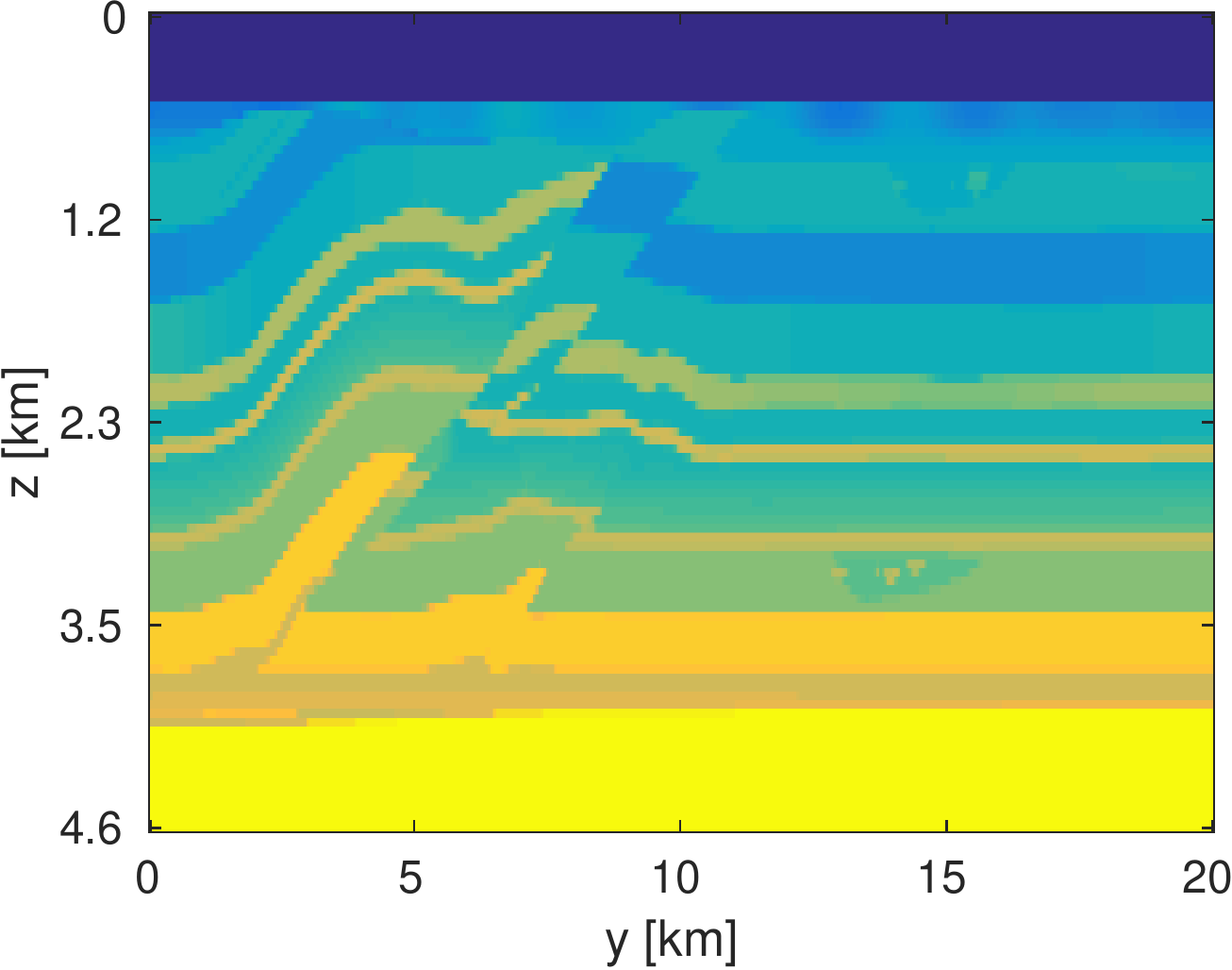}}
\subfloat[x=12500m]{\includegraphics[width=0.330\hsize]{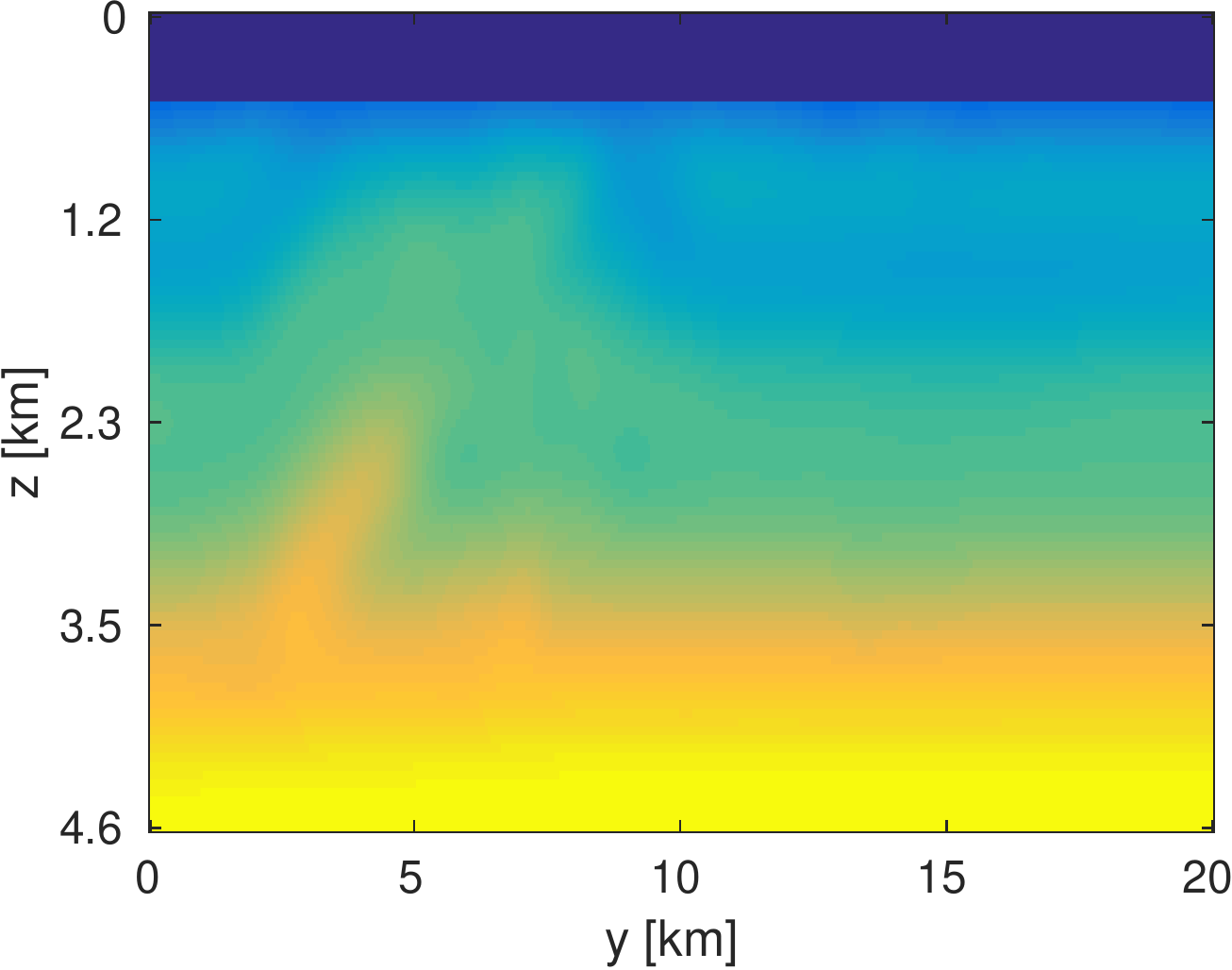}}
\subfloat[x=12500m]{\includegraphics[width=0.330\hsize]{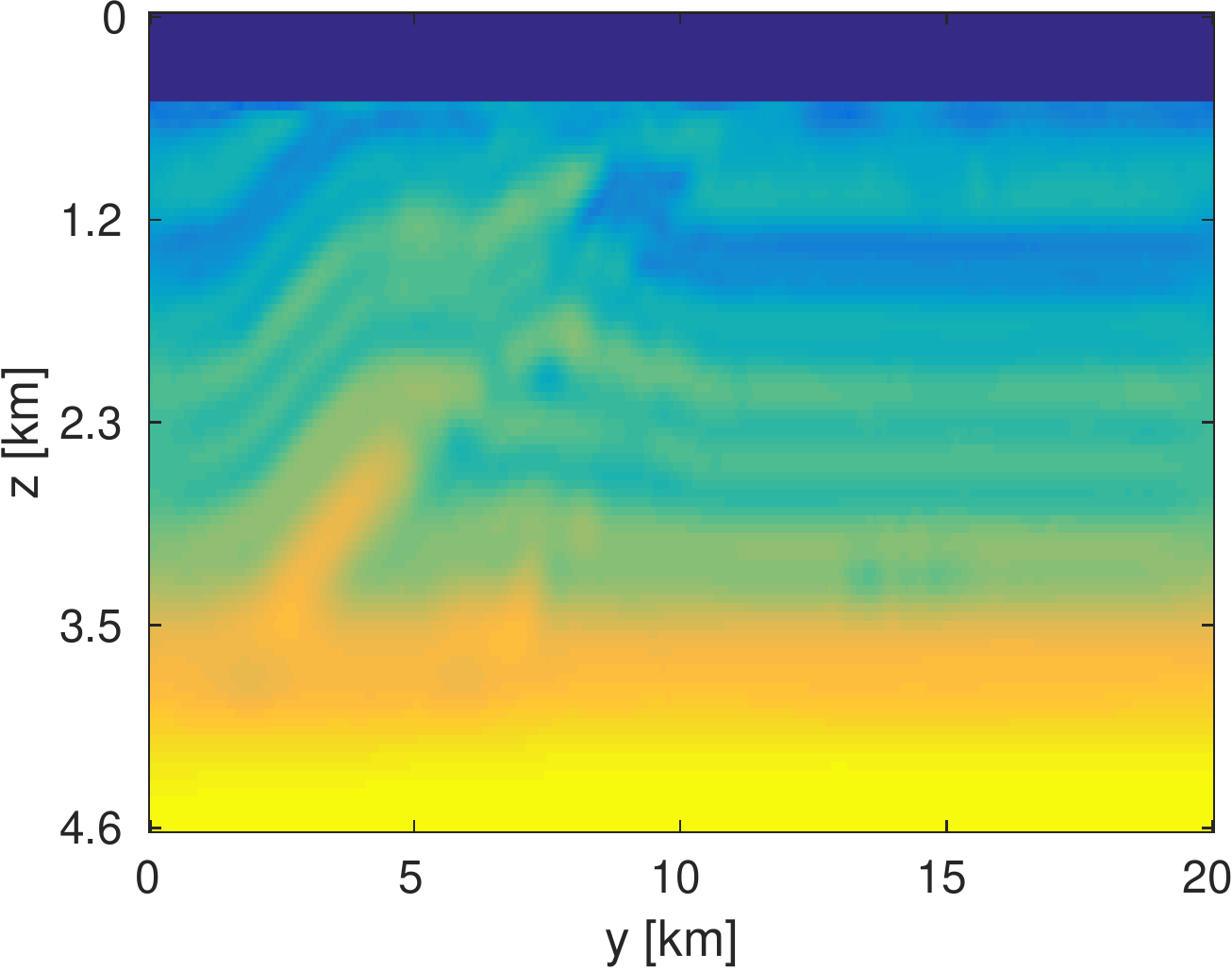}}
\\
\subfloat[y=10000m]{\includegraphics[width=0.330\hsize]{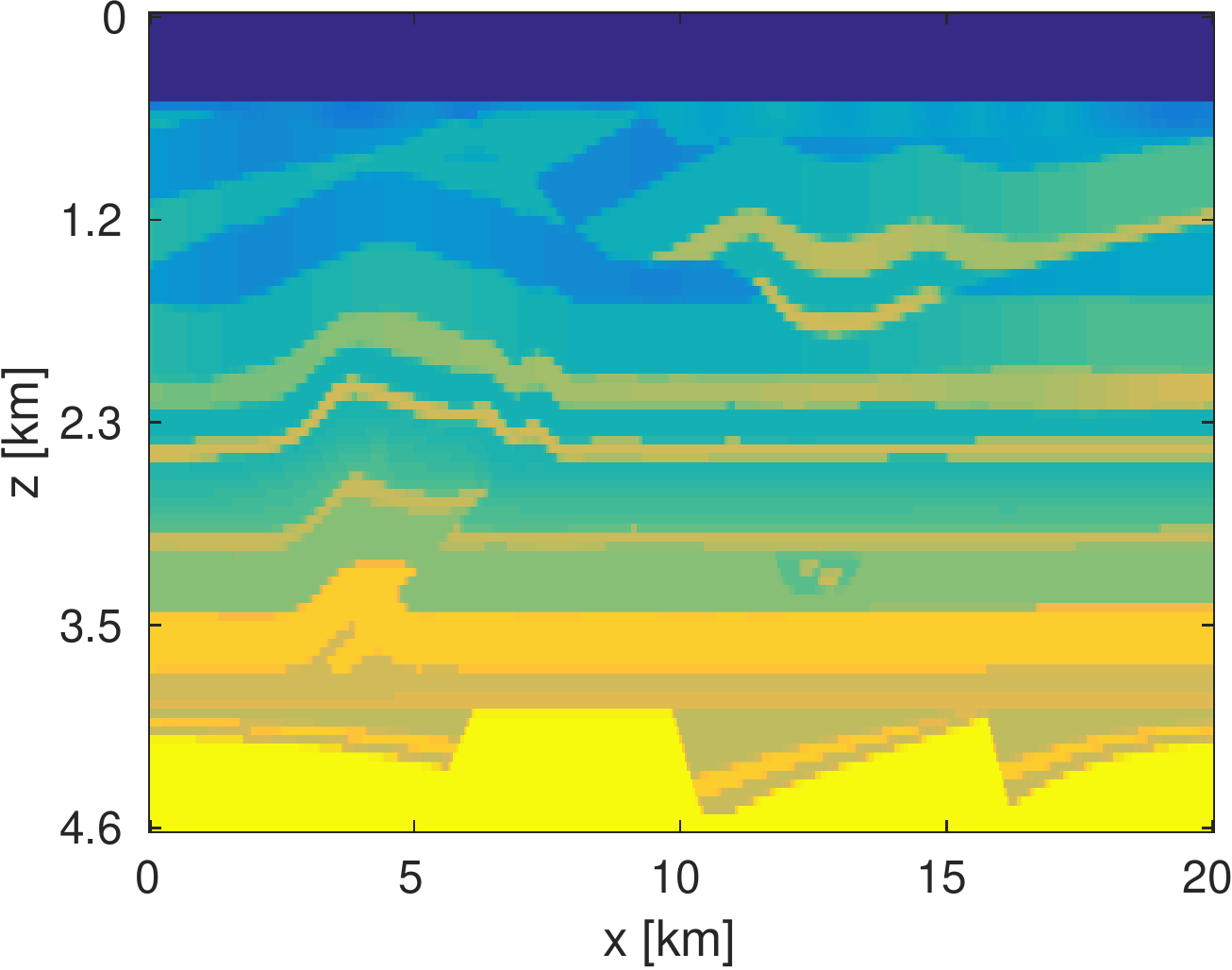}}
\subfloat[y=10000m]{\includegraphics[width=0.330\hsize]{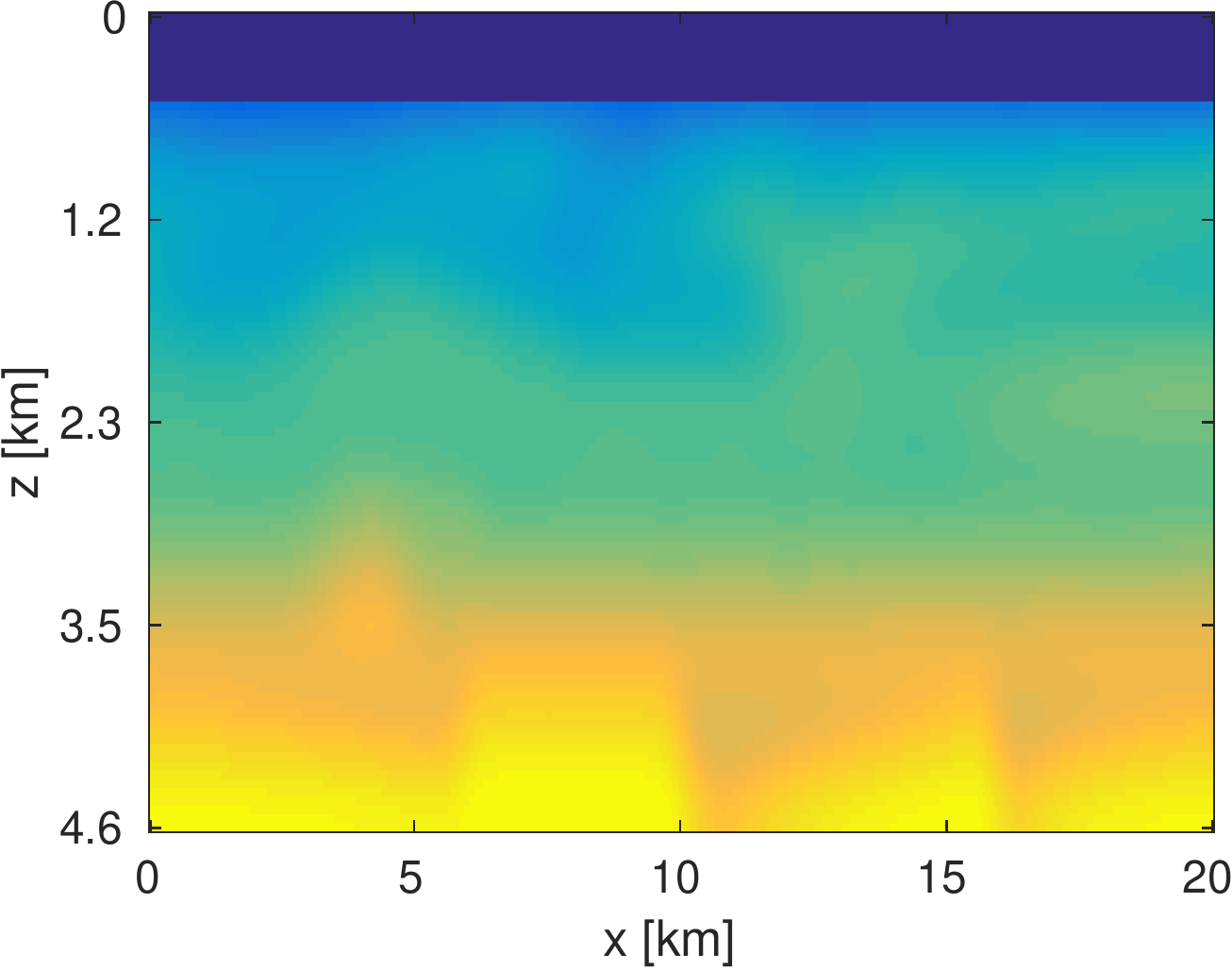}}
\subfloat[y=10000m]{\includegraphics[width=0.330\hsize]{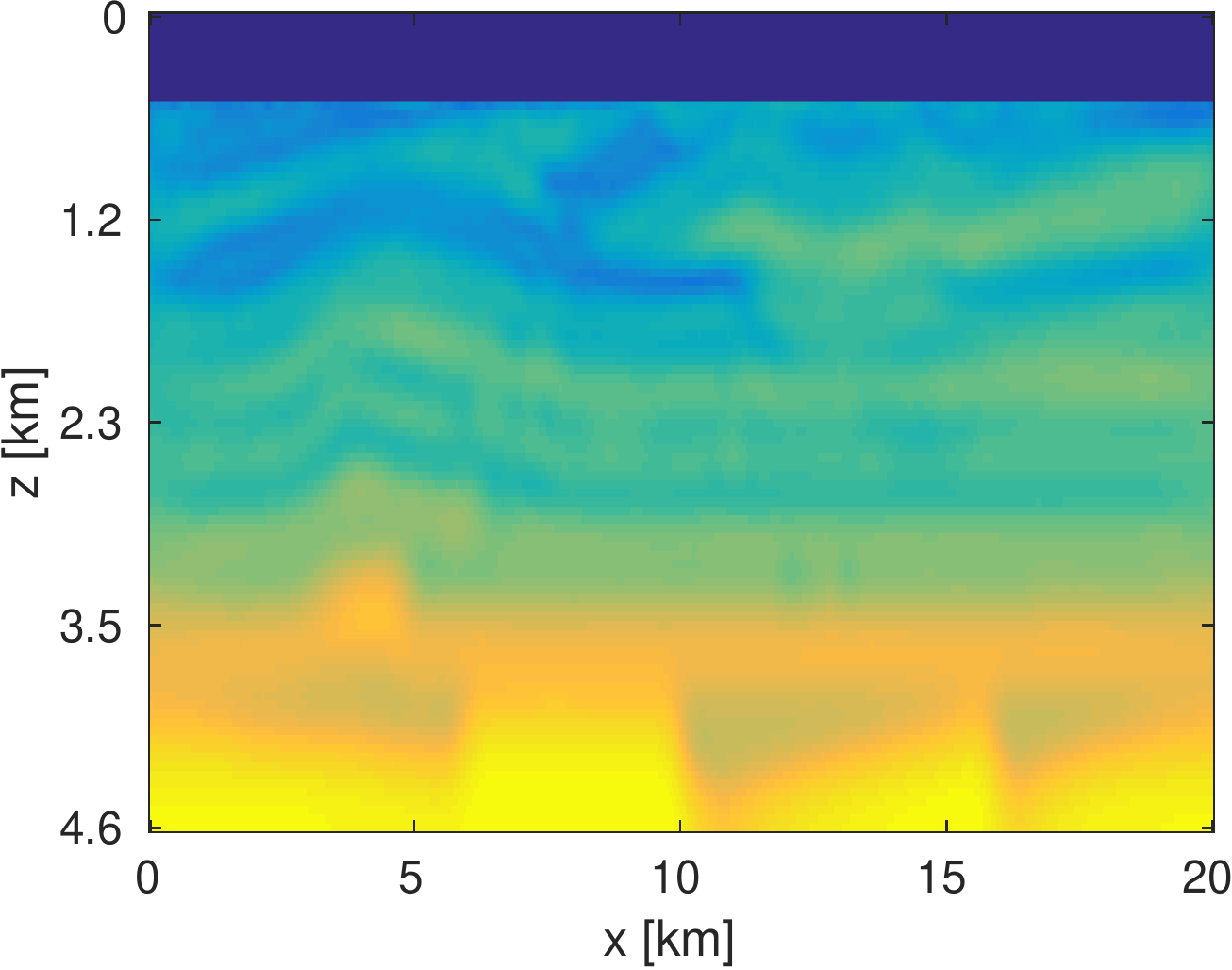}}
\caption{True model (left), initial model (middle), inverted model
(right) for a variety of fixed coordinate
slices}\label{overthrust_results_xy}
\end{figure}

\begin{figure}
\centering
\captionsetup[subfigure]{labelformat=empty}
\subfloat[z=750m]{\includegraphics[width=0.330\hsize]{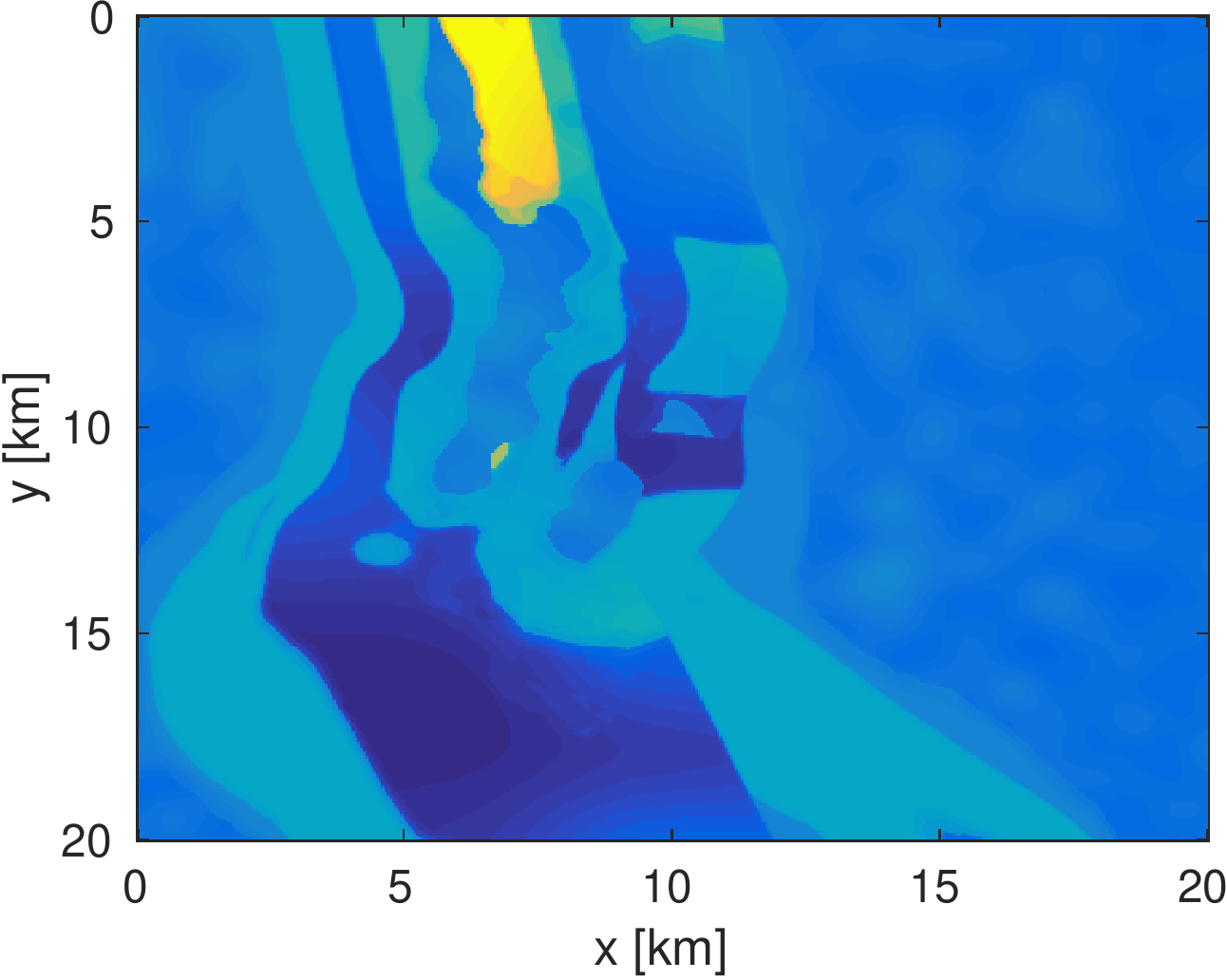}}
\subfloat[z=750m]{\includegraphics[width=0.330\hsize]{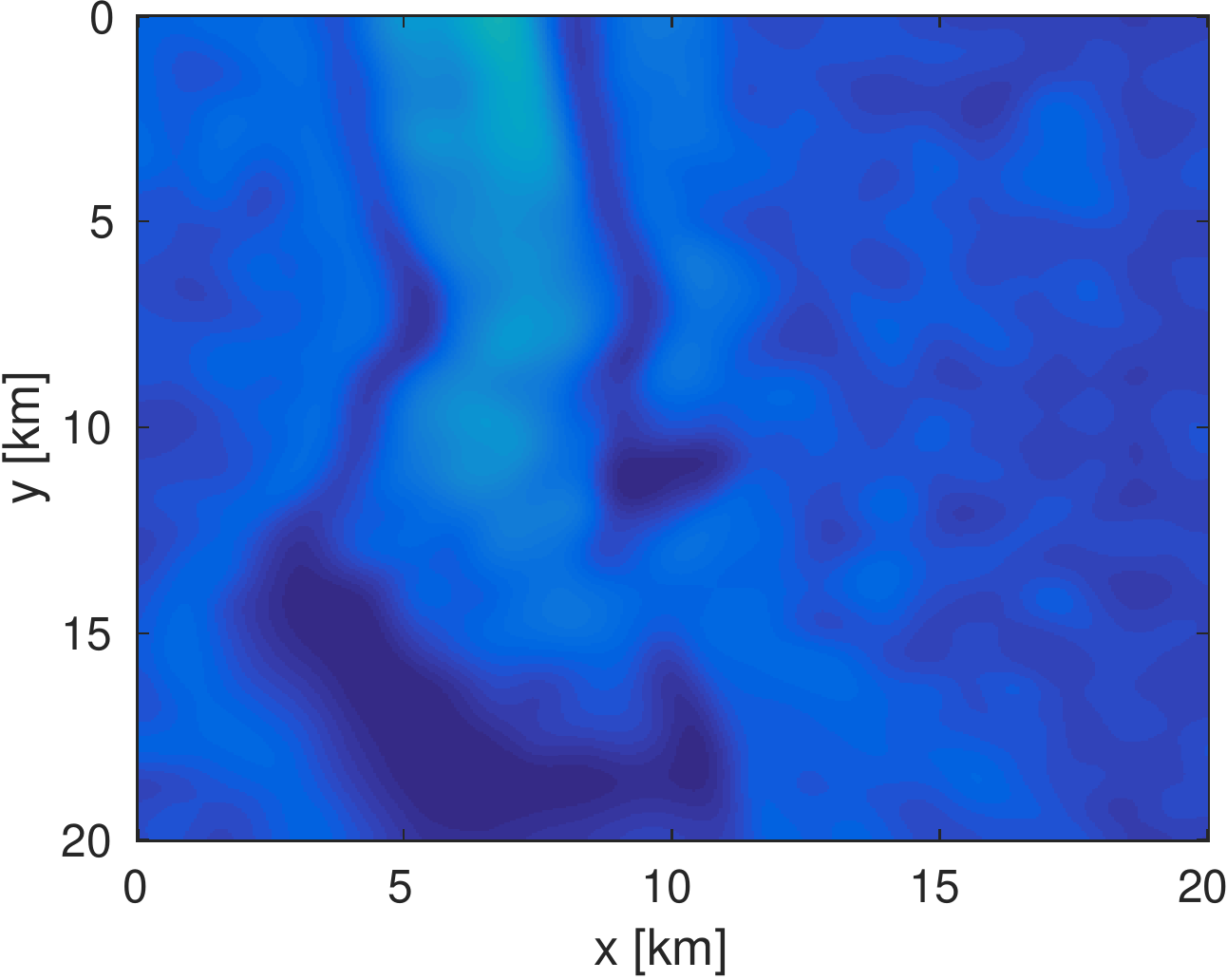}}
\subfloat[z=750m]{\includegraphics[width=0.330\hsize]{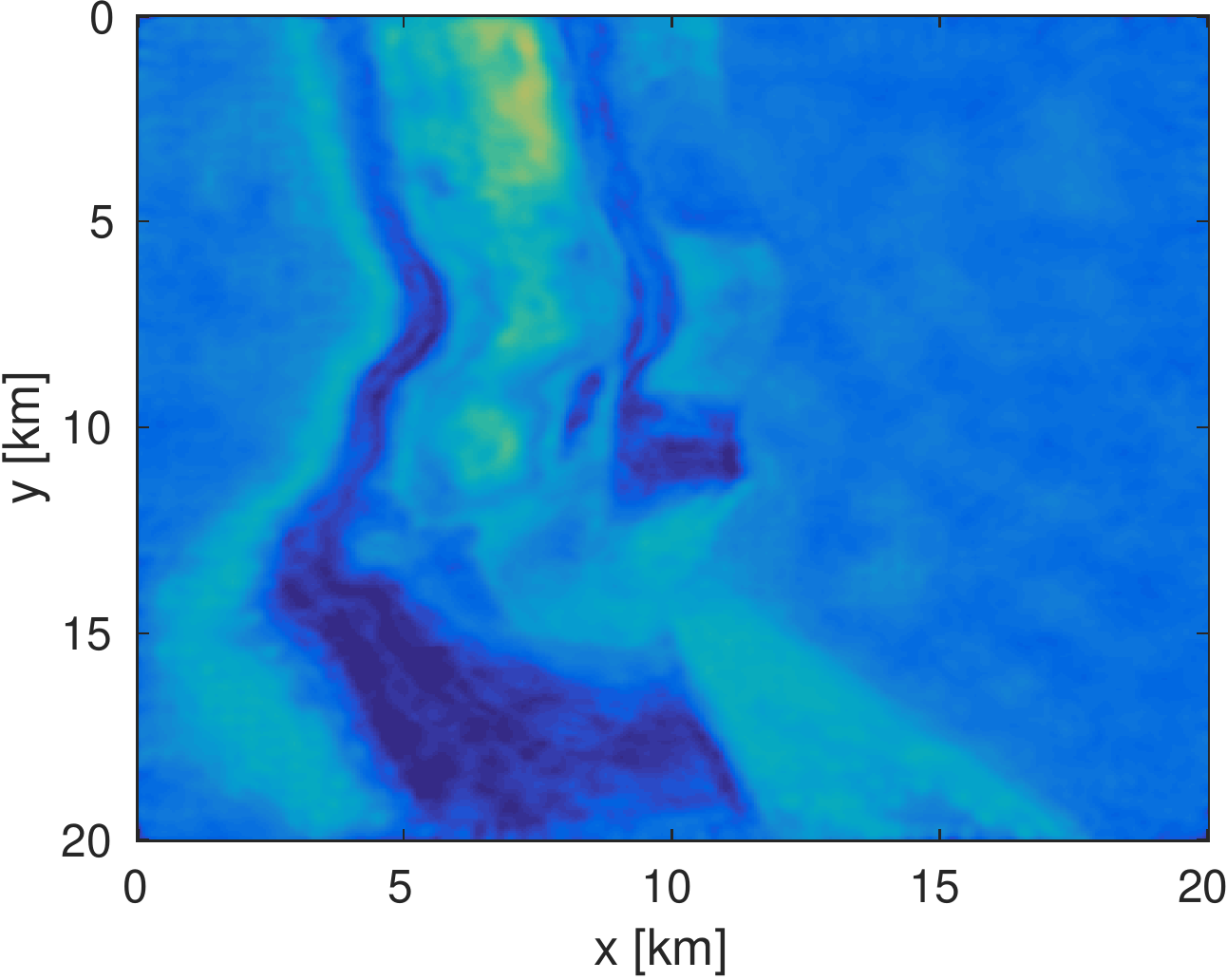}}
\\
\subfloat[z=1000m]{\includegraphics[width=0.330\hsize]{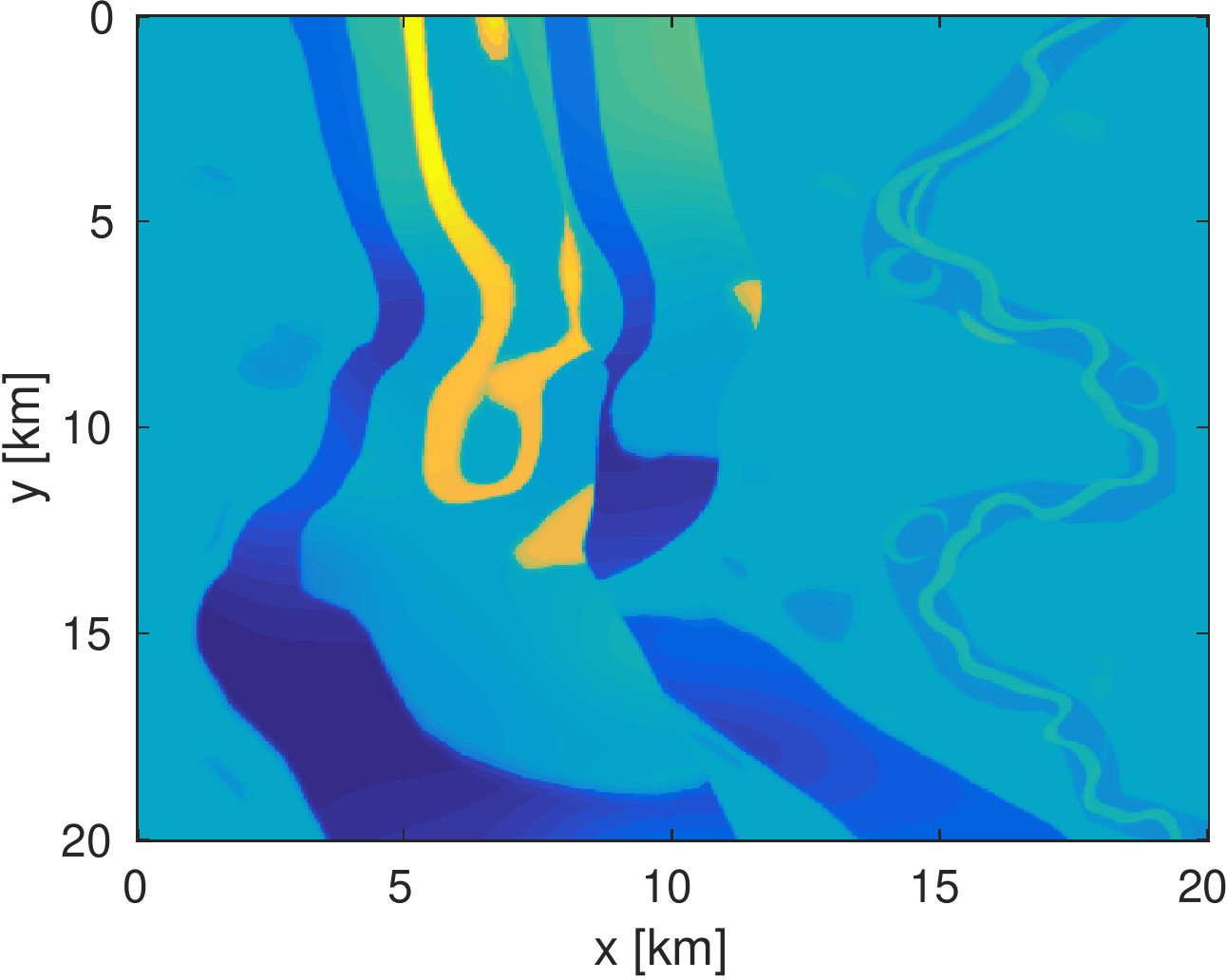}}
\subfloat[z=1000m]{\includegraphics[width=0.330\hsize]{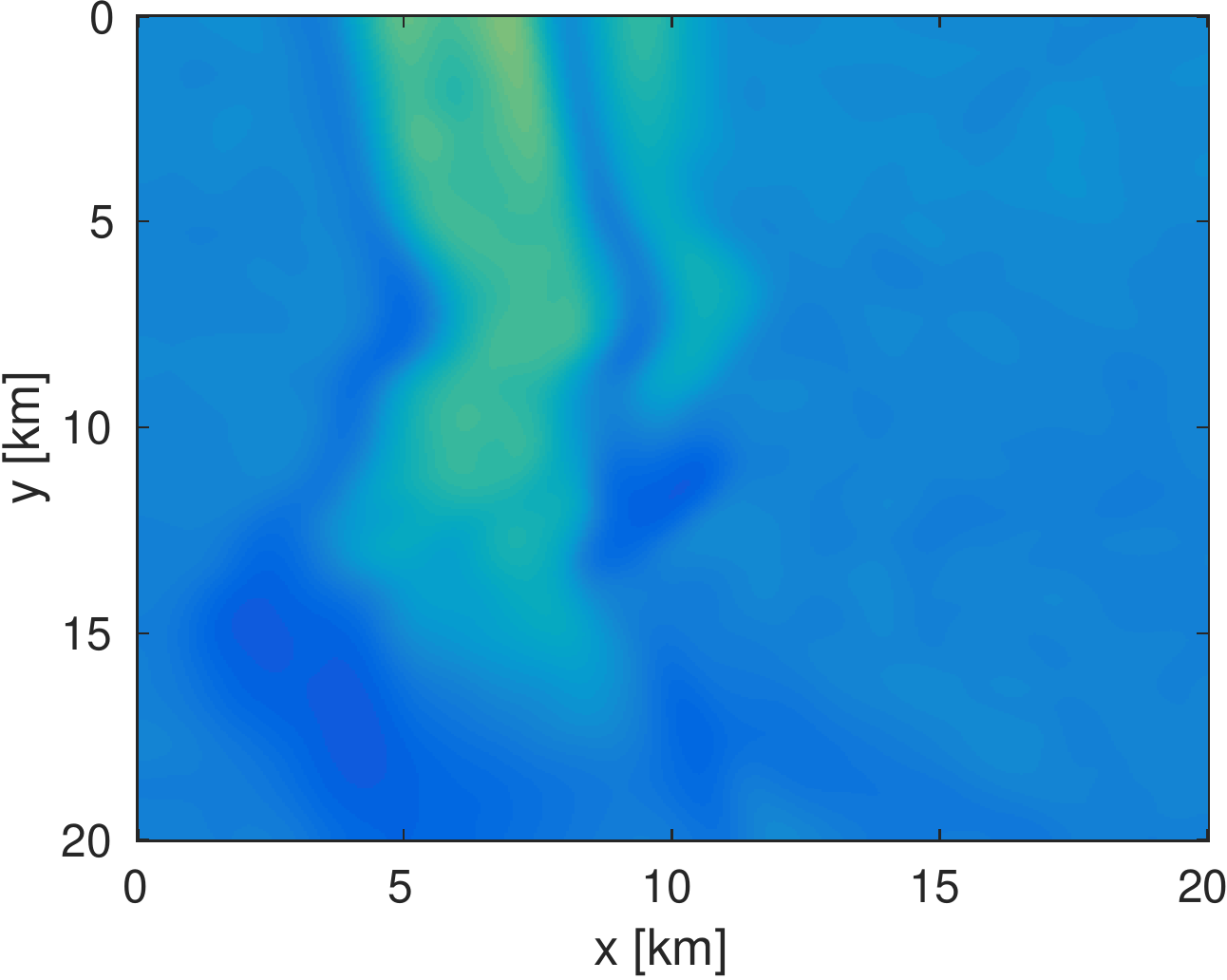}}
\subfloat[z=1000m]{\includegraphics[width=0.330\hsize]{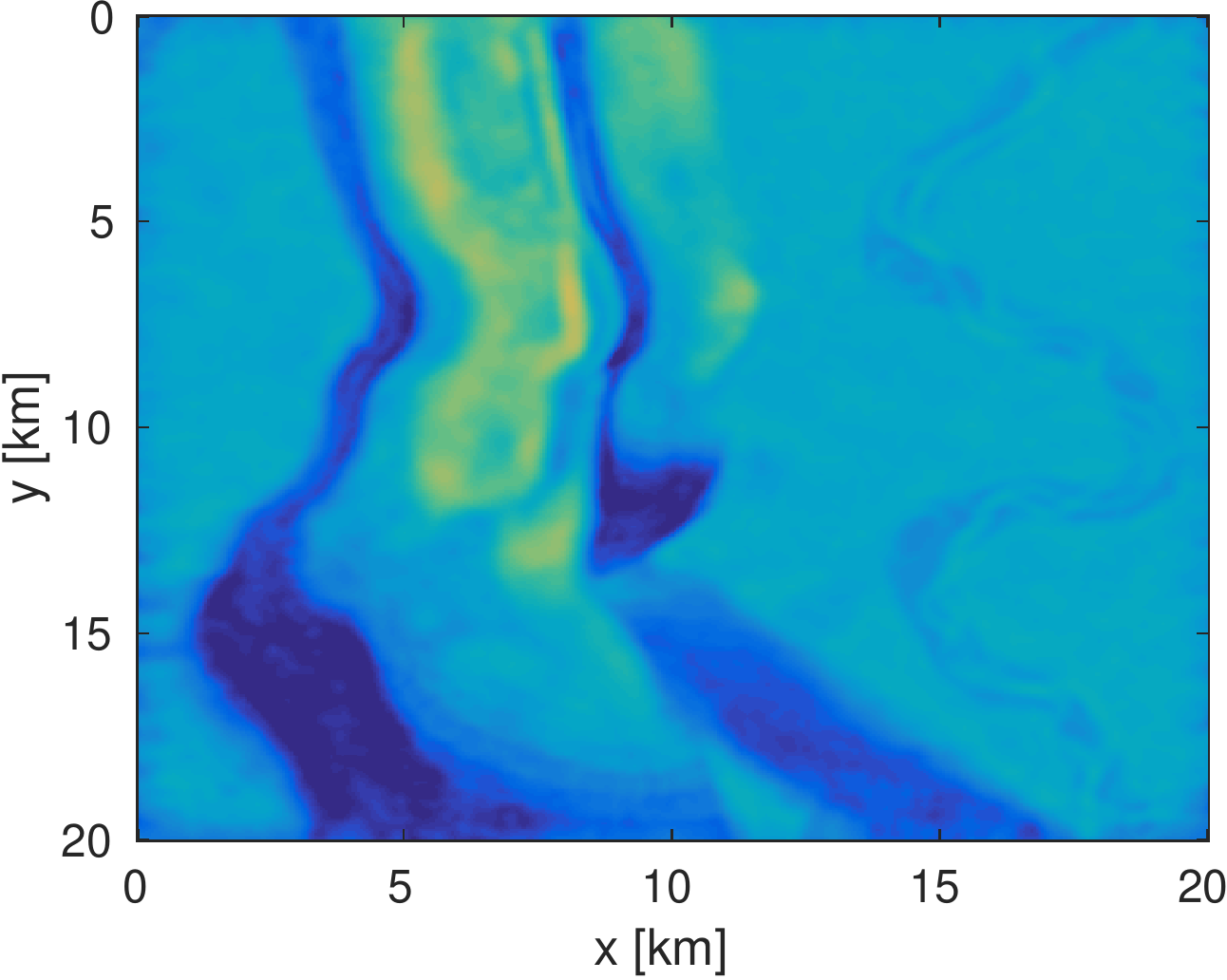}}
\\
\subfloat[z=2000m]{\includegraphics[width=0.330\hsize]{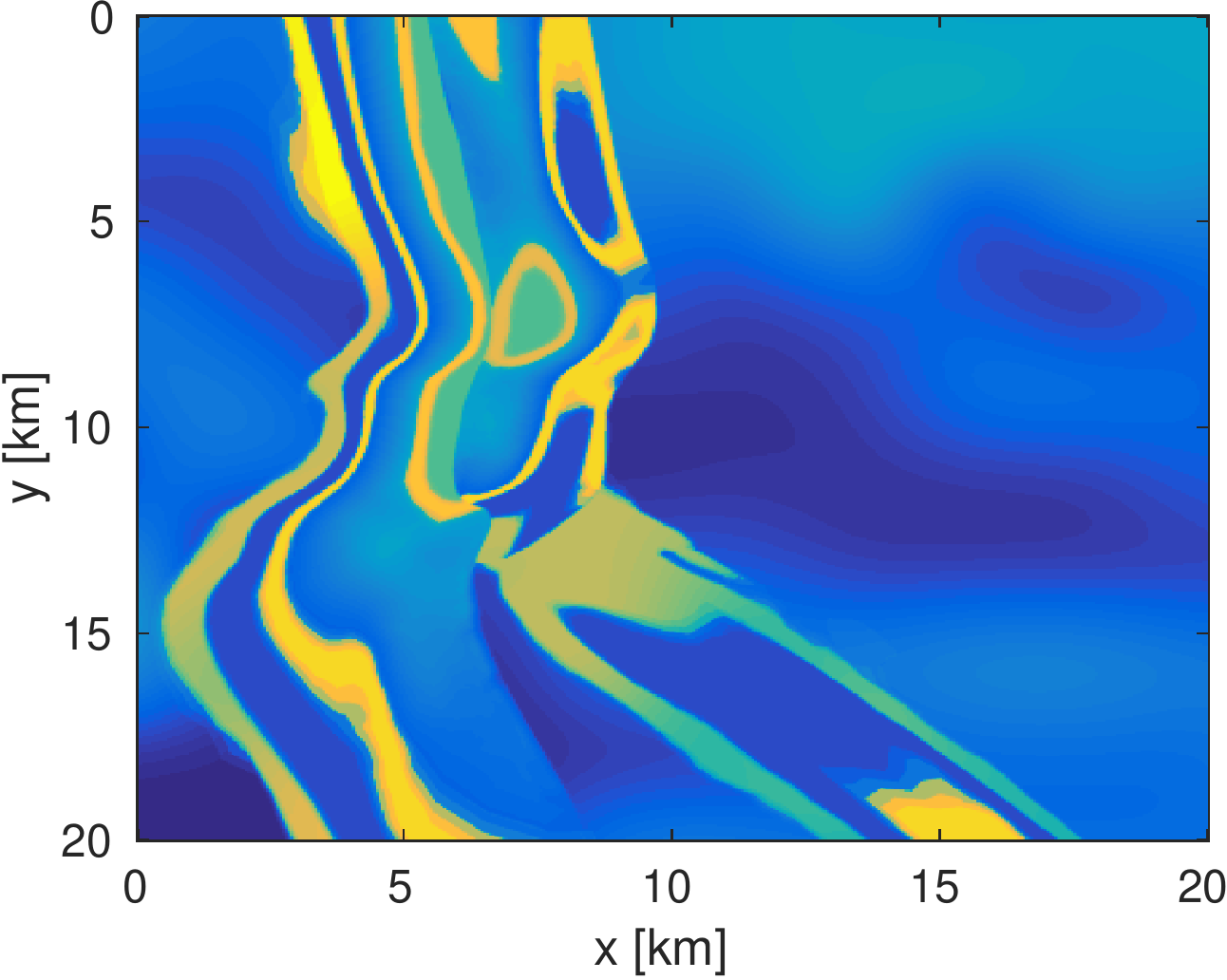}}
\subfloat[z=2000m]{\includegraphics[width=0.330\hsize]{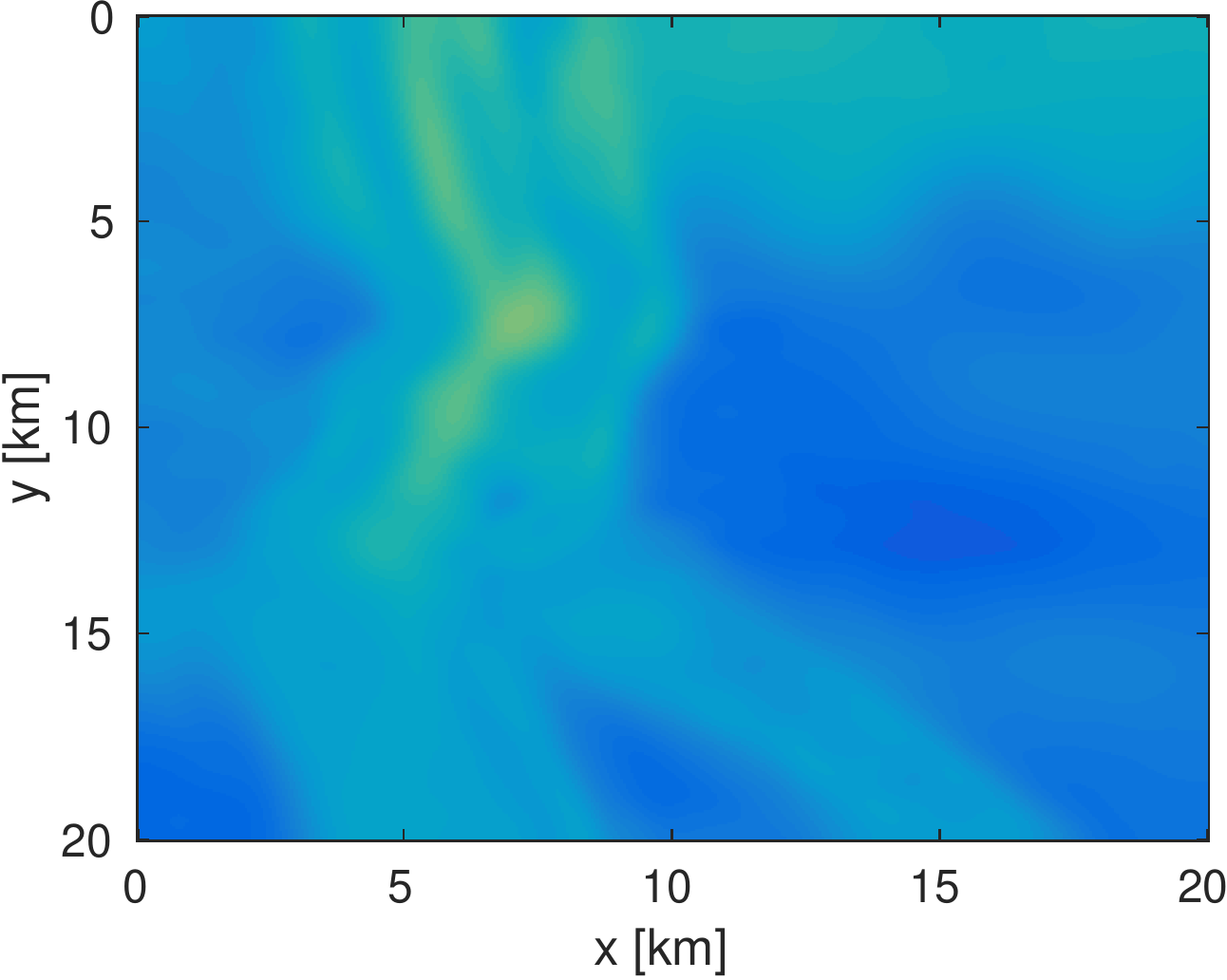}}
\subfloat[z=2000m]{\includegraphics[width=0.330\hsize]{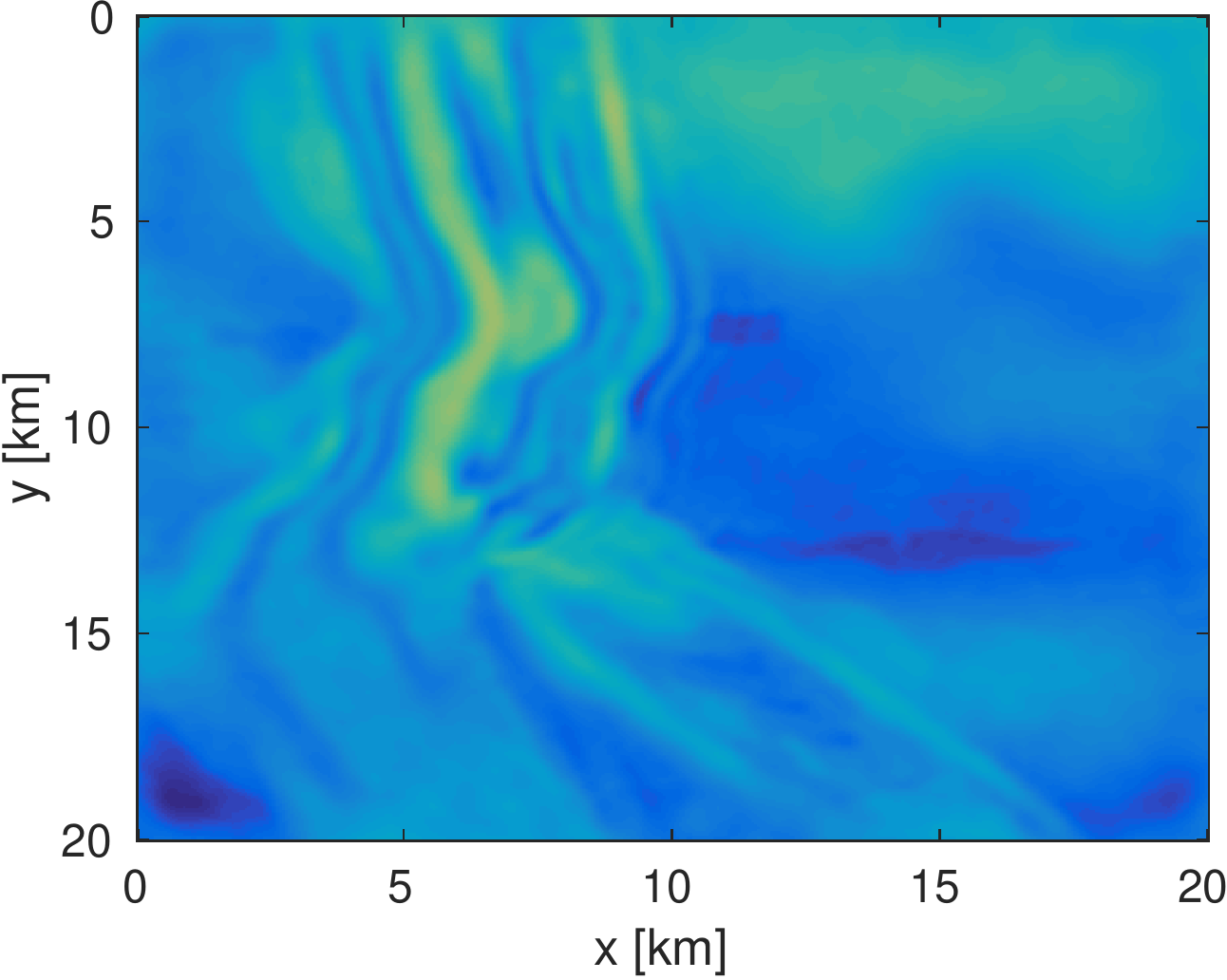}}
\caption{True model (left), initial model (middle), inverted model
(right) for a variety of fixed $z$ coordinate
slices}\label{overthrust_results_z}
\end{figure}

We apply the aforementioned algorithm to the SEG/EAGE Overthrust model,
which spans 20km x 20km x 5km and is discretized on a 50m x 50m x 50m
grid with a 500m deep water layer and minimum and maximum velocities of
$1500 m/s$ and $6000 m/s$. The ocean floor is covered with a 50 x 50
grid of sources, each with 400m spacing, and a 396 x 396 grid of
receivers, each with 50m spacing. The frequencies we use are in the
range of $3-5.5Hz$ with $0.25 Hz$ sampling, corresponding to 4s of data
in the time domain, and inverted a single frequency at a time. The
number of wavelengths in the $x,y,z$ directions vary from $(40,40,10)$
to $(73,73,18)$, respectively, with the number of points per wavelength
varying from 9.8 at the lowest frequency to 5.3 at the highest
frequency. The model is clamped to be equal to the true model in the
water layer and is otherwise allowed to vary between the maximum and
minimum velocity values. In practical problems, some care must be taken
to ensure that discretization discrepancies between the boundary of the
water and the true earth model are minimized. Our initial model is a
significantly smoothed version of the true model and we limit the number
of stochastic redraws to $T=3$, so that we are performing the same
amount of work as evaluating the objective and gradient three times with
the full data. The number of unknown parameters is 14,880,000 and the
fields are inverted on a grid with 39,859,200 points, owing to the PML
layers in each direction. We use 100 nodes with 4 Matlab workers each of
the Yemoja cluster for this computation. Each node has 128GB of RAM and
a 20-core Intel processor. We use 5 threads for each matrix-vector
product and subsample sources so that each Matlab worker solves a single
PDE at a time (i.e., $|I_k| = 400$ in the above formulation) and set the
effective number of passes through the data for each subproblem to be
one, i.e., $r_{\text{sub}}=1$. Despite the limited number of passes
through the data, our code is able to make substantial progress towards
the true model, as shown in Figures~\eqref{overthrust_results_xy}
and~\eqref{overthrust_results_z}. Unlike in the 2D case, we are limited
in our ability to invert higher frequencies in a reasonable amount of
time and therefore our inverted results do not fully resolve the fine
features, especially in the deeper parts of the model.

\begin{figure}
\centering
\captionsetup[subfigure]{labelformat=empty}
\subfloat[]{\includegraphics[width=0.500\hsize]{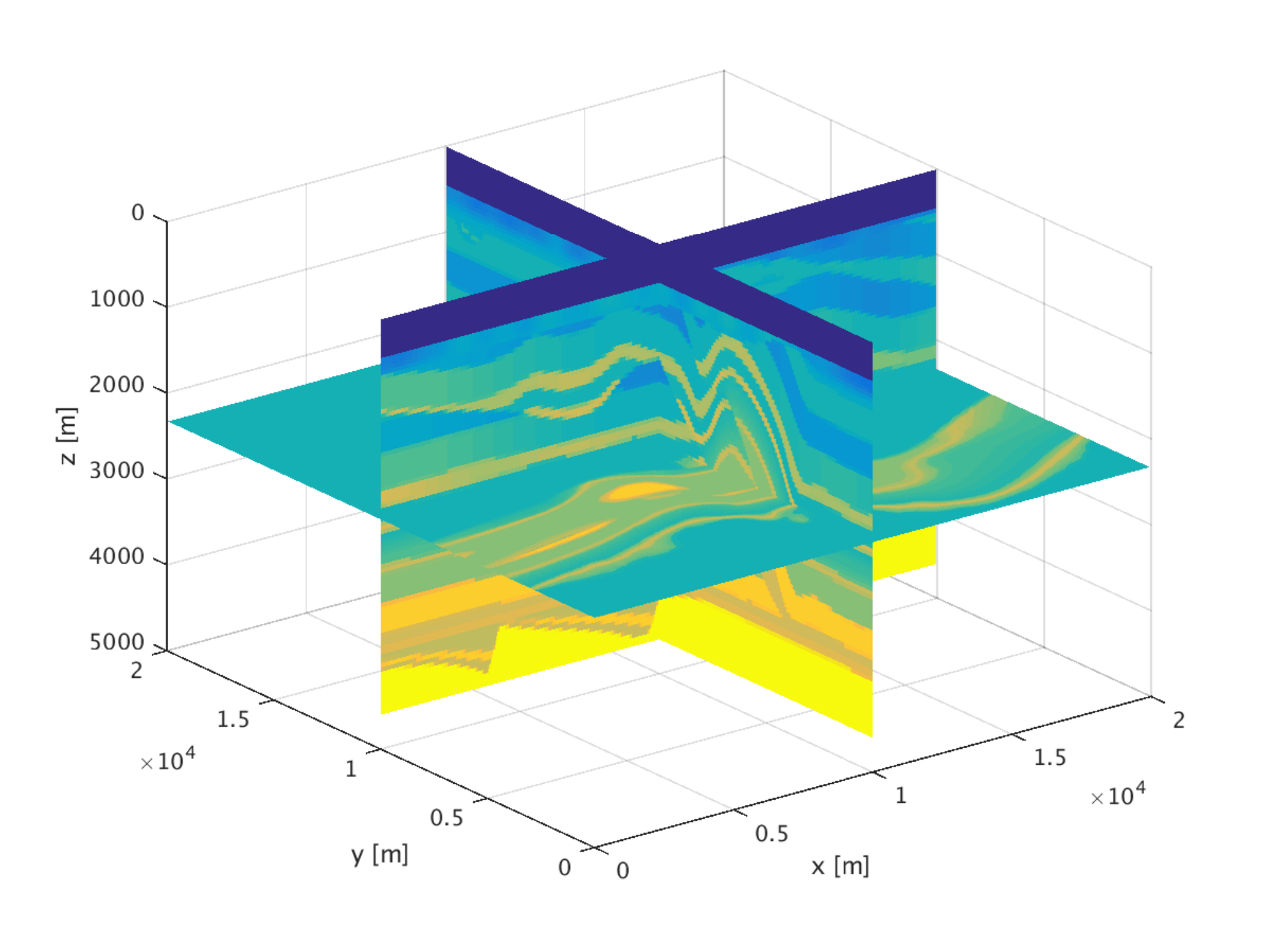}}
\subfloat[]{\includegraphics[width=0.500\hsize]{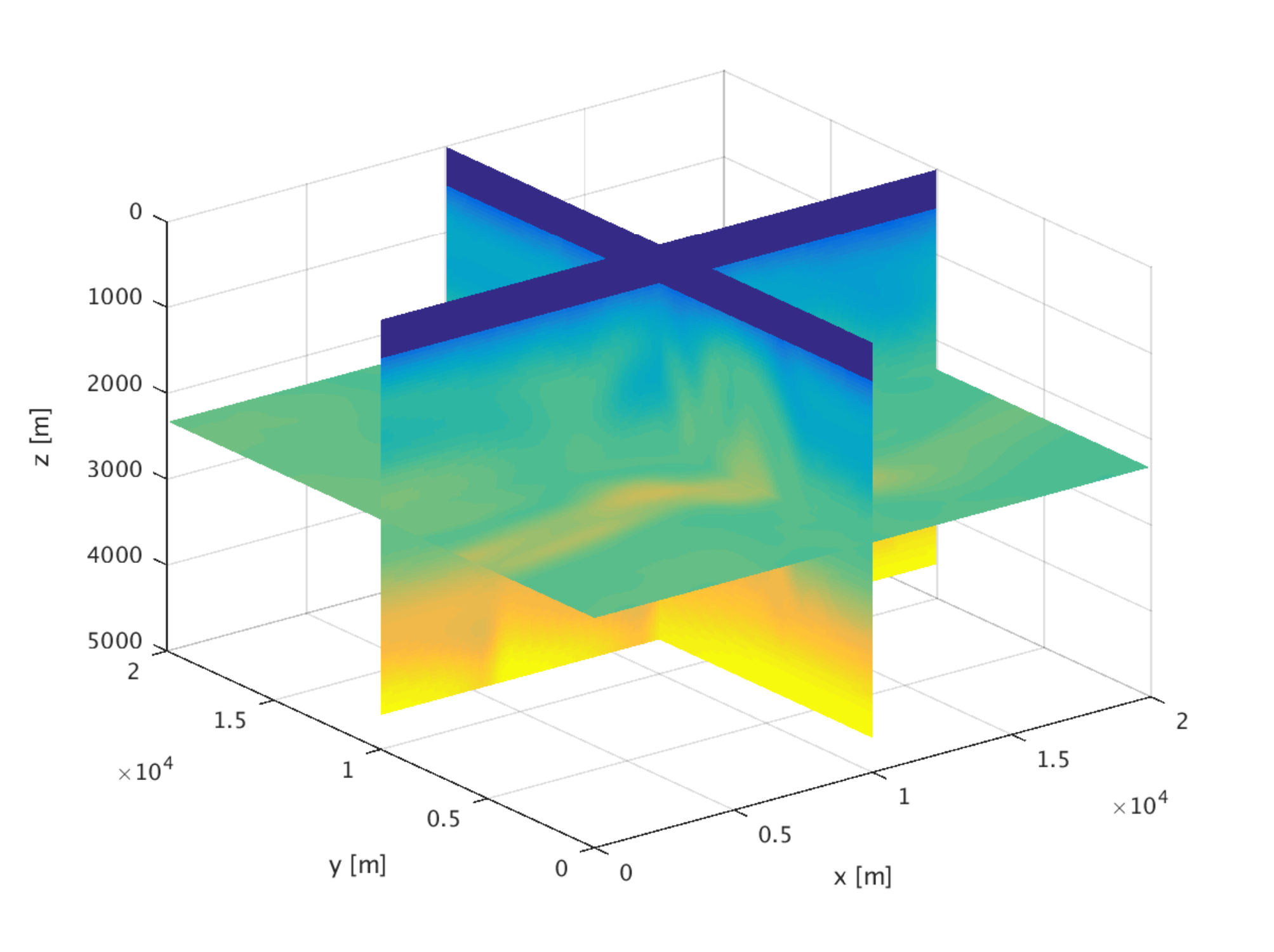}}
\caption{True (left) and initial (right) models}\label{overthrust}
\end{figure}

\section{Discussion}\label{pdediscussion}

There are a number of problem-specific constraints that have motivated
our design decisions thus far. Computing solutions to the Helmholtz
equation in particular is challenging due to the indefiniteness of the
underlying system for even moderate frequencies and the sampling
requirements of a given finite difference stencil. These challenges
preclude one from simply employing the same sparse-matrix techniques in
2D for the 3D case. Direct methods that store even partial LU
decomposition of the system matrix are infeasible from a memory
perspective, unless one allows for using multiple nodes to compute the
PDE solutions. Even in that case, one can run in to resiliency issues in
computational environments where nodes have a non-zero probability of
failure over the lifetime of the computation. Regardless of the
dimensionality, our unified interface for multiplying and dividing the
Helmholtz system with a vector abstracts away the implementation
specific details of the underlying matrix, while still allowing for high
performance. These design choices give us the ability to scale from
simple 2D problems on 4 cores to realistically sized 3D problems running
on 2000 cores with minimal engineering effort.

Although our choice of Matlab has enabled us to succinctly design and
prototype our code, there have been a few stumbling blocks as a result
of this language choice. There is an onerous licensing issue for the
Parallel Toolbox, which makes scaling to a large number of workers
costly. The Parallel Toolbox is built on MPI, which is simply
insufficient for performing large scale computations in an environment
that can be subject to disconnections and node failures and cannot be
swapped out for another parallelization scheme within Matlab itself. In
an environment where one does not have full control over the
computational hardware, such as on Amazon's cloud computing services,
this paradigm is untenable. For interfacing with the C/Fortran language,
there is a large learning curve for compiling MEX files, which are
particularly constructed C files that can be called from with Matlab.
Matlab offers its Matlab Coder product that allows one to, in principle,
compile any function in to a MEX file, and thus reap potential
performance benefits. The compilation process has its limits in terms of
functionality, however, and cannot, for instance, compile our framework
easily.

Thankfully, there have been a few efforts to help alleviate these
issues. The relatively new numerical computing language Julia has
substantially matured in the last few years, making it a viable
competitor to Matlab. Julia aims to bridge the gap between an
interpreted and a compiled language, the former being easier to
prototype in and the latter being much faster by offering a just-in-time
compiler, which balances ease of use and performance. Julia is open
source, whereas Matlab is decidedly not, allows for a much more
fine-grained control over parallelization, and has other attractive
features such as built-in interfacing to C, is strongly-typed, although
flexibly so, and has a large, active package ecosystem. We aim to
reimplement the framework described in this paper in Julia in the
future. The Devito framework \citep{lange2016devito} is another such
approach to balance the high-level mathematics and low-level performance
in PDE-constrained problems specifically through the compilation of
symbolic Python to C on-the-fly. Devito incurs some initial setup time
to process the symbolic expressions of the PDEs and compile them in to
runnable C binaries, but this overhead is negligible compared to the
cost of time-stepping solutions and only has to be performed once for a
PDE with fixed parameters. This is one possible option to speed up
matrix-vector products, or allow for user-specified order of accuracy at
runtime, if the relevant complex-valued extensions can be written. We
leave this as an option for future work.

\section{Conclusion}\label{pdeconclusion}

The designs outlined in this paper make for an inversion framework that
successfully obfuscates the inner workings of the construction,
solution, and recombination of solutions of PDE systems. As a result,
the high-level interfaces exposed to the user allow a researcher to
easily construct algorithms dealing with the outer structure of the
problem, such as stochastic subsampling or solving Newton-type methods,
rather than being hindered by the complexities of, for example, solving
linear systems or distributing computation. This hierarchical and
modular approach allows us to delineate the various components
associated to these computations in a straightforward and demonstrably
correct way, without sacrificing performance. Moreover, we have
demonstrated that this design allows us to easily swap different PDE
stencils, or even PDEs themselves, while still keeping the outer,
high-level interfaces intact. This design allows us to apply a large
number high-level algorithms to large-scale problems with minimal
amounts of effort.

With this design, we believe that we have struck the right balance
between readability and performance and, by exposing the right amount of
information at each level of the software hierarchy, researchers should
be able to use this codebase as a starting point for developing future
inversion algorithms.

\section{Acknowledgements}\label{acknowledgements}

The authors would like to thank Tristan van Leeuwan for his helpful
suggestions. This research was carried out as part of the SINBAD project
with the support of the member organizations of the SINBAD Consortium.
Curt Da Silva has been supported by the CGS D award from the National
Sciences and Engineering Research Council of Canada (NSERC) throughout
the duration of this work. The authors wish to acknowledge the SENAI
CIMATEC Supercomputing Center for Industrial Innovation, with support
from Shell and the Brazilian Authority for Oil, Gas and Biofuels (ANP),
for the provision and operation of computational facilities and the
commitment to invest in Research \& Development.

\section{Appendix}\label{appendix}

\subsection{\texorpdfstring{A `User Friendly' Guide to Basic Inverse
Problems}{A User Friendly Guide to Basic Inverse Problems}}\label{userfriendly}

For our framework, not only do we want to solve~\eqref{mainprob}
directly, but allow researchers to explore other subproblems associated
to the primary problem, such as the linearized problem
\citep{herrmann2011EAGEefmsp} and the full-Newton trust region
subproblem. As such, we are not only interested in the objective
function and gradient, but also other intermediate quantities based on
differentiating the state equation $H(m)u(m) = q$. A standard approach
to deriving these quantities is the adjoint-state approach, described
for instance in \citep{plessix2006review}, but the results we outline
below are slightly more elementary and do not make use of Lagrange
multipliers.

Rather than focusing on differentiating the expressions
in~\eqref{mainprob} directly, we find it useful to consider
\textbf{directional derivatives} various quantities and their
relationships to the gradient. That is to say, for a sufficiently smooth
function $f(m)$ that can be scalar, vector, or matrix-valued, the
directional dervative in the direction $\delta m$, denoted
$Df(m)[\delta m]$, is a linear function of $\delta m$ that satisfies
\begin{equation*}
    Df(m)[\delta m] = \lim_{t \to 0} \dfrac{f(m+t\delta m)-f(m)}{t}.
\end{equation*}
 The most important part of the directional derivative, for the purposes
of the following calculations, is that $Df(m)[\delta m]$ is the same
mathematical object as $f(m)$, i.e., if $f(m)$ is a matrix, so is
$Df(m)[\delta m]$.

For a given inner product $\langle \cdot, \cdot \rangle$, the gradient
of $f(m)$, denoted $\nabla f(m)$, is the unique vector that satisfies
\begin{equation*}
    \langle \nabla f(m), \delta m\rangle = Df(m)[\delta m] \quad \forall \delta m
\end{equation*}
 If we are not terribly worried about specifying the exact vectors
spaces in which these objects live and treat them as we'd expect them to
behave (i.e., satisfying product, chain rules, commuting with matrix
transposes, complex conjugation, linear operators, etc.), the resulting
derivations become much more manageable.

Starting from the baseline expression for the misfit $f(m)$ and
differentiating, using the chain rule, we have that
\begin{equation*}
\begin{aligned}
    f(m) &= \phi(P_r H(m)^{-1}q, d) = \phi(P_r u(m),d)\\
    Df(m)[\delta m] &= D\phi(r(m),d)[Dr(m)[\delta m]]\\
    r(m) &= P_r u(m)\\
    Dr(m)[\delta m] &= P_r Du(m)[\delta m]
\end{aligned}
\end{equation*}
 In order to determine the expression for $Du(m)[\delta m]$, we
differentiate the state equation,
\begin{equation*}
    H(m)u(m) = q,
\end{equation*}
 in the direction $\delta m$ using the product rule to obtain
\begin{equation}
\begin{split}
    DH(m)[\delta m]u(m) + H(m)Du(m)[\delta m] = 0\\
    Du(m)[\delta m] = H(m)^{-1} (-DH(m)[\delta m]u(m)).
\end{split}
\label{dU2}
\end{equation}
 For the forward modelling operator $F(m)$, $DF(m)[\delta m]$ is the
Jacobian or so-called linearized born-modelling operator in geophysical
parlance. Since, for any linear operator $A$, its transpose satisfies
\begin{equation*}
    \langle Ax, y\rangle = \langle x, A^T y\rangle
\end{equation*}
 for any appropriately sized vectors $x, y$, in order to determine the
transpose of $Dr(m)[\delta m]$, we merely take the inner product with an
arbitrary vector $y$, ``isolate'' the vector $\delta m$ on one side of
the inner product. The other side is the expression for the adjoint
operator. In this case, we have that
\begin{equation*}
\begin{aligned}
    \langle DF(m)[\delta m], y \rangle &= \langle P_r H(m)^{-1}(-DH(m)[\delta m]u(m)), y \rangle\\
    &= \langle H(m)^{-1}(-DH(m)[\delta m]u(m)),P_r^T y\rangle\\
    &= \langle DH(m)[\delta m]u(m), -H(m)^{-H} P_r^T y \rangle\\
    &= \langle T\delta m, -H(m)^{-H} P_r^T y \rangle\\
    &= \langle \delta m, T^* (-H(m)^{-H} P_r^T y) \rangle
\end{aligned}
\end{equation*}
 In order to completely specify the adjoint of $DF(m)[\delta m]$, we
need to specify the adjoint of $T\delta m = DH(m)[\delta m]u(m)$ acting
on a vector. This expression is particular to the form of the PDE with
which we're working. For instance, discretizing the constant-density
acoustic Helmholtz equation with finite differences results in
\begin{equation*}
    \nabla^2u(x) + \omega^2 m(x)u(x) = q(x)
\end{equation*}
 with particular matrices $L, A$ discretizing the Laplacian and identity
operators, respectively, yields the linear system
\begin{equation*}
    Lu + \omega^2 A\text{diag}(m)u = q.
\end{equation*}
 Therefore, we have the expression
\begin{equation}
\begin{aligned}
T \delta m &:= DH(m)[\delta m]u(m)\\
&= \omega^2 A\text{diag}(\delta m)u(m)\\
&= \omega^2 A\text{diag}(u(m))\delta m,
\end{aligned}
\label{dH}
\end{equation}
 whose adjoint is clearly
\begin{equation}
    T^* = \omega^2\text{diag}(\overline{u(m)})A^H
\label{dHadj}
\end{equation}
 with directional derivative
\begin{equation}
  DT^*[\delta m, \delta u] = \omega^2\text{diag}(\overline{\delta u})A^H.
\label{DTadj}
\end{equation}
 Our final expression for $DF(m)[\cdot]^*y$ is therefore
\begin{equation*}
    DF(m)[\cdot]^*y = \omega^2 \text{diag}(\overline{u(m)})A^H (-H(m)^{-H} P_r^T y)
\end{equation*}
 Setting $y = \nabla \phi(P_r u(m))$, $v(m) = -H(m)^{-H} P_r^T y$ yields
the familiar expression for the gradient of $f(m)$
\begin{equation*}
    \nabla f(m) = \omega^2 \text{diag}(\overline{u(m)})A^H H(m)^{-H}P_r^T (-\nabla \phi(u(m)))
\end{equation*}
 This sort of methodology can be used to symbolically differentiate more
complicated expressions as well as compute higher order derivatives of
$f(m)$. Let us write $\nabla f(m)$ more abstractly as
\begin{equation*}
\nabla f(m) = T(m,u)^* v(m),
\end{equation*}
 which will allow us to compute the Hessian as
\begin{equation*}
\nabla^2 f(m)\delta m = DT(m,u)^*[\delta m,Du[\delta m]] v(m) + T(m,u)^* Dv(m)[\delta m].
\end{equation*}
 Here, $Du[\delta m]$ is given in~\eqref{dU2} and
$DT(m,u)^*[\delta m, \delta u]$ is given in~\eqref{DTadj}, for the
Helmholtz equation. We compute $Dv(m)[\delta m]$ by differentiating
$v(m)$ as
\begin{equation*}
\begin{aligned}
Dv(m)[\delta m] &= H(m)^{-H} DH(m)[\delta m]^H v(m) -H(m)^{-H} P_r^T (\nabla^2 \phi[P_r Du(m)[\delta m]])\\
&= H(m)^{-H}( DH(m)[\delta m]^H v(m) - P_r^T (\nabla^2 \phi[P_r Du(m)[\delta m]]) )
\end{aligned}
\end{equation*}
 which completes our derivation for the Hessian-vector product.

\subsection{Derivative expressions for Waveform Reconstruction
Inversion}\label{wrideriv}

From~\eqref{wrifieldeq}, we have that the augmented wavefield $u(m)$
solves the least-squares system
\begin{equation*}
\begin{split}
\min_{u} \left \| \left[ {\begin{array}{c} P_r \\ \lambda H(m) \end{array} } \right] u - \left[ { \begin{array}{c} d \\ \lambda q \end{array} } \right] \right\|_2^2,
\end{split}
\end{equation*}
 i.e., $u(m)$ solves the normal equations
\begin{equation*}
(P_r^T P_r + \lambda^2 H(m)^H H(m))u(m) = P_r^T d + \lambda^2 H(m)^H q.
\end{equation*}
 For the objective
$\phi(m,u) = \dfrac{1}{2} \|P_r u - d\|_2^2 + \dfrac{\lambda^2}{2} \|H(m)u - q\|_2^2$,
the corresponding WRI objective is $f(m) = \phi(m,u(m))$. Owing to the
variable projection structure of this objective, the expression for
$\nabla_m f(m)$, by \citep{aravkin2012estimating}, is
\begin{equation*}
\begin{aligned}
\nabla_m f(m) &= \nabla_m \phi(m,u(m)) \\
&= \lambda^2 T(m,u(m))^* (H(m)u(m) - q),
\end{aligned}
\end{equation*}
 which is identical to the original adjoint-state formulation, except
evaluated at the wavefield $u(m)$.

The Hessian-vector product is therefore
\begin{equation*}
\nabla^2_m f(m) = DT(m,u(m))[\delta m, Du(m)[\delta m]]^* (H(m)u(m)-q) + T(m,u(m))^* (DH(m)[\delta m]u(m) + H(m)Du(m)[\delta m]).
\end{equation*}
 As previously, the expressions for $DH(m)[\delta m]$ and
$DT(m,u)[\delta m, \delta u]^*$ are implementation specific. It remains
to derive an explicit expression for $Du(m)[\delta m]$ below.

Let $G(m) = (P_r^T P_r + \lambda^2 H(m)^H H(m))$,
$r(m) = P_r^T d + \lambda^2 H(m)^H q$, so the above equation reads as
$G(m)u(m) = r(m)$.

We differentiate this equation in the direction $\delta m$ to obtain
\begin{equation*}
\begin{split}
DG(m)[\delta m] u(m) + G(m)Du(m)[\delta m] = Dr(m)[\delta m] \\
\rightarrow Du(m)[\delta m] = G(m)^{-1} (Dr(m)[\delta m] - DG(m)[\delta m]u(m)).
\end{split}
\end{equation*}
 Since
$DG(m)[\delta m] = \lambda^2 (DH(m)[\delta m]^{H} H(m) + H(m)^HDH(m)[\delta m])$
and $Dr(m)[\delta m] = \lambda^2 DH(m)[\delta m]^H q$, we have that
\begin{equation*}
Du(m)[\delta m] = \lambda^2 G(m)^{-1}( - H(m)^H DH(m)[\delta m]u(m) + DH(m)[\delta m]^H( H(m)u(m)- q) ).
\end{equation*}
%


\end{document}